\documentclass[nofootinbib,floatfix,aps,reprint,amsmath,amssymb,twocolumn,superscriptaddress,longbibliography]{revtex4-2}
\usepackage{amsmath}
\usepackage{amssymb}
\usepackage{float}
\usepackage{graphicx}
\usepackage{xr}
\usepackage[dvipsnames]{xcolor}
\usepackage[colorlinks = true,
            allcolors = {Blue}]{hyperref}
\newcommand{\nn}{\nonumber\\}
\newcommand{\rhoS}{\rho}

\makeatletter
\renewcommand\frontmatter@abstractwidth{\dimexpr\textwidth\relax}
\makeatother

\newcommand{\sect}[1]{\vspace{3mm}\noindent\textbf{\large #1}}
\newcommand{\secnl}[1]{\sect{#1}\\\noindent}
\renewcommand{\subsection}[1]{\leavevmode\\\noindent\textbf{#1.}}

\begin{document}
\title{Numerically exact open quantum systems simulations for arbitrary environments using automated compression of environments}
\author{Moritz Cygorek}
\affiliation{SUPA, Institute of Photonics and Quantum Sciences, Heriot-Watt University, Edinburgh, EH14 4AS, United Kingdom}
\author{Michael Cosacchi}
\affiliation{Theoretische Physik III, Universit\"at Bayreuth, 95440 Bayreuth, Germany}
\author{Alexei Vagov}
\affiliation{Theoretische Physik III, Universit\"at Bayreuth, 95440 Bayreuth, Germany}
\author{Vollrath~Martin~Axt}
\affiliation{Theoretische Physik III, Universit\"at Bayreuth, 95440 Bayreuth, Germany}
\author{Brendon W. Lovett}
\affiliation{SUPA, School of Physics and Astronomy, University of St Andrews, St Andrews, KY16 9SS, United Kingdom}
\author{Jonathan Keeling}
\affiliation{SUPA, School of Physics and Astronomy, University of St Andrews, St Andrews, KY16 9SS, United Kingdom}
\author{Erik M. Gauger}
\affiliation{SUPA, Institute of Photonics and Quantum Sciences, Heriot-Watt University, Edinburgh, EH14 4AS, United Kingdom}

\begin{abstract}
\textbf{
The central challenge for describing the dynamics in open quantum systems is that
the Hilbert space of typical environments is too large to be treated
exactly.
In some  cases, such as when the environment has a short memory time or only interacts weakly with the system, approximate descriptions of the system are possible.  
Beyond these, numerically exact methods exist, 
but these are typically restricted to baths with Gaussian correlations, such as non-interacting bosons.
Here we present a numerically exact method for simulating open quantum systems with arbitrary environments which consist of a set of independent degrees of freedom.
Our approach automatically reduces the large number of environmental degrees of freedom to those which are most relevant.
Specifically, we show how the process tensor---which describes the effect of the environment---can be iteratively constructed and compressed using matrix product state techniques. 
We demonstrate the power of this method by applying it to problems with bosonic, fermionic, and spin environments: electron transport, phonon effects and radiative decay in quantum dots, central spin dynamics, anharmonic environments, dispersive coupling to time-dependent lossy cavity modes, and superradiance.
The versatility and efficiency of our 
automated compression of environments (ACE) method 
provides a practical general-purpose tool for open quantum systems.
}
\end{abstract}
\maketitle

An inevitable property of quantum technologies is that quantum devices
interact with their environment~\cite{Breuer}.
This interaction gives rise to dephasing and dissipation but, if understood, it can be exploited for 
example in
environment-assisted quantum transport~\cite{Plenio2008,Rebentrost2009,Chin2010},
or even quantum information processing~\cite{Beige2000,Verstraete2009quantum}.
Because of the exponential growth of Hilbert space dimension, and the large number of environmental degrees of freedom, the direct solution of Schr\"odinger's equation for system and environment is usually infeasible.
As such, one requires practical methods that allow simulation of the dynamics
of the system, while accounting for effects of the environment~\cite{Breuer,RevModPhys_deVega,Review_stochastic,RevModPhys_jump}.

Among such approaches, those most frequently used  rely on the Born and Markov approximations, which enable one to derive time-local equations of motion for the reduced system density matrix~\cite{Breuer,Redfield}.
The Born approximation implies that the environment 
does not change significantly with time---i.e.~that system-environment
correlations are weak and transient.
While valid for weakly coupled open quantum systems, 
other environments lead to  strong
system-environment correlations~\cite{NazirReview}. 
The Markov approximation depends on the 
memory time of the environment being short compared to the time evolution of the system. 
This fails if the spectral density is highly structured, or if there is a long memory time~\cite{Breuer2016}.
Given these widespread limitations, approaches beyond the Born--Markov approximation are clearly necessary.

\begin{figure*}
\includegraphics[width=0.999\linewidth]{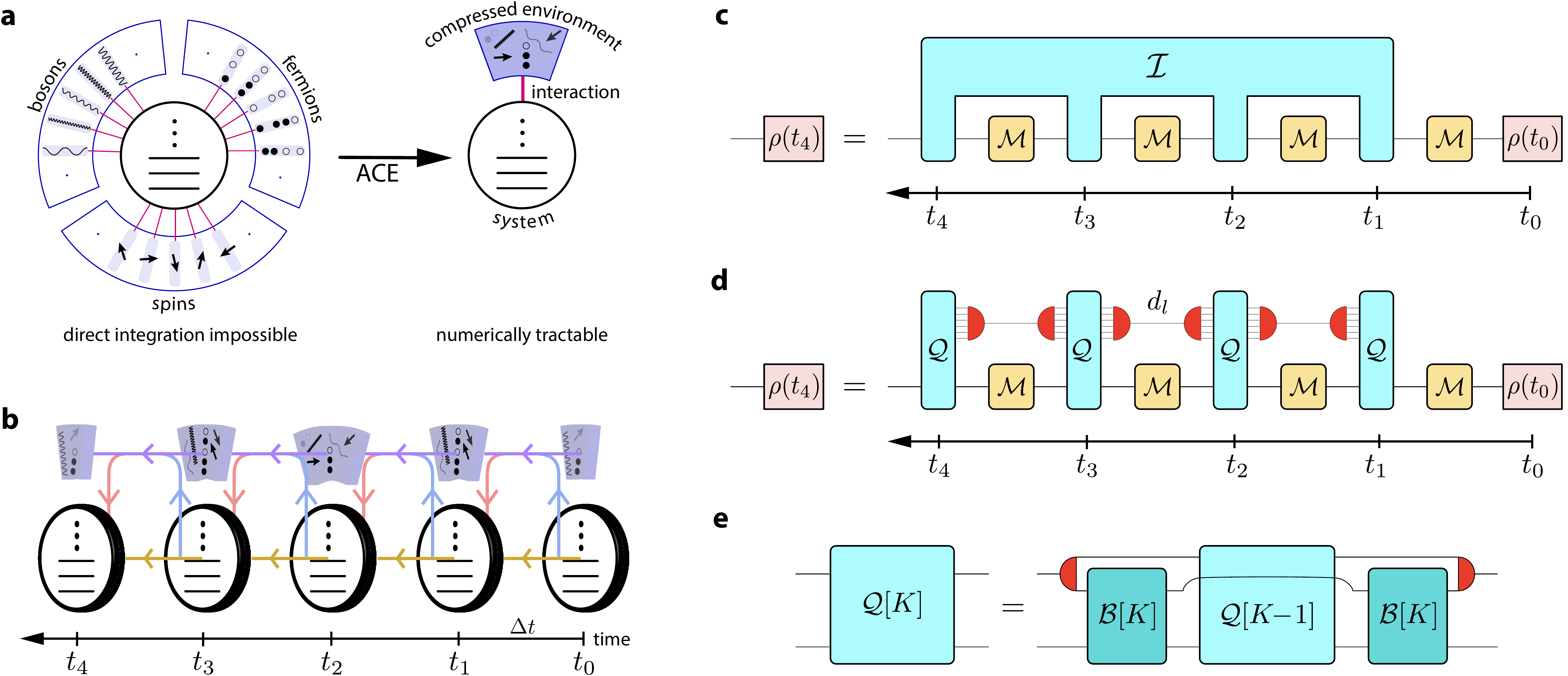}
\caption{\label{fig:sketchMPS}
\textbf{Depiction of the automated compression of environments approach.}
\textbf{a,} The identification of an efficient representation is fully automatic and does not rely on any a priori approximations or assumptions.
\textbf{b,} The time evolution of system plus its compressed environment proceeds in discrete time steps. Information flow is indicated by the coloured arrows.
\textbf{c,} Formally, the general propagation of a quantum system can be expressed with a process tensor~$\mathcal{I}$.
\textbf{d,}~Propagation with a process tensor in MPO form: this corresponds to the schematic situation depicted in panel b.
\textbf{e,}~Combination of the influence of environment mode $K$ with the 
process tensor containing the influences of modes $1,2,\dots, K-1$. 
Red semicircles indicate the effects of the MPO compression (as depicted schematically in panels a and b).
}
\end{figure*}

Numerically exact methods---where tuning convergence parameters allows one to trade off precision against computation time---do exist for some non-Markovian problems: those  where the environments have Gaussian correlations,  such as non-interacting bosonic modes.
Such methods include  hierarchical equations
of motion (HEOM)~\cite{HEOM89,HEOMreview20}, chain mapping through orthogonal polynomials~\cite{TEDOPA_PRL,Somoza2019,fTEDOPA}, or the Feynman-Vernon real-time path integral formalism~\cite{FeynmanVernon}.
In particular, the iterative form of the path integral~\cite{Makri,Makri_Iterative2,PI_cQED} and its reformulation with matrix product operators~\cite{TEMPO} have been used successfully, e.g.,
to simulate phonon effects on spectra~\cite{PI_multitime,Denning_phonon_decoupling}, 
to devise robust and high-fidelity protocols for the emission of nonclassical light~\cite{PRL_singlephoton, PRL_concurrence, ExtTEMPOKaestle}, and to model concrete experiments on optically driven quantum dots~\cite{PI_2011,PRLQuilter,dichromatic}.
Such approaches have been extended to systems with multiple environments~\cite{Nalbach_multiEnv}, to multi-level systems~\cite{PI_cQED}, and to special types of non-Gaussian baths such as quadratic coupling to bosons or fermions~\cite{Segal_quadratic_ferm_bos}.
Some methods for general environments do exist, such as correlation expansion~\cite{RossiKuhn2002}, but it is complicated to derive these equations at higher expansion order.
As such, a challenge remains: to provide general and efficient numerically exact methods which can also model non-Gaussian non-Markovian environments.

Here we provide such a method, which can be used to simulate  open quantum systems coupled to arbitrary environments (see Fig.~\ref{fig:sketchMPS}a). We demonstrate its practical application with a variety of forms of environment---bosonic, fermionic, and spins.  
Because the derivation is general the same code can be used to simulate the dynamics of a large variety of different physical systems.
At the core of our \emph{automated compression of environments} (ACE) method is the explicit microscopic construction of the process tensor~(PT)~\cite{ProcessTensor,ProcessTensor_PRA}---an object originally conceived  as a way to conceptualize correlations for a general non-Markovian environment---and a route to efficiently compress this object using matrix product operator (MPO) techniques~\cite{MPS_Schollwoeck,MPS_Orus}. 
Specifically, we provide a general and efficient algorithm to directly construct an MPO representation of the PT,  corresponding to an automated projection of the environment onto its most relevant degrees of freedom.

\sect{Results}
\subsection{Automated compression of environments}
The working principle of ACE is to represent the environment efficiently by concentrating on its most relevant degrees of freedom (cf. Fig.~\ref{fig:sketchMPS}a). These are selected automatically using MPO compression techniques and may differ from one time step to another. This procedure guarantees fully capturing the non-Markovian information flow from past time steps to later time steps via the environment
(cf. Fig.~\ref{fig:sketchMPS}b).  
We now summarise the ACE method introduced in this paper; further details are provided in the 
Methods section.
Our goal is to obtain the reduced system density matrix $\rhoS_{\nu\mu}(t)$ at a time $t$, accounting for coupling to  a given environment.
We discretise the time axis on a grid $t_l=l \Delta t$ with equal time steps
$\Delta t$ (Fig.~\ref{fig:sketchMPS}b-d); then, for a single time step, the time evolution operator $U(\Delta t)=e^{-\frac i\hbar H \Delta t}$ of the 
total system can be factorised using the Trotter expansion 
$U(\Delta t)=e^{-\frac i\hbar H_E\Delta t} e^{-\frac i\hbar H_S\Delta t}
+\mathcal{O}(\Delta t^2)$, where the total Hamiltonian $H=H_S+H_E$ is decomposed into the system Hamiltonian $H_S$ and the environment Hamiltonian $H_E$ including the system-environment coupling.
Inserting a complete set of basis states for the system and the environment
and tracing out the environment, the reduced system density matrix at
time $t_n$ can be written
\begin{align}
\rhoS_{\alpha_n}=
\sum_{\substack{\alpha_{n-1}\dots\alpha_0 \\
\tilde{\alpha}_n\dots\tilde{\alpha}_1}}
\mathcal{I}^{(\alpha_{n}\tilde{\alpha}_n)\dots(\alpha_1\tilde{\alpha}_1)}
\bigg(\prod_{l=1}^{n}
\mathcal{M}^{\tilde{\alpha}_{l}\alpha_{l-1}} \bigg)
\rhoS_{\alpha_0},
\label{eq:rho_alpha}
\end{align}
where we have defined $\alpha=(\nu,\mu)$ to combine two Hilbert space indices into a single Liouville space index.
A visual representation of Eq.~\eqref{eq:rho_alpha} is depicted in 
Fig.~\ref{fig:sketchMPS}c.
Here, $\mathcal{M}$ describes the free propagation of the system. This can be time-dependent, and can additionally include effects of Markovian baths. 
The effects of the general non-Markovian non-Gaussian
environment are captured in the quantity $\mathcal{I}$, which we refer to as 
the process tensor (PT).  This object differs slightly from the original definition of the PT~\cite{ProcessTensor_PRA}, in that we have separated out the initial state and the free system evolution.
When $\mathcal{I}$ is non-zero only
for diagonal couplings $\alpha_l=\tilde{\alpha}_l$ 
this object becomes equivalent to the Feynman-Vernon influence 
functional~\cite{FeynmanVernon}. The PT can thus be considered as a generalisation of this influence functional to the case of non-diagonal couplings.
From the explicit expression for the PT we find that it automatically has the form of an MPO:
\begin{align}
&\mathcal{I}^{(\alpha_{n},\tilde{\alpha}_{n})(\alpha_{n-1},\tilde{\alpha}_{n-1})
\dots (\alpha_{1},\tilde{\alpha}_{1})}
= \nn&
\sum_{d_{n-1}\dots d_1} 
\mathcal{Q}_{1 d_{n-1}}^{(\alpha_{n},\tilde{\alpha}_{n})} 
\mathcal{Q}_{d_{n-1} d_{n-2}}^{(\alpha_{n-1},\tilde{\alpha}_{n-1})}\dots 
\mathcal{Q}_{d_1 1}^{(\alpha_{1},\tilde{\alpha}_{1})}.
\label{eq:PT_MPR}
\end{align}
Here the dimension of the inner indices $d_l$ is very large, corresponding to a complete basis of 
environment states in Liouville space. 
This large dimension  precludes the direct  
application of Eqs.~\eqref{eq:rho_alpha} and \eqref{eq:PT_MPR} 
for typical environments.
However, the MPO form of the PT means it is in principle amenable to standard MPO compression, based on singular value decomposition as described in the Methods~\cite{MPS_Schollwoeck, MPS_Orus}. 
Such compression corresponds physically to reducing
the environment to its most relevant degrees of freedom, which, as theoretical consideration of PTs suggest~\cite{Luchnikov2019}, may be few in number.

\begin{figure*}[t!]
\includegraphics[width=0.98\linewidth]{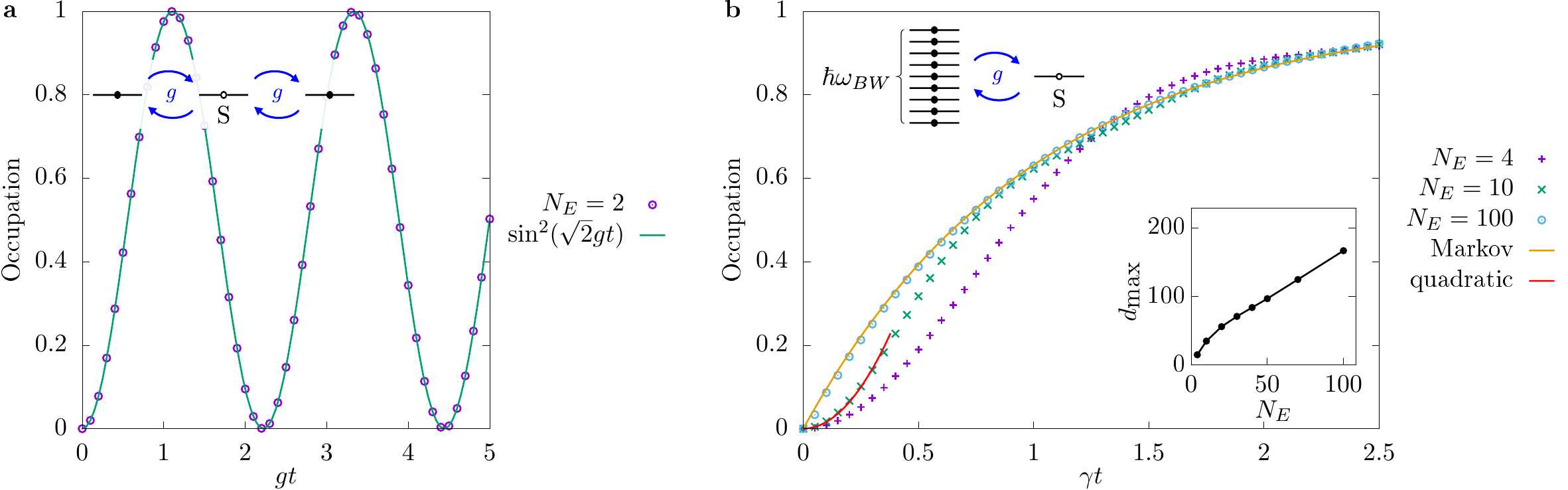}
\caption{\label{fig:leads} 
\textbf{Resonant-level model application of ACE, spanning small to infinite bath memory time.}
\textbf{a,} Dynamics of the occupations of a single localised quantum state (S) 
coupled to two resonant environment modes.
\textbf{b,} Dynamics of a quantum state coupled to 
a quasi-continuum of modes.
ACE simulations (points) are shown together with analytic solutions (lines).
In b, the analytic result in the Markov limit corresponds to an exponential transfer with the rate obtained from Fermi's golden rule. 
The result of a quadratic Taylor expansion around $t=0$ is depicted for the case $N_E=10$.
The top left insets depict the respective physical situations.
The bottom right inset in b shows the maximal inner dimension $d_\textrm{max}$ of
the PT MPO as a function of the number of environment states $N_E$ 
for constant density of states.
}
\end{figure*}
The key challenge is thus to find an efficient way to calculate the compressed form of the PT MPO, without first constructing the uncompressed PT.
This can be achieved through the ACE approach, for any problem with an environment that can be decomposed into $N_E$ 
different noninteracting degrees of freedom:
\begin{align}
H= H_S+ \sum_{k=1}^{N_E} H_E^k.
\label{eq:Htot}
\end{align}
The label $k$ can describe both the different degrees of freedom within a bath (e.g. different spins, or photon modes defined by their wave vector $\mathbf{q}$), but can also enumerate multiple environments coupled to the same system.  
In all of these cases, the PT can be constructed iteratively, by adding successively the contribution of each bath degree of freedom.
The process of combining the influence of the $K$-th degree of freedom, $\mathcal{B}[K]$, with an existing PT MPO $Q[K-1]$
is shown in Fig.~\ref{fig:sketchMPS}e. If the
resulting MPOs are compressed after each step (red semicircles), 
the inner dimension remains manageable and 
exact diagonalisation can be used for the singular value decomposition. 
This is described in more detail in the Methods section.

Once one has the compressed PT in MPO representation, this can be substituted into Eq.~\eqref{eq:rho_alpha}. 
The calculation of the reduced system density matrix then amounts to the contraction of a network of the form shown in Fig.~\ref{fig:sketchMPS}d.  If the PT MPO has a sufficiently small inner dimension, this contraction is straightforward.  
Because this algorithm can be applied in principle to arbitrary environments simply by specifying the respective environment Hamiltonians $H_E^k$, ACE allows investigations of a huge variety of different open quantum systems.
We next show how this method works in practice for a few paradigmatic example problems.

\subsection{Resonant-level model}
As a first test of ACE, we consider the archetypal problem of 
electron transport between a single localised electron state and other nearby environment states, described by the resonant-level model.  The $k$-th environment state is described by 
\begin{align}
H_E^k=&  \hbar \omega_k c^\dagger_k c_k
+\hbar g_k (c^\dagger_k c_S + c^\dagger_S c_k), 
\label{eq:hop}
\end{align}
where $c^\dagger_S (c_S)$ and $c_k^\dagger (c_k)$ create (destroy) a fermion in the localised system state and the $k$-th environment state, respectively,
$\hbar\omega_k$ is the energy of the $k$-th environment state with respect to the system state, and $g_k$ is the coupling constant, which we assume  to be independent of $k$, $g_k=g$. The free system Hamiltonian is $H_S=0$. The Hamiltonian in Eq.~\eqref{eq:hop} shows distinct behaviour depending on the number of environment modes: coherent oscillations for few modes, and irreversible decay for a broad continuum of modes.  In the following we show that ACE can automatically capture both these limits, and interpolate between them.

For a few environment modes, the dynamics is described by 
coherent oscillations at the eigenfrequencies of the coupled system and environment. 
Here, we consider the situation depicted in the inset of Fig.~\ref{fig:leads}a where
a single initially empty site is connected to two sites at the same energy
$\omega_k=0$, which are initially occupied.
In this scenario the time-dependent many-body state of the total system is
\begin{align}
|\Psi(t)\rangle=& \bigg[ 
\cos(\sqrt{2}g t)\frac{c^\dagger_1 + c^\dagger_2}{\sqrt{2}}
-i\sin(\sqrt{2}g t) c^\dagger_S 
\bigg]\frac{c^\dagger_1-c^\dagger_2}{\sqrt{2}} |0\rangle.
\label{eq:analytNE2psi}
\end{align}
In Figure~\ref{fig:leads}a, we compare the occupation $n_{S}=\sin^2(\sqrt{2}g t)$ to the results of ACE simulations for convergence parameters 
$\Delta t=0.01 g$ and $\epsilon=10^{-7}$ (see Methods).  We see the results match perfectly.
Since the oscillations are undamped, the memory time of the environment is 
infinite. Furthermore, whenever $n_S=\frac 12$,
Eq.~\eqref{eq:analytNE2psi} describes a state with maximal entanglement
between system and environment.
This demonstrates that ACE can account for infinite memory times as
well as strong system-environment correlations.

Different behaviour occurs for a quasi-continuum of 
environment states, e.g.,  metallic leads coupled to a 
quantum dot~\cite{Brandes_doubledot},
as depicted in the top left inset of Fig.~\ref{fig:leads}b. The oscillatory 
contributions of the different modes interfere destructively, suppressing oscillations.
When the continuum is broad enough, there is a short memory time and weak system-bath correlations, so the situation is well described by Markovian master equations.  These predict charge transfer to the localised state at a rate $\gamma=2\pi\hbar g^2D$, where $D=(N_E-1)/(\hbar\omega_{BW})$ is the 
density of states and $\hbar\omega_{BW}$ is the bandwidth. 
Figure~\ref{fig:leads}b shows the corresponding dynamics 
for different numbers of environment modes $N_E$ 
with a fixed density of states $D=1/(\hbar\gamma)$.
As the number of environment modes (and therefore
the bandwidth) increases, the simulations approach the 
Markovian analytic result $1-\exp(-\gamma t)$. 
For intermediate $N_E=10$, the finite bandwidth introduces a 
finite memory time $\sim 1/\omega_{BW}$. 
To check the validity of the ACE results in  this more complicated crossover regime, we also plot the analytic short-time Taylor expansion,
$n_S\approx \gamma\omega_{BW}t^2/(2\pi)$ for the case $N_E=10$.

The inset in Fig.~\ref{fig:leads}b shows the maximal inner dimension $d_\textrm{max}$ of the PT MPO as a function of the number of modes $N_E$. We see this scales linearly with the number of modes, indicating a very efficient reduction, compared to the exponential scaling of the dimension of the full environment Liouville space of up to $4^{100} \approx 1.6\times 10^{60}$ for $N_E=100$. 
A more detailed analysis of numerical convergence is given in the Supplemental Material S.2.
This simple example demonstrates that ACE is able to reproduce analytic results in all regimes from infinite memories to Markovian environments and from strong to weak system-environment correlations.

\subsection{Simultaneous coupling of quantum dots to phonons and 
electromagnetic field modes}
Our second example involves a system coupled simultaneously to two structured baths, as exemplified by a semiconductor quantum dot, coupled both to acoustic phonons and an electromagnetic environment.  The acoustic phonon modes couple via a pure-dephasing interaction:
\begin{align}
H_\textrm{ph}^\mathbf{q}=& 
\hbar \omega_\mathbf{q} b^\dagger_\mathbf{q} b_\mathbf{q} + 
\hbar \gamma_\mathbf{q} \big(b^\dagger_\mathbf{q} + b_\mathbf{q}\big) 
|X\rangle\langle X|,
\label{eq:Hphonon}
\end{align}
where $b^\dagger_\mathbf{q}$ ($b_\mathbf{q}$) creates (annihilates) a 
phonon with wave vector $\mathbf{q}$ 
and $|X\rangle$ denotes the exciton state of the quantum dot. 
If this were the only interaction, its linear 
and diagonal structure would mean it could be treated within the iterative quasi-adiabatic path integral (iQUAPI) method~\cite{Makri,PI_cQED,PI_multitime}.  We will use this below to compare the results of ACE to that of iQUAPI.
\begin{figure*}[t]
\includegraphics[width=0.85\linewidth]{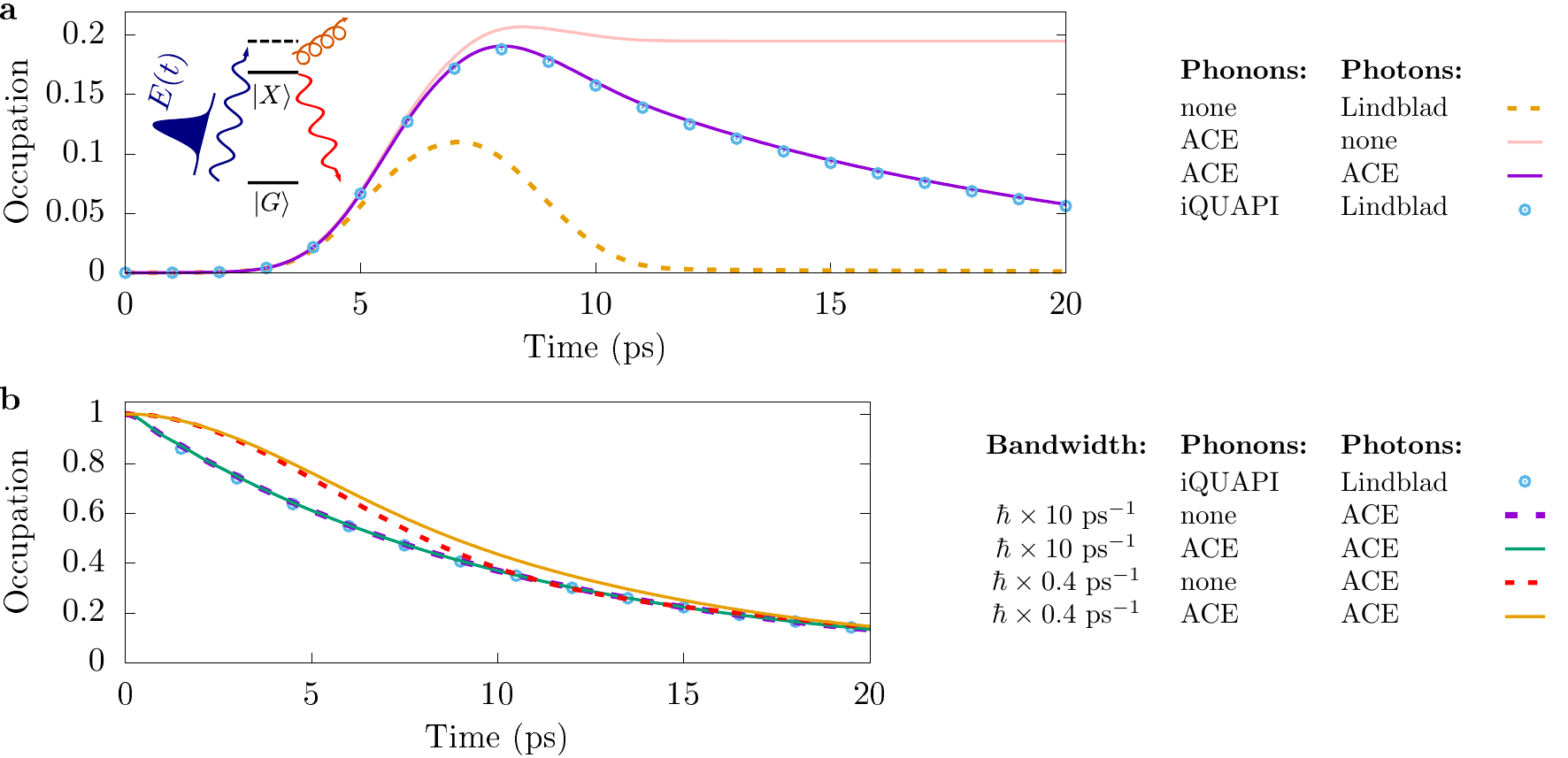}
\caption{\label{fig:QDPhonon}
\textbf{Dynamics of quantum dots embedded in (non-additive) photon and phonon environments.}
\textbf{a,} Dynamics of the exciton occupation for
phonon-assisted off-resonant excitation of a QD
driven by a Gaussian laser pulse and subject to radiative decay
according to different theoretical approaches:
The QD-phonon interaction may be disregarded (none) or treated within ACE
or iQUAPI. The coupling between the QD and the photonic modes may be 
disregarded (none), included explicitly in ACE via its Hamiltonian, or 
replaced by a Lindblad term for radiative decay.
\textbf{b,} Radiative decay of an initially occupied exciton state with and without
interactions with phonons for model photon densities of states with 
different bandwidths $\hbar\omega_{BZ}$. 
}
\end{figure*}

In addition to the bath of phonons, QDs also couple to 
the continuum of electromagnetic modes, which are responsible for radiative decay. Here the interaction with photon mode $k$ takes the Jaynes-Cummings form:
\begin{align}
H_{JC}^k=&\hbar \omega_k a^\dagger_k a_k 
+ \hbar g_k \big( 
a^\dagger_k |G\rangle\langle X| + a_k |X\rangle\langle G|\big),
\label{eq:HJC}
\end{align}
where $a^\dagger_k$ ($a_k$) is the bosonic creation (annihilation) operator
for a photon in mode $k$. 

There are several ways of including both baths in simulations:
First, for unstructured (i.e. Markovian) photon environments, the Born--Markov approximation holds, so we can account for the radiative decay as a Lindblad term, $\kappa \mathcal{L}\big[|G\rangle\langle X|,\rhoS\big]$ where
\begin{multline}
\mathcal{L}\big[|G\rangle\langle X|,\rhoS\big]
\\\equiv
|G\rangle\langle X| \rhoS |X\rangle\langle G| -\tfrac 12
\big(|X\rangle\langle X|\rhoS + \rhoS|X\rangle\langle X|\big).
\end{multline}
In both ACE and iQUAPI~\cite{Andi_nonHamil}, such Markovian dissipation can be 
included into the free system Liouville propagator $\mathcal{M}$.
Due to the flexibility of ACE, we can also describe the radiative decay microscopically by including both the phonon and electromagnetic environments in the PT.  This has the advantage that it automatically captures possible 
non-additive effects of the simultaneous coupling to multiple baths~\cite{Nagy2011,Mitchison2018,Nazir_Nonadditivity}, and also allows one to extend to structured electromagnetic environments.

In Fig.~\ref{fig:QDPhonon}a, we show how the occupation of a QD responds to  off-resonant excitation by a Gaussian laser pulse.  This drive corresponds to the following time-dependent Hamiltonian in the rotating frame of the laser:
\begin{align}
H_S=-\hbar\delta|X\rangle\langle X| +\tfrac\hbar2 \Omega(t) 
\big(|X\rangle\langle G| + |G\rangle\langle X|\big),
\end{align}
where $\delta$ is the laser detuning and $\Omega(t)$ is a Gaussian envelope
centred at $t_0=7$ ps with pulse duration $\tau_\textrm{FWHM}=5$~ps. 
The QD simultaneously interacts with the phonon and photon baths, which are treated within different theoretical approaches.  In this figure we assume a flat electromagnetic environment, so all approaches should work equally well. The simulation parameters are summarised in the Methods section.

In the absence of QD-phonon interactions, the exciton is only occupied transiently during the pulse,  as absorption is suppressed by the detuning of the laser from the exciton energy.
Including phonons within ACE but disregarding radiative decay entirely
results in a nonzero stationary exciton 
occupation, as the detuning may be bridged by phonon emission. 
Including both phonons and photons, one sees absorption followed by radiative decay.   Identical results are found for this case for both ACE---treating the electromagnetic environment microscopically---and for iQUAPI with photon decay $\kappa \mathcal{L}\big[|G\rangle\langle X|,\rhoS\big]$.
As such, we both further confirm the capabilities of ACE, and see that---as may be anticipated---for 
an unstructured photon environment, no cross-action between the coupling to
photon and phonon baths can be identified.

As already noted, ACE is also able to treat situations with
non-additive environments, as is relevant for structured photonic 
environments like waveguides or  
microcavities~\cite{Hughes_structured,Busch_structured}.
Figure~\ref{fig:QDPhonon}b shows the decay of an initially occupied exciton state (with $H_S=0)$ where, in addition to the non-Markovian phonon bath, we use a photon bath with a finite bandwidth $\hbar\omega_{BW}$.
For large bandwidths, no cross-interaction between the couplings to the two
baths is found (and so the results again match iQUAPI with Lindbladian photon loss). For small bandwidths $\omega_{BW}=0.4$ ps$^{-1}$, the photon
environment obtains a memory time $\tau\sim 1/\omega_{BW}$ of the same
order of magnitude as the phonon environment. 
As a result the two baths couple non-additively, as can be seen by the fact that the coupling to phonons significantly influences the decay of excitations into the electromagnetic modes.

\subsection{Spin dynamics}
\begin{figure}
\includegraphics[width=0.99\linewidth]{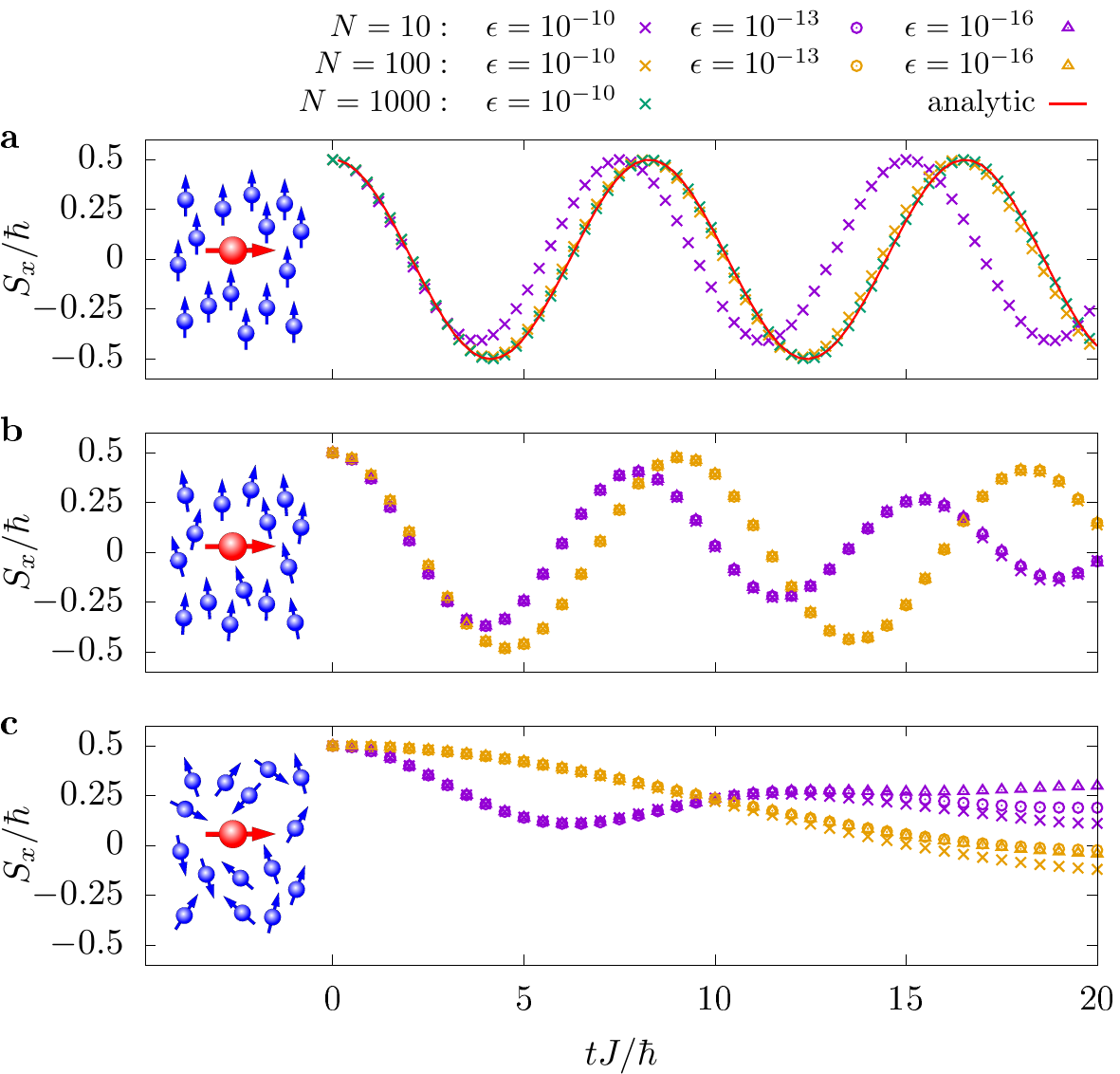}
\caption{\label{fig:spins}
\textbf{Central spin model for different degrees of spin bath polarisation.}
Dynamics of a central spin (red) initially prepared along the $x$-axis
in a bath of $N$ spins (blue) as depicted in the insets. 
The $x$-component of a central spin is shown for situations 
where the bath spins are fully polarised \textbf{a,} partially polarised \textbf{b,} 
or unpolarised \textbf{c}. The number of environment spins
$N$ is varied keeping the sum of the couplings $\sum_k J_k=J$ constant.
Colours correspond to different numbers of environment spins $N$ while point types correspond to different values of the MPO compression threshold $\epsilon$.
}
\end{figure}

Our third example concerns the spin dynamics in the presence of a spin environment~\cite{SpinInterface,Jelezko_PRB17}.
Besides demonstrating the applicability of ACE to non-Gaussian spin environments, this example also identifies the limits on efficient environment compression.
We consider a central spin coupled to a bath of environment spins by a Heisenberg interaction 
\begin{align}
H_E^k =& \frac{J_k}{\hbar^2}\, \hat{\mathbf{S}}\cdot \hat{\mathbf{s}}_k,
\end{align}
where $\hat{\mathbf{S}}$ and $\hat{\mathbf{s}}_k$ are the 
spin-$\frac 12$ operators
of the central spin and the $k$-th environment spin, respectively---see inset of Fig.~\ref{fig:spins}.
In the following we choose the coupling constants $J_k=J/N$, where $N$ is the number of environment spins and $J$ defines the energy scale of the coupling.  We set $H_S=0$ and initially prepare the system spin in the state with maximal $\langle S_x\rangle$. We then explore how the initial degree of polarisation of the environment affects the system dynamics, and the ability to efficiently compress the environment.

First, we focus on the situation where the environment spins
are completely polarised along the $z$-axis. 
The respective dynamics of $\langle S_x\rangle$ is depicted in 
Fig.~\ref{fig:spins}a for different numbers of environment spins 
$N=10$, $N=100$, and $N=1000$
and for convergence parameters
$\Delta t=0.01 \hbar/J$ and $\epsilon=10^{-10}$.
The Heisenberg coupling 
leads to a coherent precession of the system and environment spins 
about each other. In the limit $N\to\infty$, there is no back-action
on the environment so the environment remains in its
initial state. The dynamics is then equivalent to a precession about
a constant effective magnetic field, for which 
$\langle S_x\rangle=(\hbar/2)\cos\big[(tJ)/(2\hbar)\big]$. 
We see the ACE simulations for $N=1000$ approach this limit. 
It is noteworthy that for all $N$ 
the inner dimension of the PT MPO remains 4, corresponding to the Liouville
space dimension of a single spin $\frac 12$. This is
because all environment spins behave 
identically, so the environment can be replaced by a single effective
spin. 

We next explore the limitations of compression of the environment, by considering randomised initial conditions for the environment spins.
In Fig.~\ref{fig:spins}b and c we present ACE simulations 
with $N=10$ and $N=100$ environment spins 
for different values of the MPO truncation threshold $\epsilon$.
In Fig.~\ref{fig:spins}b the bath is partially polarised: we randomly select  pure spin states from an isotropic distribution and filter these
with a rejection probability 
$1-\exp\big[b \big(s^z_k/\hbar-\tfrac{1}{2} \big)\big]$.
Here, $b=(g \mu_B B)/(k_B T)$ is a Boltzmann factor, taken as $b=20$ for Fig.~\ref{fig:spins}b.
In Fig.~\ref{fig:spins}c we instead use a uniform distribution (i.e. $b=0$).
In both cases a dephasing of the central spin is visible.
However, for the unpolarised case, the spin 
dynamics for different $\epsilon$ start to diverge at long times.
The slow convergence with $\epsilon$ in this situation is a consequence of the
intrinsic incompressibility of the environment degrees of freedom.
That is, because each environment spin reacts differently to the 
system spin, the joint PT cannot be compressed efficiently.
Furthermore, environment spins can become correlated via an effective 
interaction mediated by the system, and without an external
magnetic field the environment states are highly degenerate.
Consequently, there is no clear physical constraint on the accessible
environment Hilbert space.
 In the partially polarised case, the 
environment can be compressed more efficiently, so that the ACE simulations 
show a better convergence.

\subsection{Anharmonic environments}
\begin{figure}[t]
\includegraphics[width=\linewidth]{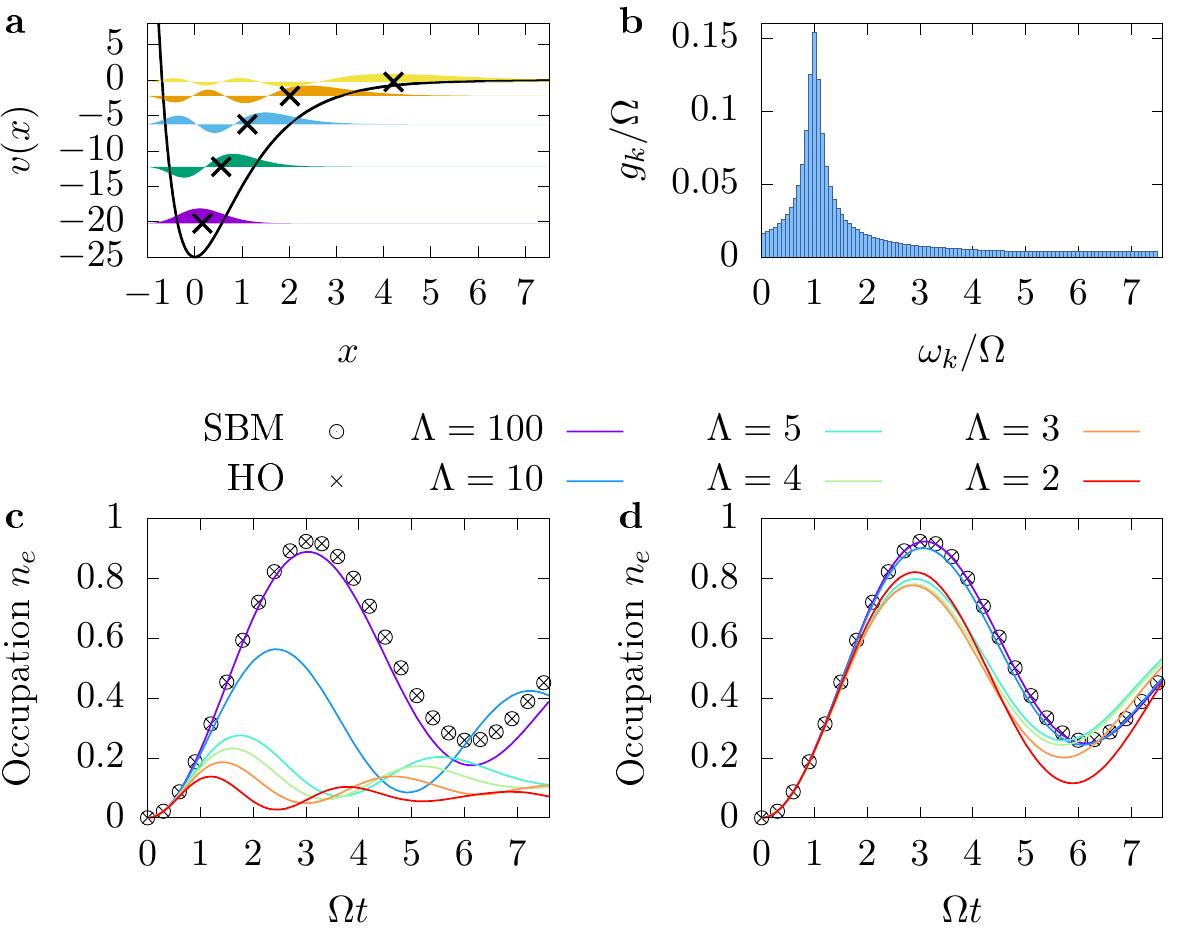}
\caption{\label{fig:morse}
\textbf{Two-level system coupled to a bath of anharmonic modes.}
\textbf{a}: Morse potential, Eq.~\eqref{eq:Morse}, with parameter $\Lambda=5$ and its bound eigenstates obtained numerically.
Crosses mark the average position $\langle i | \hat{x}| i\rangle$ for each  eigenstate. 
\textbf{b}: Coupling coefficients $g_k$ corresponding to a Lorentzian spectral density of environment modes.
\textbf{c}: ACE simulations with $M= \min\{5, \Lambda\}$ environment levels for: the spin-boson model (SBM),
harmonic oscillator (HO) modes obtained by the finite differences method,
and finite differences solutions of the Morse potential for different
$\Lambda$.
\textbf{d}: Analogous calculations to \textbf{c} but where energy shifts
due to non-zero $\langle i | \hat{x}| i\rangle$ have been subtracted.}
\end{figure}
While a bath of harmonic oscillators forms a Gaussian environment, which can be addressed by a multitude of existing numerically exact methods, anharmonic environment modes have so far been out of reach.
Anharmonicities are found in practice, e.g., in vibrational modes of molecules with a finite number of bound vibrational states, commonly modelled by a Morse potential~\cite{Morse_deVega} 
\begin{align}
\label{eq:Morse}
v(x)=&\Lambda^2\Big( e^{-2x} - 2e^{-x}\Big),
\end{align}
where $\Lambda$ controls the depth of the potential and number of bound states.
Here, we use the Morse potential as a demonstration of simulating environment modes with arbitrary potentials $v(x)$.

As described in more detail in the Supplementary Material S.4, we first use a finite differences method to numerically find the eigenstates of a single uncoupled environment mode, before introducing coupling to the system.
For example, the bound eigenstates of the Morse potential for $\Lambda=5$ are depicted in Fig.~\ref{fig:morse}\textbf{a}.
Keeping only the $M$ lowest energy eigenstates and choosing a system-environment coupling proportional to the environment position operator, we find that for environment mode $k$
\label{eq:HEmorse}
\begin{align}
H_E^k=& \sum_{j=0}^{M-1} \hbar\omega_k \tilde{E}_j\sigma^k_{jj} +%
\hbar g_k   \sum_{i,j=0}^{M-1}\!
\sqrt{2}\langle i|\tilde{x}|j\rangle
\sigma^k_{ij} \,  |e\rangle\langle e|,
\end{align}
where $\tilde{E}_j$ and $\langle i|\tilde{x}|j\rangle$ are scaled so that the spin-boson model Hamiltonian is recovered when  $v(x)$ is  the harmonic oscillator potential.

ACE simulations are performed for 
$H_S=\tfrac{\hbar}{2}\Omega \big(|e\rangle\langle g|+|g\rangle\langle e|\big)$, describing a continuously driven system performing Rabi oscillations damped by the anharmonic environment. 
We choose a set of $\omega_k$ and $g_k$ that correspond to a Lorentzian spectral density as shown in Fig.~\ref{fig:morse}\textbf{b}; for other parameters see Supplementary Material S.4.
The resulting excited state occupations $n_e$ are shown in Fig.~\ref{fig:morse}\textbf{c}.

As a validity check, we first apply the above method to a harmonic potential, and recover exactly the dynamics of the spin-boson model.
On moving to  Morse potential environments, we find significant differences, especially for small $\Lambda$.
Much of the difference is due to the asymmetry of the Morse potential, leading to an average position $\langle i|\tilde{x}|i\rangle$ that increases for higher excited states, indicated as crosses in Fig.~\ref{fig:morse}\textbf{a}. This enters in $H_E$ via the system-environment coupling and results in an energy shift of the $|e\rangle$ system state.
To better identify intrinsic effects of anharmonicity,  Fig.~\ref{fig:morse}\textbf{d} shows ACE results where this shift has been subtracted. For small $\Lambda$, one sees effects of the anharmonicity of the Morse potential, while for large $\Lambda$ the anharmonicity becomes negligible and the result of the Gaussian simulations is recovered.

\secnl{Discussion}
We have presented a novel, numerically exact, efficient, and versatile method: \emph{automated compression of environments} (ACE), which makes it possible to simulate the dynamics of $N$-level quantum systems coupled to arbitrary environments directly from the microscopic system-environment coupling Hamiltonian.
We have illustrated the power of this method with examples of electron transport, the simultaneous interaction of a QD with phonon
and photon modes, spin dynamics, and anharmonic environments.  In the Supplementary Material~S.5, we provide an example exploring superradiant decay, illustrating that ACE can handle higher-dimensional system Hilbert spaces. Supplementary Material~S.6 further contains an example of simulations of dispersive system-environment couplings as well as time-dependent driving and non-Hamiltonian loss terms acting directly on the environment.
We have shown that ACE reproduces exact results in limiting cases, and can interpolate between infinite and short memory scenarios within the same algorithm.
In particular, non-Markovian effects, system-environment correlations, and non-Gaussian baths are fully accounted for.

A fundamental restriction of ACE is that the environment must decompose into a set of separate modes without interactions between these modes.  
However, most typical models of open system environments satisfy this requirement.
Moreover, recent work by~\cite{ye2021constructing} shows that, adapting a method of~\cite{Banuls2009}, one can extend tensor network methods to models where bath modes have nearest-neighbour interactions.
Some environments have particular features that enable more specialised methods to be used, and these can be more efficient than the general method ACE. For example, Gaussian baths with a broad continuum of modes have short memory times at high temperature, and then iQUAPI~\cite{Makri} extremely efficient. In contrast, for environments consisting of only a few discrete modes, ACE outperforms methods based on Gaussian path integrals (see Supplementary Material S.3). 
For spectral densities with several peaks on top of a broad background,
the construction of a PT for Gaussian environments in Ref.~\onlinecite{ProcessTensor} can be readily combined with ACE to enable a hybrid approach within the common process tensor framework. 

However, the unique feature of ACE is its generality. It can be used in situations where no specialised methods are available,  and no new derivations or modifications of the algorithm are required when a new system or environment are considered.
Due to its numerical exactness, ACE can serve as a benchmark for approximate methods which may provide a more tangible interpretation of physical processes,
or serve a ``turnkey solution''  to simulate concrete experiments.
These features make ACE a valuable general-purpose tool for 
open quantum systems.

\bibliography{references.bib}

\begin{thebibliography}{51}%
\makeatletter
\providecommand \@ifxundefined [1]{%
 \@ifx{#1\undefined}
}%
\providecommand \@ifnum [1]{%
 \ifnum #1\expandafter \@firstoftwo
 \else \expandafter \@secondoftwo
 \fi
}%
\providecommand \@ifx [1]{%
 \ifx #1\expandafter \@firstoftwo
 \else \expandafter \@secondoftwo
 \fi
}%
\providecommand \natexlab [1]{#1}%
\providecommand \enquote  [1]{``#1''}%
\providecommand \bibnamefont  [1]{#1}%
\providecommand \bibfnamefont [1]{#1}%
\providecommand \citenamefont [1]{#1}%
\providecommand \href@noop [0]{\@secondoftwo}%
\providecommand \href [0]{\begingroup \@sanitize@url \@href}%
\providecommand \@href[1]{\@@startlink{#1}\@@href}%
\providecommand \@@href[1]{\endgroup#1\@@endlink}%
\providecommand \@sanitize@url [0]{\catcode `\\12\catcode `\$12\catcode
  `\&12\catcode `\#12\catcode `\^12\catcode `\_12\catcode `\%12\relax}%
\providecommand \@@startlink[1]{}%
\providecommand \@@endlink[0]{}%
\providecommand \url  [0]{\begingroup\@sanitize@url \@url }%
\providecommand \@url [1]{\endgroup\@href {#1}{\urlprefix }}%
\providecommand \urlprefix  [0]{URL }%
\providecommand \Eprint [0]{\href }%
\providecommand \doibase [0]{https://doi.org/}%
\providecommand \selectlanguage [0]{\@gobble}%
\providecommand \bibinfo  [0]{\@secondoftwo}%
\providecommand \bibfield  [0]{\@secondoftwo}%
\providecommand \translation [1]{[#1]}%
\providecommand \BibitemOpen [0]{}%
\providecommand \bibitemStop [0]{}%
\providecommand \bibitemNoStop [0]{.\EOS\space}%
\providecommand \EOS [0]{\spacefactor3000\relax}%
\providecommand \BibitemShut  [1]{\csname bibitem#1\endcsname}%
\let\auto@bib@innerbib\@empty
\bibitem [{\citenamefont {Breuer}\ and\ \citenamefont
  {Petruccione}(2002)}]{Breuer}%
  \BibitemOpen
  \bibfield  {author} {\bibinfo {author} {\bibfnamefont {H.-P.}\ \bibnamefont
  {Breuer}}\ and\ \bibinfo {author} {\bibfnamefont {F.}~\bibnamefont
  {Petruccione}},\ }\href@noop {} {\emph {\bibinfo {title} {The Theory of Open
  Quantum Systems}}}\ (\bibinfo  {publisher} {Oxford University Press,
  Oxford},\ \bibinfo {year} {2002})\BibitemShut {NoStop}%
\bibitem [{\citenamefont {Plenio}\ and\ \citenamefont
  {Huelga}(2008)}]{Plenio2008}%
  \BibitemOpen
  \bibfield  {author} {\bibinfo {author} {\bibfnamefont {M.~B.}\ \bibnamefont
  {Plenio}}\ and\ \bibinfo {author} {\bibfnamefont {S.~F.}\ \bibnamefont
  {Huelga}},\ }\bibfield  {title} {\bibinfo {title} {Dephasing-assisted
  transport: quantum networks and biomolecules},\ }\href
  {https://doi.org/10.1088/1367-2630/10/11/113019} {\bibfield  {journal}
  {\bibinfo  {journal} {New J. Phys.}\ }\textbf {\bibinfo {volume} {10}},\
  \bibinfo {pages} {113019} (\bibinfo {year} {2008})}\BibitemShut {NoStop}%
\bibitem [{\citenamefont {Rebentrost}\ \emph {et~al.}(2009)\citenamefont
  {Rebentrost}, \citenamefont {Mohseni}, \citenamefont {Kassal}, \citenamefont
  {Lloyd},\ and\ \citenamefont {Aspuru-Guzik}}]{Rebentrost2009}%
  \BibitemOpen
  \bibfield  {author} {\bibinfo {author} {\bibfnamefont {P.}~\bibnamefont
  {Rebentrost}}, \bibinfo {author} {\bibfnamefont {M.}~\bibnamefont {Mohseni}},
  \bibinfo {author} {\bibfnamefont {I.}~\bibnamefont {Kassal}}, \bibinfo
  {author} {\bibfnamefont {S.}~\bibnamefont {Lloyd}},\ and\ \bibinfo {author}
  {\bibfnamefont {A.}~\bibnamefont {Aspuru-Guzik}},\ }\bibfield  {title}
  {\bibinfo {title} {Environment-assisted quantum transport},\ }\href
  {https://doi.org/10.1088/1367-2630/11/3/033003} {\bibfield  {journal}
  {\bibinfo  {journal} {New J. Phys.}\ }\textbf {\bibinfo {volume} {11}},\
  \bibinfo {pages} {033003} (\bibinfo {year} {2009})}\BibitemShut {NoStop}%
\bibitem [{\citenamefont {Chin}\ \emph {et~al.}(2010)\citenamefont {Chin},
  \citenamefont {Datta}, \citenamefont {Caruso}, \citenamefont {Huelga},\ and\
  \citenamefont {Plenio}}]{Chin2010}%
  \BibitemOpen
  \bibfield  {author} {\bibinfo {author} {\bibfnamefont {A.~W.}\ \bibnamefont
  {Chin}}, \bibinfo {author} {\bibfnamefont {A.}~\bibnamefont {Datta}},
  \bibinfo {author} {\bibfnamefont {F.}~\bibnamefont {Caruso}}, \bibinfo
  {author} {\bibfnamefont {S.~F.}\ \bibnamefont {Huelga}},\ and\ \bibinfo
  {author} {\bibfnamefont {M.~B.}\ \bibnamefont {Plenio}},\ }\bibfield  {title}
  {\bibinfo {title} {Noise-assisted energy transfer in quantum networks and
  light-harvesting complexes},\ }\href
  {https://doi.org/10.1088/1367-2630/12/6/065002} {\bibfield  {journal}
  {\bibinfo  {journal} {New J. Phys.}\ }\textbf {\bibinfo {volume} {12}},\
  \bibinfo {pages} {065002} (\bibinfo {year} {2010})}\BibitemShut {NoStop}%
\bibitem [{\citenamefont {Beige}\ \emph {et~al.}(2000)\citenamefont {Beige},
  \citenamefont {Braun}, \citenamefont {Tregenna},\ and\ \citenamefont
  {Knight}}]{Beige2000}%
  \BibitemOpen
  \bibfield  {author} {\bibinfo {author} {\bibfnamefont {A.}~\bibnamefont
  {Beige}}, \bibinfo {author} {\bibfnamefont {D.}~\bibnamefont {Braun}},
  \bibinfo {author} {\bibfnamefont {B.}~\bibnamefont {Tregenna}},\ and\
  \bibinfo {author} {\bibfnamefont {P.~L.}\ \bibnamefont {Knight}},\ }\bibfield
   {title} {\bibinfo {title} {Quantum computing using dissipation to remain in
  a decoherence-free subspace},\ }\href
  {https://doi.org/10.1103/PhysRevLett.85.1762} {\bibfield  {journal} {\bibinfo
   {journal} {Phys. Rev. Lett.}\ }\textbf {\bibinfo {volume} {85}},\ \bibinfo
  {pages} {1762} (\bibinfo {year} {2000})}\BibitemShut {NoStop}%
\bibitem [{\citenamefont {Verstraete}\ \emph {et~al.}(2009)\citenamefont
  {Verstraete}, \citenamefont {Wolf},\ and\ \citenamefont
  {Cirac}}]{Verstraete2009quantum}%
  \BibitemOpen
  \bibfield  {author} {\bibinfo {author} {\bibfnamefont {F.}~\bibnamefont
  {Verstraete}}, \bibinfo {author} {\bibfnamefont {M.~M.}\ \bibnamefont
  {Wolf}},\ and\ \bibinfo {author} {\bibfnamefont {J.~I.}\ \bibnamefont
  {Cirac}},\ }\bibfield  {title} {\bibinfo {title} {Quantum computation and
  quantum-state engineering driven by dissipation},\ }\href
  {https://doi.org/10.1038/nphys1342} {\bibfield  {journal} {\bibinfo
  {journal} {Nat. Phys.}\ }\textbf {\bibinfo {volume} {5}},\ \bibinfo {pages}
  {633} (\bibinfo {year} {2009})}\BibitemShut {NoStop}%
\bibitem [{\citenamefont {de~Vega}\ and\ \citenamefont
  {Alonso}(2017)}]{RevModPhys_deVega}%
  \BibitemOpen
  \bibfield  {author} {\bibinfo {author} {\bibfnamefont {I.}~\bibnamefont
  {de~Vega}}\ and\ \bibinfo {author} {\bibfnamefont {D.}~\bibnamefont
  {Alonso}},\ }\bibfield  {title} {\bibinfo {title} {Dynamics of non-markovian
  open quantum systems},\ }\href {https://doi.org/10.1103/RevModPhys.89.015001}
  {\bibfield  {journal} {\bibinfo  {journal} {Rev. Mod. Phys.}\ }\textbf
  {\bibinfo {volume} {89}},\ \bibinfo {pages} {015001} (\bibinfo {year}
  {2017})}\BibitemShut {NoStop}%
\bibitem [{\citenamefont {Tanimura}(2006)}]{Review_stochastic}%
  \BibitemOpen
  \bibfield  {author} {\bibinfo {author} {\bibfnamefont {Y.}~\bibnamefont
  {Tanimura}},\ }\bibfield  {title} {\bibinfo {title} {{Stochastic Liouville,
  Langevin, Fokker–Planck, and Master Equation Approaches to Quantum
  Dissipative Systems}},\ }\href {https://doi.org/10.1143/JPSJ.75.082001}
  {\bibfield  {journal} {\bibinfo  {journal} {J. Phys. Soc. Jpn.}\ }\textbf
  {\bibinfo {volume} {75}},\ \bibinfo {pages} {082001} (\bibinfo {year}
  {2006})}\BibitemShut {NoStop}%
\bibitem [{\citenamefont {Plenio}\ and\ \citenamefont
  {Knight}(1998)}]{RevModPhys_jump}%
  \BibitemOpen
  \bibfield  {author} {\bibinfo {author} {\bibfnamefont {M.~B.}\ \bibnamefont
  {Plenio}}\ and\ \bibinfo {author} {\bibfnamefont {P.~L.}\ \bibnamefont
  {Knight}},\ }\bibfield  {title} {\bibinfo {title} {The quantum-jump approach
  to dissipative dynamics in quantum optics},\ }\href
  {https://doi.org/10.1103/RevModPhys.70.101} {\bibfield  {journal} {\bibinfo
  {journal} {Rev. Mod. Phys.}\ }\textbf {\bibinfo {volume} {70}},\ \bibinfo
  {pages} {101} (\bibinfo {year} {1998})}\BibitemShut {NoStop}%
\bibitem [{\citenamefont {Redfield}(1965)}]{Redfield}%
  \BibitemOpen
  \bibfield  {author} {\bibinfo {author} {\bibfnamefont {A.~G.}\ \bibnamefont
  {Redfield}},\ }\bibfield  {title} {\bibinfo {title} {The theory of relaxation
  processes},\ }in\ \href
  {https://doi.org/https://doi.org/10.1016/B978-1-4832-3114-3.50007-6} {\emph
  {\bibinfo {booktitle} {Advances in Magnetic Resonance}}},\ \bibinfo {series}
  {Advances in Magnetic and Optical Resonance}, Vol.~\bibinfo {volume} {1},\
  \bibinfo {editor} {edited by\ \bibinfo {editor} {\bibfnamefont {J.~S.}\
  \bibnamefont {Waugh}}}\ (\bibinfo  {publisher} {Academic Press},\ \bibinfo
  {year} {1965})\ pp.\ \bibinfo {pages} {1 -- 32}\BibitemShut {NoStop}%
\bibitem [{\citenamefont {Nazir}\ and\ \citenamefont
  {McCutcheon}(2016)}]{NazirReview}%
  \BibitemOpen
  \bibfield  {author} {\bibinfo {author} {\bibfnamefont {A.}~\bibnamefont
  {Nazir}}\ and\ \bibinfo {author} {\bibfnamefont {D.~P.~S.}\ \bibnamefont
  {McCutcheon}},\ }\bibfield  {title} {\bibinfo {title} {Modelling
  exciton{\textendash}phonon interactions in optically driven quantum dots},\
  }\href {https://doi.org/10.1088/0953-8984/28/10/103002} {\bibfield  {journal}
  {\bibinfo  {journal} {J. Phys.: Condens. Matter}\ }\textbf {\bibinfo {volume}
  {28}},\ \bibinfo {pages} {103002} (\bibinfo {year} {2016})}\BibitemShut
  {NoStop}%
\bibitem [{\citenamefont {Breuer}\ \emph {et~al.}(2016)\citenamefont {Breuer},
  \citenamefont {Laine}, \citenamefont {Piilo},\ and\ \citenamefont
  {Vacchini}}]{Breuer2016}%
  \BibitemOpen
  \bibfield  {author} {\bibinfo {author} {\bibfnamefont {H.-P.}\ \bibnamefont
  {Breuer}}, \bibinfo {author} {\bibfnamefont {E.-M.}\ \bibnamefont {Laine}},
  \bibinfo {author} {\bibfnamefont {J.}~\bibnamefont {Piilo}},\ and\ \bibinfo
  {author} {\bibfnamefont {B.}~\bibnamefont {Vacchini}},\ }\bibfield  {title}
  {\bibinfo {title} {Colloquium: Non-markovian dynamics in open quantum
  systems},\ }\href {https://doi.org/10.1103/RevModPhys.88.021002} {\bibfield
  {journal} {\bibinfo  {journal} {Rev. Mod. Phys.}\ }\textbf {\bibinfo {volume}
  {88}},\ \bibinfo {pages} {021002} (\bibinfo {year} {2016})}\BibitemShut
  {NoStop}%
\bibitem [{\citenamefont {Tanimura}\ and\ \citenamefont {Kubo}(1989)}]{HEOM89}%
  \BibitemOpen
  \bibfield  {author} {\bibinfo {author} {\bibfnamefont {Y.}~\bibnamefont
  {Tanimura}}\ and\ \bibinfo {author} {\bibfnamefont {R.}~\bibnamefont
  {Kubo}},\ }\bibfield  {title} {\bibinfo {title} {Time evolution of a quantum
  system in contact with a nearly {Gaussian-Markoffian} noise bath},\ }\href
  {https://doi.org/10.1143/JPSJ.58.101} {\bibfield  {journal} {\bibinfo
  {journal} {J. Phys. Soc. Jpn.}\ }\textbf {\bibinfo {volume} {58}},\ \bibinfo
  {pages} {101} (\bibinfo {year} {1989})}\BibitemShut {NoStop}%
\bibitem [{\citenamefont {Tanimura}(2020)}]{HEOMreview20}%
  \BibitemOpen
  \bibfield  {author} {\bibinfo {author} {\bibfnamefont {Y.}~\bibnamefont
  {Tanimura}},\ }\bibfield  {title} {\bibinfo {title} {Numerically “exact”
  approach to open quantum dynamics: The hierarchical equations of motion
  {(HEOM)}},\ }\href {https://doi.org/10.1063/5.0011599} {\bibfield  {journal}
  {\bibinfo  {journal} {J. Chem. Phys.}\ }\textbf {\bibinfo {volume} {153}},\
  \bibinfo {pages} {020901} (\bibinfo {year} {2020})}\BibitemShut {NoStop}%
\bibitem [{\citenamefont {Prior}\ \emph {et~al.}(2010)\citenamefont {Prior},
  \citenamefont {Chin}, \citenamefont {Huelga},\ and\ \citenamefont
  {Plenio}}]{TEDOPA_PRL}%
  \BibitemOpen
  \bibfield  {author} {\bibinfo {author} {\bibfnamefont {J.}~\bibnamefont
  {Prior}}, \bibinfo {author} {\bibfnamefont {A.~W.}\ \bibnamefont {Chin}},
  \bibinfo {author} {\bibfnamefont {S.~F.}\ \bibnamefont {Huelga}},\ and\
  \bibinfo {author} {\bibfnamefont {M.~B.}\ \bibnamefont {Plenio}},\ }\bibfield
   {title} {\bibinfo {title} {Efficient simulation of strong system-environment
  interactions},\ }\href {https://doi.org/10.1103/PhysRevLett.105.050404}
  {\bibfield  {journal} {\bibinfo  {journal} {Phys. Rev. Lett.}\ }\textbf
  {\bibinfo {volume} {105}},\ \bibinfo {pages} {050404} (\bibinfo {year}
  {2010})}\BibitemShut {NoStop}%
\bibitem [{\citenamefont {Somoza}\ \emph {et~al.}(2019)\citenamefont {Somoza},
  \citenamefont {Marty}, \citenamefont {Lim}, \citenamefont {Huelga},\ and\
  \citenamefont {Plenio}}]{Somoza2019}%
  \BibitemOpen
  \bibfield  {author} {\bibinfo {author} {\bibfnamefont {A.~D.}\ \bibnamefont
  {Somoza}}, \bibinfo {author} {\bibfnamefont {O.}~\bibnamefont {Marty}},
  \bibinfo {author} {\bibfnamefont {J.}~\bibnamefont {Lim}}, \bibinfo {author}
  {\bibfnamefont {S.~F.}\ \bibnamefont {Huelga}},\ and\ \bibinfo {author}
  {\bibfnamefont {M.~B.}\ \bibnamefont {Plenio}},\ }\bibfield  {title}
  {\bibinfo {title} {Dissipation-assisted matrix product factorization},\
  }\href {https://doi.org/10.1103/PhysRevLett.123.100502} {\bibfield  {journal}
  {\bibinfo  {journal} {Phys. Rev. Lett.}\ }\textbf {\bibinfo {volume} {123}},\
  \bibinfo {pages} {100502} (\bibinfo {year} {2019})}\BibitemShut {NoStop}%
\bibitem [{\citenamefont {N\"u\ss{}eler}\ \emph {et~al.}(2020)\citenamefont
  {N\"u\ss{}eler}, \citenamefont {Dhand}, \citenamefont {Huelga},\ and\
  \citenamefont {Plenio}}]{fTEDOPA}%
  \BibitemOpen
  \bibfield  {author} {\bibinfo {author} {\bibfnamefont {A.}~\bibnamefont
  {N\"u\ss{}eler}}, \bibinfo {author} {\bibfnamefont {I.}~\bibnamefont
  {Dhand}}, \bibinfo {author} {\bibfnamefont {S.~F.}\ \bibnamefont {Huelga}},\
  and\ \bibinfo {author} {\bibfnamefont {M.~B.}\ \bibnamefont {Plenio}},\
  }\bibfield  {title} {\bibinfo {title} {Efficient simulation of open quantum
  systems coupled to a fermionic bath},\ }\href
  {https://doi.org/10.1103/PhysRevB.101.155134} {\bibfield  {journal} {\bibinfo
   {journal} {Phys. Rev. B}\ }\textbf {\bibinfo {volume} {101}},\ \bibinfo
  {pages} {155134} (\bibinfo {year} {2020})}\BibitemShut {NoStop}%
\bibitem [{\citenamefont {Feynman}\ and\ \citenamefont
  {Vernon}(1963)}]{FeynmanVernon}%
  \BibitemOpen
  \bibfield  {author} {\bibinfo {author} {\bibfnamefont {R.}~\bibnamefont
  {Feynman}}\ and\ \bibinfo {author} {\bibfnamefont {F.}~\bibnamefont
  {Vernon}},\ }\bibfield  {title} {\bibinfo {title} {The theory of a general
  quantum system interacting with a linear dissipative system},\ }\href
  {https://doi.org/https://doi.org/10.1016/0003-4916(63)90068-X} {\bibfield
  {journal} {\bibinfo  {journal} {Ann. Phys. (N.Y.)}\ }\textbf {\bibinfo
  {volume} {24}},\ \bibinfo {pages} {118 } (\bibinfo {year}
  {1963})}\BibitemShut {NoStop}%
\bibitem [{\citenamefont {Makri}\ and\ \citenamefont
  {Makarov}(1995{\natexlab{a}})}]{Makri}%
  \BibitemOpen
  \bibfield  {author} {\bibinfo {author} {\bibfnamefont {N.}~\bibnamefont
  {Makri}}\ and\ \bibinfo {author} {\bibfnamefont {D.~E.}\ \bibnamefont
  {Makarov}},\ }\bibfield  {title} {\bibinfo {title} {{Tensor propagator for
  iterative quantum time evolution of reduced density matrices. I. Theory}},\
  }\href {https://doi.org/10.1063/1.469508} {\bibfield  {journal} {\bibinfo
  {journal} {J. Chem. Phys.}\ }\textbf {\bibinfo {volume} {102}},\ \bibinfo
  {pages} {4600} (\bibinfo {year} {1995}{\natexlab{a}})}\BibitemShut {NoStop}%
\bibitem [{\citenamefont {Makri}\ and\ \citenamefont
  {Makarov}(1995{\natexlab{b}})}]{Makri_Iterative2}%
  \BibitemOpen
  \bibfield  {author} {\bibinfo {author} {\bibfnamefont {N.}~\bibnamefont
  {Makri}}\ and\ \bibinfo {author} {\bibfnamefont {D.~E.}\ \bibnamefont
  {Makarov}},\ }\bibfield  {title} {\bibinfo {title} {{Tensor propagator for
  iterative quantum time evolution of reduced density matrices. II. Numerical
  methodology}},\ }\href {https://doi.org/10.1063/1.469509} {\bibfield
  {journal} {\bibinfo  {journal} {J. Chem. Phys.}\ }\textbf {\bibinfo {volume}
  {102}},\ \bibinfo {pages} {4611} (\bibinfo {year}
  {1995}{\natexlab{b}})}\BibitemShut {NoStop}%
\bibitem [{\citenamefont {Cygorek}\ \emph {et~al.}(2017)\citenamefont
  {Cygorek}, \citenamefont {Barth}, \citenamefont {Ungar}, \citenamefont
  {Vagov},\ and\ \citenamefont {Axt}}]{PI_cQED}%
  \BibitemOpen
  \bibfield  {author} {\bibinfo {author} {\bibfnamefont {M.}~\bibnamefont
  {Cygorek}}, \bibinfo {author} {\bibfnamefont {A.~M.}\ \bibnamefont {Barth}},
  \bibinfo {author} {\bibfnamefont {F.}~\bibnamefont {Ungar}}, \bibinfo
  {author} {\bibfnamefont {A.}~\bibnamefont {Vagov}},\ and\ \bibinfo {author}
  {\bibfnamefont {V.~M.}\ \bibnamefont {Axt}},\ }\bibfield  {title} {\bibinfo
  {title} {Nonlinear cavity feeding and unconventional photon statistics in
  solid-state cavity qed revealed by many-level real-time path-integral
  calculations},\ }\href {https://doi.org/10.1103/PhysRevB.96.201201}
  {\bibfield  {journal} {\bibinfo  {journal} {Phys. Rev. B}\ }\textbf {\bibinfo
  {volume} {96}},\ \bibinfo {pages} {201201} (\bibinfo {year}
  {2017})}\BibitemShut {NoStop}%
\bibitem [{\citenamefont {Strathearn}\ \emph {et~al.}(2018)\citenamefont
  {Strathearn}, \citenamefont {Kirton}, \citenamefont {Kilda}, \citenamefont
  {Keeling},\ and\ \citenamefont {Lovett}}]{TEMPO}%
  \BibitemOpen
  \bibfield  {author} {\bibinfo {author} {\bibfnamefont {A.}~\bibnamefont
  {Strathearn}}, \bibinfo {author} {\bibfnamefont {P.}~\bibnamefont {Kirton}},
  \bibinfo {author} {\bibfnamefont {D.}~\bibnamefont {Kilda}}, \bibinfo
  {author} {\bibfnamefont {J.}~\bibnamefont {Keeling}},\ and\ \bibinfo {author}
  {\bibfnamefont {B.~W.}\ \bibnamefont {Lovett}},\ }\bibfield  {title}
  {\bibinfo {title} {Efficient non-markovian quantum dynamics using
  time-evolving matrix product operators},\ }\href
  {https://doi.org/10.1038/s41467-018-05617-3} {\bibfield  {journal} {\bibinfo
  {journal} {Nat. Commun.}\ }\textbf {\bibinfo {volume} {9}},\ \bibinfo {pages}
  {3322} (\bibinfo {year} {2018})}\BibitemShut {NoStop}%
\bibitem [{\citenamefont {Cosacchi}\ \emph {et~al.}(2018)\citenamefont
  {Cosacchi}, \citenamefont {Cygorek}, \citenamefont {Ungar}, \citenamefont
  {Barth}, \citenamefont {Vagov},\ and\ \citenamefont {Axt}}]{PI_multitime}%
  \BibitemOpen
  \bibfield  {author} {\bibinfo {author} {\bibfnamefont {M.}~\bibnamefont
  {Cosacchi}}, \bibinfo {author} {\bibfnamefont {M.}~\bibnamefont {Cygorek}},
  \bibinfo {author} {\bibfnamefont {F.}~\bibnamefont {Ungar}}, \bibinfo
  {author} {\bibfnamefont {A.~M.}\ \bibnamefont {Barth}}, \bibinfo {author}
  {\bibfnamefont {A.}~\bibnamefont {Vagov}},\ and\ \bibinfo {author}
  {\bibfnamefont {V.~M.}\ \bibnamefont {Axt}},\ }\bibfield  {title} {\bibinfo
  {title} {Path-integral approach for nonequilibrium multitime correlation
  functions of open quantum systems coupled to markovian and non-markovian
  environments},\ }\href {https://doi.org/10.1103/PhysRevB.98.125302}
  {\bibfield  {journal} {\bibinfo  {journal} {Phys. Rev. B}\ }\textbf {\bibinfo
  {volume} {98}},\ \bibinfo {pages} {125302} (\bibinfo {year}
  {2018})}\BibitemShut {NoStop}%
\bibitem [{\citenamefont {{Denning}}\ \emph {et~al.}(2020)\citenamefont
  {{Denning}}, \citenamefont {{Bundgaard-Nielsen}},\ and\ \citenamefont {{M\o
  rk}}}]{Denning_phonon_decoupling}%
  \BibitemOpen
  \bibfield  {author} {\bibinfo {author} {\bibfnamefont {E.~V.}\ \bibnamefont
  {{Denning}}}, \bibinfo {author} {\bibfnamefont {M.}~\bibnamefont
  {{Bundgaard-Nielsen}}},\ and\ \bibinfo {author} {\bibfnamefont
  {J.}~\bibnamefont {{M\o rk}}},\ }\bibfield  {title} {\bibinfo {title}
  {{Electron-phonon decoupling due to strong light-matter interactions}},\
  }\Eprint {https://arxiv.org/abs/2007.14719} {2007.14719}  (\bibinfo {year}
  {2020}),\ \bibinfo {note} {preprint}\BibitemShut {NoStop}%
\bibitem [{\citenamefont {Cosacchi}\ \emph {et~al.}(2019)\citenamefont
  {Cosacchi}, \citenamefont {Ungar}, \citenamefont {Cygorek}, \citenamefont
  {Vagov},\ and\ \citenamefont {Axt}}]{PRL_singlephoton}%
  \BibitemOpen
  \bibfield  {author} {\bibinfo {author} {\bibfnamefont {M.}~\bibnamefont
  {Cosacchi}}, \bibinfo {author} {\bibfnamefont {F.}~\bibnamefont {Ungar}},
  \bibinfo {author} {\bibfnamefont {M.}~\bibnamefont {Cygorek}}, \bibinfo
  {author} {\bibfnamefont {A.}~\bibnamefont {Vagov}},\ and\ \bibinfo {author}
  {\bibfnamefont {V.~M.}\ \bibnamefont {Axt}},\ }\bibfield  {title} {\bibinfo
  {title} {Emission-frequency separated high quality single-photon sources
  enabled by phonons},\ }\href {https://doi.org/10.1103/PhysRevLett.123.017403}
  {\bibfield  {journal} {\bibinfo  {journal} {Phys. Rev. Lett.}\ }\textbf
  {\bibinfo {volume} {123}},\ \bibinfo {pages} {017403} (\bibinfo {year}
  {2019})}\BibitemShut {NoStop}%
\bibitem [{\citenamefont {Seidelmann}\ \emph {et~al.}(2019)\citenamefont
  {Seidelmann}, \citenamefont {Ungar}, \citenamefont {Barth}, \citenamefont
  {Vagov}, \citenamefont {Axt}, \citenamefont {Cygorek},\ and\ \citenamefont
  {Kuhn}}]{PRL_concurrence}%
  \BibitemOpen
  \bibfield  {author} {\bibinfo {author} {\bibfnamefont {T.}~\bibnamefont
  {Seidelmann}}, \bibinfo {author} {\bibfnamefont {F.}~\bibnamefont {Ungar}},
  \bibinfo {author} {\bibfnamefont {A.~M.}\ \bibnamefont {Barth}}, \bibinfo
  {author} {\bibfnamefont {A.}~\bibnamefont {Vagov}}, \bibinfo {author}
  {\bibfnamefont {V.~M.}\ \bibnamefont {Axt}}, \bibinfo {author} {\bibfnamefont
  {M.}~\bibnamefont {Cygorek}},\ and\ \bibinfo {author} {\bibfnamefont
  {T.}~\bibnamefont {Kuhn}},\ }\bibfield  {title} {\bibinfo {title}
  {Phonon-induced enhancement of photon entanglement in quantum dot-cavity
  systems},\ }\href {https://doi.org/10.1103/PhysRevLett.123.137401} {\bibfield
   {journal} {\bibinfo  {journal} {Phys. Rev. Lett.}\ }\textbf {\bibinfo
  {volume} {123}},\ \bibinfo {pages} {137401} (\bibinfo {year}
  {2019})}\BibitemShut {NoStop}%
\bibitem [{\citenamefont {{Kaestle}}\ \emph {et~al.}(2020)\citenamefont
  {{Kaestle}}, \citenamefont {{Finsterhoelzl}}, \citenamefont {{Knorr}},\ and\
  \citenamefont {{Carmele}}}]{ExtTEMPOKaestle}%
  \BibitemOpen
  \bibfield  {author} {\bibinfo {author} {\bibfnamefont {O.}~\bibnamefont
  {{Kaestle}}}, \bibinfo {author} {\bibfnamefont {R.}~\bibnamefont
  {{Finsterhoelzl}}}, \bibinfo {author} {\bibfnamefont {A.}~\bibnamefont
  {{Knorr}}},\ and\ \bibinfo {author} {\bibfnamefont {A.}~\bibnamefont
  {{Carmele}}},\ }\href@noop {} {\bibinfo {title} {{Protected quantum
  correlations in multiple non-Markovian system-reservoir dynamics}}} (\bibinfo
  {year} {2020}),\ \bibinfo {note} {preprint},\ \Eprint
  {https://arxiv.org/abs/2011.05071} {2011.05071} \BibitemShut {NoStop}%
\bibitem [{\citenamefont {Vagov}\ \emph {et~al.}(2011)\citenamefont {Vagov},
  \citenamefont {Croitoru}, \citenamefont {Gl\"assl}, \citenamefont {Axt},\
  and\ \citenamefont {Kuhn}}]{PI_2011}%
  \BibitemOpen
  \bibfield  {author} {\bibinfo {author} {\bibfnamefont {A.}~\bibnamefont
  {Vagov}}, \bibinfo {author} {\bibfnamefont {M.~D.}\ \bibnamefont {Croitoru}},
  \bibinfo {author} {\bibfnamefont {M.}~\bibnamefont {Gl\"assl}}, \bibinfo
  {author} {\bibfnamefont {V.~M.}\ \bibnamefont {Axt}},\ and\ \bibinfo {author}
  {\bibfnamefont {T.}~\bibnamefont {Kuhn}},\ }\bibfield  {title} {\bibinfo
  {title} {Real-time path integrals for quantum dots: Quantum dissipative
  dynamics with superohmic environment coupling},\ }\href
  {https://doi.org/10.1103/PhysRevB.83.094303} {\bibfield  {journal} {\bibinfo
  {journal} {Phys. Rev. B}\ }\textbf {\bibinfo {volume} {83}},\ \bibinfo
  {pages} {094303} (\bibinfo {year} {2011})}\BibitemShut {NoStop}%
\bibitem [{\citenamefont {Quilter}\ \emph {et~al.}(2015)\citenamefont
  {Quilter}, \citenamefont {Brash}, \citenamefont {Liu}, \citenamefont
  {Gl\"assl}, \citenamefont {Barth}, \citenamefont {Axt}, \citenamefont
  {Ramsay}, \citenamefont {Skolnick},\ and\ \citenamefont {Fox}}]{PRLQuilter}%
  \BibitemOpen
  \bibfield  {author} {\bibinfo {author} {\bibfnamefont {J.~H.}\ \bibnamefont
  {Quilter}}, \bibinfo {author} {\bibfnamefont {A.~J.}\ \bibnamefont {Brash}},
  \bibinfo {author} {\bibfnamefont {F.}~\bibnamefont {Liu}}, \bibinfo {author}
  {\bibfnamefont {M.}~\bibnamefont {Gl\"assl}}, \bibinfo {author}
  {\bibfnamefont {A.~M.}\ \bibnamefont {Barth}}, \bibinfo {author}
  {\bibfnamefont {V.~M.}\ \bibnamefont {Axt}}, \bibinfo {author} {\bibfnamefont
  {A.~J.}\ \bibnamefont {Ramsay}}, \bibinfo {author} {\bibfnamefont {M.~S.}\
  \bibnamefont {Skolnick}},\ and\ \bibinfo {author} {\bibfnamefont {A.~M.}\
  \bibnamefont {Fox}},\ }\bibfield  {title} {\bibinfo {title} {Phonon-assisted
  population inversion of a single $\mathrm{InGaAs}/\mathrm{GaAs}$ quantum dot
  by pulsed laser excitation},\ }\href
  {https://doi.org/10.1103/PhysRevLett.114.137401} {\bibfield  {journal}
  {\bibinfo  {journal} {Phys. Rev. Lett.}\ }\textbf {\bibinfo {volume} {114}},\
  \bibinfo {pages} {137401} (\bibinfo {year} {2015})}\BibitemShut {NoStop}%
\bibitem [{\citenamefont {{Koong}}\ \emph {et~al.}(2020)\citenamefont
  {{Koong}}, \citenamefont {{Scerri}}, \citenamefont {{Rambach}}, \citenamefont
  {{Cygorek}}, \citenamefont {{Brotons-Gisbert}}, \citenamefont {{Picard}},
  \citenamefont {{Ma}}, \citenamefont {{Park}}, \citenamefont {{Song}},
  \citenamefont {{Gauger}},\ and\ \citenamefont {{Gerardot}}}]{dichromatic}%
  \BibitemOpen
  \bibfield  {author} {\bibinfo {author} {\bibfnamefont {Z.~X.}\ \bibnamefont
  {{Koong}}}, \bibinfo {author} {\bibfnamefont {E.}~\bibnamefont {{Scerri}}},
  \bibinfo {author} {\bibfnamefont {M.}~\bibnamefont {{Rambach}}}, \bibinfo
  {author} {\bibfnamefont {M.}~\bibnamefont {{Cygorek}}}, \bibinfo {author}
  {\bibfnamefont {M.}~\bibnamefont {{Brotons-Gisbert}}}, \bibinfo {author}
  {\bibfnamefont {R.}~\bibnamefont {{Picard}}}, \bibinfo {author}
  {\bibfnamefont {Y.}~\bibnamefont {{Ma}}}, \bibinfo {author} {\bibfnamefont
  {S.~I.}\ \bibnamefont {{Park}}}, \bibinfo {author} {\bibfnamefont {J.~D.}\
  \bibnamefont {{Song}}}, \bibinfo {author} {\bibfnamefont {E.~M.}\
  \bibnamefont {{Gauger}}},\ and\ \bibinfo {author} {\bibfnamefont {B.~D.}\
  \bibnamefont {{Gerardot}}},\ }\bibfield  {title} {\bibinfo {title} {Coherent
  dynamics in quantum emitters under dichromatic excitation},\ }\Eprint
  {https://arxiv.org/abs/2009.02121} {2009.02121}  (\bibinfo {year} {2020}),\
  \bibinfo {note} {preprint}\BibitemShut {NoStop}%
\bibitem [{\citenamefont {Palm}\ and\ \citenamefont
  {Nalbach}(2018)}]{Nalbach_multiEnv}%
  \BibitemOpen
  \bibfield  {author} {\bibinfo {author} {\bibfnamefont {T.}~\bibnamefont
  {Palm}}\ and\ \bibinfo {author} {\bibfnamefont {P.}~\bibnamefont {Nalbach}},\
  }\bibfield  {title} {\bibinfo {title} {Quasi-adiabatic path integral approach
  for quantum systems under the influence of multiple non-commuting
  fluctuations},\ }\href {https://doi.org/10.1063/1.5051652} {\bibfield
  {journal} {\bibinfo  {journal} {The Journal of Chemical Physics}\ }\textbf
  {\bibinfo {volume} {149}},\ \bibinfo {pages} {214103} (\bibinfo {year}
  {2018})}\BibitemShut {NoStop}%
\bibitem [{\citenamefont {Simine}\ and\ \citenamefont
  {Segal}(2013)}]{Segal_quadratic_ferm_bos}%
  \BibitemOpen
  \bibfield  {author} {\bibinfo {author} {\bibfnamefont {L.}~\bibnamefont
  {Simine}}\ and\ \bibinfo {author} {\bibfnamefont {D.}~\bibnamefont {Segal}},\
  }\bibfield  {title} {\bibinfo {title} {Path-integral simulations with
  fermionic and bosonic reservoirs: Transport and dissipation in molecular
  electronic junctions},\ }\href {https://doi.org/10.1063/1.4808108} {\bibfield
   {journal} {\bibinfo  {journal} {The Journal of Chemical Physics}\ }\textbf
  {\bibinfo {volume} {138}},\ \bibinfo {pages} {214111} (\bibinfo {year}
  {2013})}\BibitemShut {NoStop}%
\bibitem [{\citenamefont {Rossi}\ and\ \citenamefont
  {Kuhn}(2002)}]{RossiKuhn2002}%
  \BibitemOpen
  \bibfield  {author} {\bibinfo {author} {\bibfnamefont {F.}~\bibnamefont
  {Rossi}}\ and\ \bibinfo {author} {\bibfnamefont {T.}~\bibnamefont {Kuhn}},\
  }\bibfield  {title} {\bibinfo {title} {Theory of ultrafast phenomena in
  photoexcited semiconductors},\ }\href
  {https://doi.org/10.1103/RevModPhys.74.895} {\bibfield  {journal} {\bibinfo
  {journal} {Rev. Mod. Phys.}\ }\textbf {\bibinfo {volume} {74}},\ \bibinfo
  {pages} {895} (\bibinfo {year} {2002})}\BibitemShut {NoStop}%
\bibitem [{\citenamefont {J\o{}rgensen}\ and\ \citenamefont
  {Pollock}(2019)}]{ProcessTensor}%
  \BibitemOpen
  \bibfield  {author} {\bibinfo {author} {\bibfnamefont {M.~R.}\ \bibnamefont
  {J\o{}rgensen}}\ and\ \bibinfo {author} {\bibfnamefont {F.~A.}\ \bibnamefont
  {Pollock}},\ }\bibfield  {title} {\bibinfo {title} {Exploiting the causal
  tensor network structure of quantum processes to efficiently simulate
  non-markovian path integrals},\ }\href
  {https://doi.org/10.1103/PhysRevLett.123.240602} {\bibfield  {journal}
  {\bibinfo  {journal} {Phys. Rev. Lett.}\ }\textbf {\bibinfo {volume} {123}},\
  \bibinfo {pages} {240602} (\bibinfo {year} {2019})}\BibitemShut {NoStop}%
\bibitem [{\citenamefont {Pollock}\ \emph {et~al.}(2018)\citenamefont
  {Pollock}, \citenamefont {Rodr\'{\i}guez-Rosario}, \citenamefont
  {Frauenheim}, \citenamefont {Paternostro},\ and\ \citenamefont
  {Modi}}]{ProcessTensor_PRA}%
  \BibitemOpen
  \bibfield  {author} {\bibinfo {author} {\bibfnamefont {F.~A.}\ \bibnamefont
  {Pollock}}, \bibinfo {author} {\bibfnamefont {C.}~\bibnamefont
  {Rodr\'{\i}guez-Rosario}}, \bibinfo {author} {\bibfnamefont {T.}~\bibnamefont
  {Frauenheim}}, \bibinfo {author} {\bibfnamefont {M.}~\bibnamefont
  {Paternostro}},\ and\ \bibinfo {author} {\bibfnamefont {K.}~\bibnamefont
  {Modi}},\ }\bibfield  {title} {\bibinfo {title} {Non-markovian quantum
  processes: Complete framework and efficient characterization},\ }\href
  {https://doi.org/10.1103/PhysRevA.97.012127} {\bibfield  {journal} {\bibinfo
  {journal} {Phys. Rev. A}\ }\textbf {\bibinfo {volume} {97}},\ \bibinfo
  {pages} {012127} (\bibinfo {year} {2018})}\BibitemShut {NoStop}%
\bibitem [{\citenamefont {Schollwöck}(2011)}]{MPS_Schollwoeck}%
  \BibitemOpen
  \bibfield  {author} {\bibinfo {author} {\bibfnamefont {U.}~\bibnamefont
  {Schollwöck}},\ }\bibfield  {title} {\bibinfo {title} {The density-matrix
  renormalization group in the age of matrix product states},\ }\href
  {https://doi.org/10.1016/j.aop.2010.09.012} {\bibfield  {journal} {\bibinfo
  {journal} {Ann. Phys. (N.Y.)}\ }\textbf {\bibinfo {volume} {326}},\ \bibinfo
  {pages} {96 } (\bibinfo {year} {2011})}\BibitemShut {NoStop}%
\bibitem [{\citenamefont {Orús}(2014)}]{MPS_Orus}%
  \BibitemOpen
  \bibfield  {author} {\bibinfo {author} {\bibfnamefont {R.}~\bibnamefont
  {Orús}},\ }\bibfield  {title} {\bibinfo {title} {A practical introduction to
  tensor networks: Matrix product states and projected entangled pair states},\
  }\href {https://doi.org/https://doi.org/10.1016/j.aop.2014.06.013} {\bibfield
   {journal} {\bibinfo  {journal} {Ann. Phys. (N.Y.)}\ }\textbf {\bibinfo
  {volume} {349}},\ \bibinfo {pages} {117 } (\bibinfo {year}
  {2014})}\BibitemShut {NoStop}%
\bibitem [{\citenamefont {Luchnikov}\ \emph {et~al.}(2019)\citenamefont
  {Luchnikov}, \citenamefont {Vintskevich}, \citenamefont {Ouerdane},\ and\
  \citenamefont {Filippov}}]{Luchnikov2019}%
  \BibitemOpen
  \bibfield  {author} {\bibinfo {author} {\bibfnamefont {I.~A.}\ \bibnamefont
  {Luchnikov}}, \bibinfo {author} {\bibfnamefont {S.~V.}\ \bibnamefont
  {Vintskevich}}, \bibinfo {author} {\bibfnamefont {H.}~\bibnamefont
  {Ouerdane}},\ and\ \bibinfo {author} {\bibfnamefont {S.~N.}\ \bibnamefont
  {Filippov}},\ }\bibfield  {title} {\bibinfo {title} {Simulation complexity of
  open quantum dynamics: Connection with tensor networks},\ }\href
  {https://doi.org/10.1103/PhysRevLett.122.160401} {\bibfield  {journal}
  {\bibinfo  {journal} {Phys. Rev. Lett.}\ }\textbf {\bibinfo {volume} {122}},\
  \bibinfo {pages} {160401} (\bibinfo {year} {2019})}\BibitemShut {NoStop}%
\bibitem [{\citenamefont {Brandes}\ and\ \citenamefont
  {Kramer}(1999)}]{Brandes_doubledot}%
  \BibitemOpen
  \bibfield  {author} {\bibinfo {author} {\bibfnamefont {T.}~\bibnamefont
  {Brandes}}\ and\ \bibinfo {author} {\bibfnamefont {B.}~\bibnamefont
  {Kramer}},\ }\bibfield  {title} {\bibinfo {title} {Spontaneous emission of
  phonons by coupled quantum dots},\ }\href
  {https://doi.org/10.1103/PhysRevLett.83.3021} {\bibfield  {journal} {\bibinfo
   {journal} {Phys. Rev. Lett.}\ }\textbf {\bibinfo {volume} {83}},\ \bibinfo
  {pages} {3021} (\bibinfo {year} {1999})}\BibitemShut {NoStop}%
\bibitem [{\citenamefont {Barth}\ \emph {et~al.}(2016)\citenamefont {Barth},
  \citenamefont {Vagov},\ and\ \citenamefont {Axt}}]{Andi_nonHamil}%
  \BibitemOpen
  \bibfield  {author} {\bibinfo {author} {\bibfnamefont {A.~M.}\ \bibnamefont
  {Barth}}, \bibinfo {author} {\bibfnamefont {A.}~\bibnamefont {Vagov}},\ and\
  \bibinfo {author} {\bibfnamefont {V.~M.}\ \bibnamefont {Axt}},\ }\bibfield
  {title} {\bibinfo {title} {Path-integral description of combined
  {Hamiltonian} and non-{Hamiltonian} dynamics in quantum dissipative
  systems},\ }\href {https://doi.org/10.1103/PhysRevB.94.125439} {\bibfield
  {journal} {\bibinfo  {journal} {Phys. Rev. B}\ }\textbf {\bibinfo {volume}
  {94}},\ \bibinfo {pages} {125439} (\bibinfo {year} {2016})}\BibitemShut
  {NoStop}%
\bibitem [{\citenamefont {Nagy}\ \emph {et~al.}(2011)\citenamefont {Nagy},
  \citenamefont {Szirmai},\ and\ \citenamefont {Domokos}}]{Nagy2011}%
  \BibitemOpen
  \bibfield  {author} {\bibinfo {author} {\bibfnamefont {D.}~\bibnamefont
  {Nagy}}, \bibinfo {author} {\bibfnamefont {G.}~\bibnamefont {Szirmai}},\ and\
  \bibinfo {author} {\bibfnamefont {P.}~\bibnamefont {Domokos}},\ }\bibfield
  {title} {\bibinfo {title} {Critical exponent of a quantum-noise-driven phase
  transition: The open-system dicke model},\ }\href
  {https://doi.org/10.1103/PhysRevA.84.043637} {\bibfield  {journal} {\bibinfo
  {journal} {Phys. Rev. A}\ }\textbf {\bibinfo {volume} {84}},\ \bibinfo
  {pages} {043637} (\bibinfo {year} {2011})}\BibitemShut {NoStop}%
\bibitem [{\citenamefont {Mitchison}\ and\ \citenamefont
  {Plenio}(2018)}]{Mitchison2018}%
  \BibitemOpen
  \bibfield  {author} {\bibinfo {author} {\bibfnamefont {M.~T.}\ \bibnamefont
  {Mitchison}}\ and\ \bibinfo {author} {\bibfnamefont {M.~B.}\ \bibnamefont
  {Plenio}},\ }\bibfield  {title} {\bibinfo {title} {Non-additive dissipation
  in open quantum networks out of equilibrium},\ }\href
  {https://doi.org/10.1088/1367-2630/aa9f70} {\bibfield  {journal} {\bibinfo
  {journal} {New J. Phys.}\ }\textbf {\bibinfo {volume} {20}},\ \bibinfo
  {pages} {033005} (\bibinfo {year} {2018})}\BibitemShut {NoStop}%
\bibitem [{\citenamefont {Maguire}\ \emph {et~al.}(2019)\citenamefont
  {Maguire}, \citenamefont {Iles-Smith},\ and\ \citenamefont
  {Nazir}}]{Nazir_Nonadditivity}%
  \BibitemOpen
  \bibfield  {author} {\bibinfo {author} {\bibfnamefont {H.}~\bibnamefont
  {Maguire}}, \bibinfo {author} {\bibfnamefont {J.}~\bibnamefont
  {Iles-Smith}},\ and\ \bibinfo {author} {\bibfnamefont {A.}~\bibnamefont
  {Nazir}},\ }\bibfield  {title} {\bibinfo {title} {Environmental nonadditivity
  and {Franck--Condon} physics in nonequilibrium quantum systems},\ }\href
  {https://doi.org/10.1103/PhysRevLett.123.093601} {\bibfield  {journal}
  {\bibinfo  {journal} {Phys. Rev. Lett.}\ }\textbf {\bibinfo {volume} {123}},\
  \bibinfo {pages} {093601} (\bibinfo {year} {2019})}\BibitemShut {NoStop}%
\bibitem [{\citenamefont {Roy-Choudhury}\ and\ \citenamefont
  {Hughes}(2015)}]{Hughes_structured}%
  \BibitemOpen
  \bibfield  {author} {\bibinfo {author} {\bibfnamefont {K.}~\bibnamefont
  {Roy-Choudhury}}\ and\ \bibinfo {author} {\bibfnamefont {S.}~\bibnamefont
  {Hughes}},\ }\bibfield  {title} {\bibinfo {title} {Spontaneous emission from
  a quantum dot in a structured photonic reservoir: phonon-mediated breakdown
  of {Fermi}'s golden rule},\ }\href {https://doi.org/10.1364/OPTICA.2.000434}
  {\bibfield  {journal} {\bibinfo  {journal} {Optica}\ }\textbf {\bibinfo
  {volume} {2}},\ \bibinfo {pages} {434} (\bibinfo {year} {2015})}\BibitemShut
  {NoStop}%
\bibitem [{\citenamefont {Hoeppe}\ \emph {et~al.}(2012)\citenamefont {Hoeppe},
  \citenamefont {Wolff}, \citenamefont {K\"uchenmeister}, \citenamefont
  {Niegemann}, \citenamefont {Drescher}, \citenamefont {Benner},\ and\
  \citenamefont {Busch}}]{Busch_structured}%
  \BibitemOpen
  \bibfield  {author} {\bibinfo {author} {\bibfnamefont {U.}~\bibnamefont
  {Hoeppe}}, \bibinfo {author} {\bibfnamefont {C.}~\bibnamefont {Wolff}},
  \bibinfo {author} {\bibfnamefont {J.}~\bibnamefont {K\"uchenmeister}},
  \bibinfo {author} {\bibfnamefont {J.}~\bibnamefont {Niegemann}}, \bibinfo
  {author} {\bibfnamefont {M.}~\bibnamefont {Drescher}}, \bibinfo {author}
  {\bibfnamefont {H.}~\bibnamefont {Benner}},\ and\ \bibinfo {author}
  {\bibfnamefont {K.}~\bibnamefont {Busch}},\ }\bibfield  {title} {\bibinfo
  {title} {Direct observation of non-markovian radiation dynamics in 3d bulk
  photonic crystals},\ }\href {https://doi.org/10.1103/PhysRevLett.108.043603}
  {\bibfield  {journal} {\bibinfo  {journal} {Phys. Rev. Lett.}\ }\textbf
  {\bibinfo {volume} {108}},\ \bibinfo {pages} {043603} (\bibinfo {year}
  {2012})}\BibitemShut {NoStop}%
\bibitem [{\citenamefont {Gangloff}\ \emph {et~al.}(2019)\citenamefont
  {Gangloff}, \citenamefont {{\'E}thier-Majcher}, \citenamefont {Lang},
  \citenamefont {Denning}, \citenamefont {Bodey}, \citenamefont {Jackson},
  \citenamefont {Clarke}, \citenamefont {Hugues}, \citenamefont {Le~Gall},\
  and\ \citenamefont {Atat{\"u}re}}]{SpinInterface}%
  \BibitemOpen
  \bibfield  {author} {\bibinfo {author} {\bibfnamefont {D.~A.}\ \bibnamefont
  {Gangloff}}, \bibinfo {author} {\bibfnamefont {G.}~\bibnamefont
  {{\'E}thier-Majcher}}, \bibinfo {author} {\bibfnamefont {C.}~\bibnamefont
  {Lang}}, \bibinfo {author} {\bibfnamefont {E.~V.}\ \bibnamefont {Denning}},
  \bibinfo {author} {\bibfnamefont {J.~H.}\ \bibnamefont {Bodey}}, \bibinfo
  {author} {\bibfnamefont {D.~M.}\ \bibnamefont {Jackson}}, \bibinfo {author}
  {\bibfnamefont {E.}~\bibnamefont {Clarke}}, \bibinfo {author} {\bibfnamefont
  {M.}~\bibnamefont {Hugues}}, \bibinfo {author} {\bibfnamefont
  {C.}~\bibnamefont {Le~Gall}},\ and\ \bibinfo {author} {\bibfnamefont
  {M.}~\bibnamefont {Atat{\"u}re}},\ }\bibfield  {title} {\bibinfo {title}
  {Quantum interface of an electron and a nuclear ensemble},\ }\href
  {https://doi.org/10.1126/science.aaw2906} {\bibfield  {journal} {\bibinfo
  {journal} {Science}\ }\textbf {\bibinfo {volume} {364}},\ \bibinfo {pages}
  {62} (\bibinfo {year} {2019})}\BibitemShut {NoStop}%
\bibitem [{\citenamefont {Scheuer}\ \emph {et~al.}(2017)\citenamefont
  {Scheuer}, \citenamefont {Schwartz}, \citenamefont {M\"uller}, \citenamefont
  {Chen}, \citenamefont {Dhand}, \citenamefont {Plenio}, \citenamefont
  {Naydenov},\ and\ \citenamefont {Jelezko}}]{Jelezko_PRB17}%
  \BibitemOpen
  \bibfield  {author} {\bibinfo {author} {\bibfnamefont {J.}~\bibnamefont
  {Scheuer}}, \bibinfo {author} {\bibfnamefont {I.}~\bibnamefont {Schwartz}},
  \bibinfo {author} {\bibfnamefont {S.}~\bibnamefont {M\"uller}}, \bibinfo
  {author} {\bibfnamefont {Q.}~\bibnamefont {Chen}}, \bibinfo {author}
  {\bibfnamefont {I.}~\bibnamefont {Dhand}}, \bibinfo {author} {\bibfnamefont
  {M.~B.}\ \bibnamefont {Plenio}}, \bibinfo {author} {\bibfnamefont
  {B.}~\bibnamefont {Naydenov}},\ and\ \bibinfo {author} {\bibfnamefont
  {F.}~\bibnamefont {Jelezko}},\ }\bibfield  {title} {\bibinfo {title} {Robust
  techniques for polarization and detection of nuclear spin ensembles},\ }\href
  {https://doi.org/10.1103/PhysRevB.96.174436} {\bibfield  {journal} {\bibinfo
  {journal} {Phys. Rev. B}\ }\textbf {\bibinfo {volume} {96}},\ \bibinfo
  {pages} {174436} (\bibinfo {year} {2017})}\BibitemShut {NoStop}%
\bibitem [{\citenamefont {Bramberger}\ and\ \citenamefont
  {De~Vega}(2020)}]{Morse_deVega}%
  \BibitemOpen
  \bibfield  {author} {\bibinfo {author} {\bibfnamefont {M.}~\bibnamefont
  {Bramberger}}\ and\ \bibinfo {author} {\bibfnamefont {I.}~\bibnamefont
  {De~Vega}},\ }\bibfield  {title} {\bibinfo {title} {Dephasing dynamics of an
  impurity coupled to an anharmonic environment},\ }\href
  {https://doi.org/10.1103/PhysRevA.101.012101} {\bibfield  {journal} {\bibinfo
   {journal} {Phys. Rev. A}\ }\textbf {\bibinfo {volume} {101}},\ \bibinfo
  {pages} {012101} (\bibinfo {year} {2020})}\BibitemShut {NoStop}%
\bibitem [{\citenamefont {Ye}\ and\ \citenamefont
  {Chan}(2021)}]{ye2021constructing}%
  \BibitemOpen
  \bibfield  {author} {\bibinfo {author} {\bibfnamefont {E.}~\bibnamefont
  {Ye}}\ and\ \bibinfo {author} {\bibfnamefont {G.~K.-L.}\ \bibnamefont
  {Chan}},\ }\bibfield  {title} {\bibinfo {title} {Constructing tensor network
  influence functionals for general quantum dynamics},\ }\href
  {https://doi.org/10.1063/5.0047260} {\bibfield  {journal} {\bibinfo
  {journal} {The Journal of Chemical Physics}\ }\textbf {\bibinfo {volume}
  {155}},\ \bibinfo {pages} {044104} (\bibinfo {year} {2021})}\BibitemShut
  {NoStop}%
\bibitem [{\citenamefont {Ba\~nuls}\ \emph {et~al.}(2009)\citenamefont
  {Ba\~nuls}, \citenamefont {Hastings}, \citenamefont {Verstraete},\ and\
  \citenamefont {Cirac}}]{Banuls2009}%
  \BibitemOpen
  \bibfield  {author} {\bibinfo {author} {\bibfnamefont {M.~C.}\ \bibnamefont
  {Ba\~nuls}}, \bibinfo {author} {\bibfnamefont {M.~B.}\ \bibnamefont
  {Hastings}}, \bibinfo {author} {\bibfnamefont {F.}~\bibnamefont
  {Verstraete}},\ and\ \bibinfo {author} {\bibfnamefont {J.~I.}\ \bibnamefont
  {Cirac}},\ }\bibfield  {title} {\bibinfo {title} {Matrix product states for
  dynamical simulation of infinite chains},\ }\href
  {https://doi.org/10.1103/PhysRevLett.102.240603} {\bibfield  {journal}
  {\bibinfo  {journal} {Phys. Rev. Lett.}\ }\textbf {\bibinfo {volume} {102}},\
  \bibinfo {pages} {240603} (\bibinfo {year} {2009})}\BibitemShut {NoStop}%
\bibitem [{\citenamefont {Krummheuer}\ \emph {et~al.}(2005)\citenamefont
  {Krummheuer}, \citenamefont {Axt}, \citenamefont {Kuhn}, \citenamefont
  {D'Amico},\ and\ \citenamefont {Rossi}}]{Krummheuer}%
  \BibitemOpen
  \bibfield  {author} {\bibinfo {author} {\bibfnamefont {B.}~\bibnamefont
  {Krummheuer}}, \bibinfo {author} {\bibfnamefont {V.~M.}\ \bibnamefont {Axt}},
  \bibinfo {author} {\bibfnamefont {T.}~\bibnamefont {Kuhn}}, \bibinfo {author}
  {\bibfnamefont {I.}~\bibnamefont {D'Amico}},\ and\ \bibinfo {author}
  {\bibfnamefont {F.}~\bibnamefont {Rossi}},\ }\bibfield  {title} {\bibinfo
  {title} {Pure dephasing and phonon dynamics in gaas- and gan-based quantum
  dot structures: Interplay between material parameters and geometry},\ }\href
  {https://doi.org/10.1103/PhysRevB.71.235329} {\bibfield  {journal} {\bibinfo
  {journal} {Phys. Rev. B}\ }\textbf {\bibinfo {volume} {71}},\ \bibinfo
  {pages} {235329} (\bibinfo {year} {2005})}\BibitemShut {NoStop}%
\end{thebibliography}%


\begin{thebibliography}{12}%
\makeatletter
\providecommand \@ifxundefined [1]{%
 \@ifx{#1\undefined}
}%
\providecommand \@ifnum [1]{%
 \ifnum #1\expandafter \@firstoftwo
 \else \expandafter \@secondoftwo
 \fi
}%
\providecommand \@ifx [1]{%
 \ifx #1\expandafter \@firstoftwo
 \else \expandafter \@secondoftwo
 \fi
}%
\providecommand \natexlab [1]{#1}%
\providecommand \enquote  [1]{``#1''}%
\providecommand \bibnamefont  [1]{#1}%
\providecommand \bibfnamefont [1]{#1}%
\providecommand \citenamefont [1]{#1}%
\providecommand \href@noop [0]{\@secondoftwo}%
\providecommand \href [0]{\begingroup \@sanitize@url \@href}%
\providecommand \@href[1]{\@@startlink{#1}\@@href}%
\providecommand \@@href[1]{\endgroup#1\@@endlink}%
\providecommand \@sanitize@url [0]{\catcode `\\12\catcode `\$12\catcode
  `\&12\catcode `\#12\catcode `\^12\catcode `\_12\catcode `\%12\relax}%
\providecommand \@@startlink[1]{}%
\providecommand \@@endlink[0]{}%
\providecommand \url  [0]{\begingroup\@sanitize@url \@url }%
\providecommand \@url [1]{\endgroup\@href {#1}{\urlprefix }}%
\providecommand \urlprefix  [0]{URL }%
\providecommand \Eprint [0]{\href }%
\providecommand \doibase [0]{https://doi.org/}%
\providecommand \selectlanguage [0]{\@gobble}%
\providecommand \bibinfo  [0]{\@secondoftwo}%
\providecommand \bibfield  [0]{\@secondoftwo}%
\providecommand \translation [1]{[#1]}%
\providecommand \BibitemOpen [0]{}%
\providecommand \bibitemStop [0]{}%
\providecommand \bibitemNoStop [0]{.\EOS\space}%
\providecommand \EOS [0]{\spacefactor3000\relax}%
\providecommand \BibitemShut  [1]{\csname bibitem#1\endcsname}%
\let\auto@bib@innerbib\@empty
\bibitem [{\citenamefont {Childs}\ \emph {et~al.}(2021)\citenamefont {Childs},
  \citenamefont {Su}, \citenamefont {Tran}, \citenamefont {Wiebe},\ and\
  \citenamefont {Zhu}}]{Childs2021:Theory}%
  \BibitemOpen
  \bibfield  {author} {\bibinfo {author} {\bibfnamefont {A.~M.}\ \bibnamefont
  {Childs}}, \bibinfo {author} {\bibfnamefont {Y.}~\bibnamefont {Su}}, \bibinfo
  {author} {\bibfnamefont {M.~C.}\ \bibnamefont {Tran}}, \bibinfo {author}
  {\bibfnamefont {N.}~\bibnamefont {Wiebe}},\ and\ \bibinfo {author}
  {\bibfnamefont {S.}~\bibnamefont {Zhu}},\ }\bibfield  {title} {\bibinfo
  {title} {Theory of trotter error with commutator scaling},\ }\href
  {https://doi.org/10.1103/PhysRevX.11.011020} {\bibfield  {journal} {\bibinfo
  {journal} {Phys. Rev. X}\ }\textbf {\bibinfo {volume} {11}},\ \bibinfo
  {pages} {011020} (\bibinfo {year} {2021})}\BibitemShut {NoStop}%
\bibitem [{\citenamefont {Eckart}\ and\ \citenamefont
  {Young}(1936)}]{eckart1936approximation}%
  \BibitemOpen
  \bibfield  {author} {\bibinfo {author} {\bibfnamefont {C.}~\bibnamefont
  {Eckart}}\ and\ \bibinfo {author} {\bibfnamefont {G.}~\bibnamefont {Young}},\
  }\bibfield  {title} {\bibinfo {title} {The approximation of one matrix by
  another of lower rank},\ }\href@noop {} {\bibfield  {journal} {\bibinfo
  {journal} {Psychometrika}\ }\textbf {\bibinfo {volume} {1}},\ \bibinfo
  {pages} {211} (\bibinfo {year} {1936})}\BibitemShut {NoStop}%
\bibitem [{\citenamefont {Ye}\ and\ \citenamefont
  {Chan}(2021)}]{ye2021constructing}%
  \BibitemOpen
  \bibfield  {author} {\bibinfo {author} {\bibfnamefont {E.}~\bibnamefont
  {Ye}}\ and\ \bibinfo {author} {\bibfnamefont {G.~K.-L.}\ \bibnamefont
  {Chan}},\ }\bibfield  {title} {\bibinfo {title} {Constructing tensor network
  influence functionals for general quantum dynamics},\ }\href
  {https://doi.org/10.1063/5.0047260} {\bibfield  {journal} {\bibinfo
  {journal} {The Journal of Chemical Physics}\ }\textbf {\bibinfo {volume}
  {155}},\ \bibinfo {pages} {044104} (\bibinfo {year} {2021})}\BibitemShut
  {NoStop}%
\bibitem [{\citenamefont {Cygorek}(2021)}]{aceCode}%
  \BibitemOpen
  \bibfield  {author} {\bibinfo {author} {\bibfnamefont {M.}~\bibnamefont
  {Cygorek}},\ }\href {https://doi.org/10.5281/zenodo.5214128} {\bibinfo
  {title} {{Automated Compression of Environments (ACE)}}},\ \bibinfo
  {howpublished} {http://dx.doi.org/10.5281/zenodo.5214128} (\bibinfo {year}
  {2021})\BibitemShut {NoStop}%
\bibitem [{\citenamefont {J\o{}rgensen}\ and\ \citenamefont
  {Pollock}(2019)}]{ProcessTensor}%
  \BibitemOpen
  \bibfield  {author} {\bibinfo {author} {\bibfnamefont {M.~R.}\ \bibnamefont
  {J\o{}rgensen}}\ and\ \bibinfo {author} {\bibfnamefont {F.~A.}\ \bibnamefont
  {Pollock}},\ }\bibfield  {title} {\bibinfo {title} {Exploiting the causal
  tensor network structure of quantum processes to efficiently simulate
  non-markovian path integrals},\ }\href
  {https://doi.org/10.1103/PhysRevLett.123.240602} {\bibfield  {journal}
  {\bibinfo  {journal} {Phys. Rev. Lett.}\ }\textbf {\bibinfo {volume} {123}},\
  \bibinfo {pages} {240602} (\bibinfo {year} {2019})}\BibitemShut {NoStop}%
\bibitem [{\citenamefont {Strathearn}\ \emph {et~al.}(2018)\citenamefont
  {Strathearn}, \citenamefont {Kirton}, \citenamefont {Kilda}, \citenamefont
  {Keeling},\ and\ \citenamefont {Lovett}}]{TEMPO}%
  \BibitemOpen
  \bibfield  {author} {\bibinfo {author} {\bibfnamefont {A.}~\bibnamefont
  {Strathearn}}, \bibinfo {author} {\bibfnamefont {P.}~\bibnamefont {Kirton}},
  \bibinfo {author} {\bibfnamefont {D.}~\bibnamefont {Kilda}}, \bibinfo
  {author} {\bibfnamefont {J.}~\bibnamefont {Keeling}},\ and\ \bibinfo {author}
  {\bibfnamefont {B.~W.}\ \bibnamefont {Lovett}},\ }\bibfield  {title}
  {\bibinfo {title} {Efficient non-markovian quantum dynamics using
  time-evolving matrix product operators},\ }\href
  {https://doi.org/10.1038/s41467-018-05617-3} {\bibfield  {journal} {\bibinfo
  {journal} {Nat. Commun.}\ }\textbf {\bibinfo {volume} {9}},\ \bibinfo {pages}
  {3322} (\bibinfo {year} {2018})}\BibitemShut {NoStop}%
\bibitem [{\citenamefont {Makri}\ and\ \citenamefont {Makarov}(1995)}]{Makri}%
  \BibitemOpen
  \bibfield  {author} {\bibinfo {author} {\bibfnamefont {N.}~\bibnamefont
  {Makri}}\ and\ \bibinfo {author} {\bibfnamefont {D.~E.}\ \bibnamefont
  {Makarov}},\ }\bibfield  {title} {\bibinfo {title} {{Tensor propagator for
  iterative quantum time evolution of reduced density matrices. I. Theory}},\
  }\href {https://doi.org/10.1063/1.469508} {\bibfield  {journal} {\bibinfo
  {journal} {J. Chem. Phys.}\ }\textbf {\bibinfo {volume} {102}},\ \bibinfo
  {pages} {4600} (\bibinfo {year} {1995})}\BibitemShut {NoStop}%
\bibitem [{\citenamefont {Morse}(1929)}]{Morse1936}%
  \BibitemOpen
  \bibfield  {author} {\bibinfo {author} {\bibfnamefont {P.~M.}\ \bibnamefont
  {Morse}},\ }\bibfield  {title} {\bibinfo {title} {Diatomic molecules
  according to the wave mechanics. ii. vibrational levels},\ }\href
  {https://doi.org/10.1103/PhysRev.34.57} {\bibfield  {journal} {\bibinfo
  {journal} {Phys. Rev.}\ }\textbf {\bibinfo {volume} {34}},\ \bibinfo {pages}
  {57} (\bibinfo {year} {1929})}\BibitemShut {NoStop}%
\bibitem [{\citenamefont {Bramberger}\ and\ \citenamefont
  {De~Vega}(2020)}]{Morse_deVega}%
  \BibitemOpen
  \bibfield  {author} {\bibinfo {author} {\bibfnamefont {M.}~\bibnamefont
  {Bramberger}}\ and\ \bibinfo {author} {\bibfnamefont {I.}~\bibnamefont
  {De~Vega}},\ }\bibfield  {title} {\bibinfo {title} {Dephasing dynamics of an
  impurity coupled to an anharmonic environment},\ }\href
  {https://doi.org/10.1103/PhysRevA.101.012101} {\bibfield  {journal} {\bibinfo
   {journal} {Phys. Rev. A}\ }\textbf {\bibinfo {volume} {101}},\ \bibinfo
  {pages} {012101} (\bibinfo {year} {2020})}\BibitemShut {NoStop}%
\bibitem [{\citenamefont {Gross}\ and\ \citenamefont
  {Haroche}(1982)}]{suppl_Superradiance_Haroche}%
  \BibitemOpen
  \bibfield  {author} {\bibinfo {author} {\bibfnamefont {M.}~\bibnamefont
  {Gross}}\ and\ \bibinfo {author} {\bibfnamefont {S.}~\bibnamefont
  {Haroche}},\ }\bibfield  {title} {\bibinfo {title} {Superradiance: An essay
  on the theory of collective spontaneous emission},\ }\href
  {https://doi.org/https://doi.org/10.1016/0370-1573(82)90102-8} {\bibfield
  {journal} {\bibinfo  {journal} {Physics Reports}\ }\textbf {\bibinfo {volume}
  {93}},\ \bibinfo {pages} {301 } (\bibinfo {year} {1982})}\BibitemShut
  {NoStop}%
\bibitem [{\citenamefont {Cygorek}\ \emph {et~al.}(2017)\citenamefont
  {Cygorek}, \citenamefont {Barth}, \citenamefont {Ungar}, \citenamefont
  {Vagov},\ and\ \citenamefont {Axt}}]{suppl_PI_cQED}%
  \BibitemOpen
  \bibfield  {author} {\bibinfo {author} {\bibfnamefont {M.}~\bibnamefont
  {Cygorek}}, \bibinfo {author} {\bibfnamefont {A.~M.}\ \bibnamefont {Barth}},
  \bibinfo {author} {\bibfnamefont {F.}~\bibnamefont {Ungar}}, \bibinfo
  {author} {\bibfnamefont {A.}~\bibnamefont {Vagov}},\ and\ \bibinfo {author}
  {\bibfnamefont {V.~M.}\ \bibnamefont {Axt}},\ }\bibfield  {title} {\bibinfo
  {title} {Nonlinear cavity feeding and unconventional photon statistics in
  solid-state cavity qed revealed by many-level real-time path-integral
  calculations},\ }\href {https://doi.org/10.1103/PhysRevB.96.201201}
  {\bibfield  {journal} {\bibinfo  {journal} {Phys. Rev. B}\ }\textbf {\bibinfo
  {volume} {96}},\ \bibinfo {pages} {201201} (\bibinfo {year}
  {2017})}\BibitemShut {NoStop}%
\bibitem [{\citenamefont {Blais}\ \emph {et~al.}(2004)\citenamefont {Blais},
  \citenamefont {Huang}, \citenamefont {Wallraff}, \citenamefont {Girvin},\
  and\ \citenamefont {Schoelkopf}}]{Blais2004}%
  \BibitemOpen
  \bibfield  {author} {\bibinfo {author} {\bibfnamefont {A.}~\bibnamefont
  {Blais}}, \bibinfo {author} {\bibfnamefont {R.-S.}\ \bibnamefont {Huang}},
  \bibinfo {author} {\bibfnamefont {A.}~\bibnamefont {Wallraff}}, \bibinfo
  {author} {\bibfnamefont {S.~M.}\ \bibnamefont {Girvin}},\ and\ \bibinfo
  {author} {\bibfnamefont {R.~J.}\ \bibnamefont {Schoelkopf}},\ }\bibfield
  {title} {\bibinfo {title} {Cavity quantum electrodynamics for superconducting
  electrical circuits: An architecture for quantum computation},\ }\href
  {https://doi.org/10.1103/PhysRevA.69.062320} {\bibfield  {journal} {\bibinfo
  {journal} {Phys. Rev. A}\ }\textbf {\bibinfo {volume} {69}},\ \bibinfo
  {pages} {062320} (\bibinfo {year} {2004})}\BibitemShut {NoStop}%
\end{thebibliography}%

\begin{widetext}
\end{widetext}
\sect{Methods}
\subsection{Derivation of the process tensor}
\label{derivation}
We consider an arbitrary open quantum system specified by the Hamiltonian $H=H_S+H_E$,
where $H_S$ is the free system Hamiltonian without coupling to the environment.  For simplicity of notation we assume a time-independent Hamiltonian in the following, but generalisation to the time-dependent case is straightforward.
The time evolution of the system density operator $\hat{\rho}_S$ 
can be obtained from the time evolution operator $U(t)$ of
the total system, including the environment, by tracing out the environment to give:
\begin{align}
\hat{\rho}_S(t)=& \textrm{Tr}_E\Big[ U(t) 
\big(\hat{\rho}_S(0) \otimes \hat{\rho}_E(0)\big)
U^\dagger (t) \Big].
\end{align}

We discretise the time evolution operator 
$U(t)=\prod_{l=1}^n U(\Delta t)$ on a time grid $t_l=l \Delta t$, $l=1\ldots n$ 
and apply a Trotter decomposition 
$
U(\Delta t)=
e^{-\frac i\hbar H_E \Delta t}\;
e^{-\frac i\hbar H_S \Delta t} 
+\mathcal{O}(\Delta t^2).
$
Next, we introduce a complete 
basis for the system ($\nu$ or $\mu$) as well as for the
full environment ($\xi$ or $\eta$).
We then introduce the matrix elements
\begin{align}
A^{\nu_{l} \tilde{\nu}_{l}}_{\xi_{l} \xi_{l-1}}=&
\langle \nu_{l}, \xi_{l}| e^{-\frac i\hbar H_E\Delta t}| 
\tilde{\nu}_{l}, \xi_{l-1}\rangle,
\\
M^{\tilde{\nu}_{l} \nu_{l-1}}=&\langle \tilde{\nu}_{l}| 
e^{-\frac i\hbar H_S\Delta t} | \nu_{l-1}\rangle,
\end{align}
and, using calligraphic symbols, their counterparts in Liouville space: 
\begin{align}
\mathcal{A}^{(\nu_l,\mu_l),(\tilde{\nu}_l,\tilde{\mu}_l)}_
{(\xi_l,\eta_l),(\xi_{l-1},\eta_{l-1})}:=&
A^{\nu_{l} \tilde{\nu}_{l}}_{\xi_{l} \xi_{l-1}}
A^{\mu_{l} \tilde{\mu}_{l}*}_{\eta_{l} \eta_{l-1}} 
\\
\mathcal{M}^{\tilde{\nu}_{l}\nu_{l-1}}_{\tilde{\mu}_{l}\mu_{l-1}} :=&
M^{\tilde{\nu}_{l}\nu_{l-1}}M^{\tilde{\mu}_{l}\mu_{l-1}*}.
\end{align}
The reduced system density matrix at time step $t_n=n\Delta t$ can then be expressed
as
\begin{align}
\rhoS_{\nu_n\mu_n}= 
\sum_{\substack{\nu_{n-1}\dots\nu_0 \\
\tilde{\nu}_n\dots\tilde{\nu}_1 \\  \mu_{n-1}\dots \mu_0 \\
\tilde{\mu}_n\dots \tilde{\mu}_1}}
I^{(\nu_{n}\tilde{\nu}_n)\dots(\nu_1\tilde{\nu}_1)}
_{(\mu_{n}\tilde{\mu}_n)\dots(\mu_1\tilde{\mu}_1)}
\bigg(\prod_{l=1}^{n}
\mathcal{M}^{\tilde{\nu}_{l}\nu_{l-1}}_{\tilde{\mu}_{l}\mu_{l-1}} 
\bigg)
\rhoS_{\nu_0\mu_0},
\label{eq:rho_numu}
\end{align}
where
\begin{align}
I^{(\nu_{n}\tilde{\nu}_n)\dots(\nu_1\tilde{\nu}_1)}
_{(\mu_{n}\tilde{\mu}_n)\dots(\mu_1\tilde{\mu}_1)}
= 
\!\!\sum_{\substack{\xi_n\dots \xi_0 \\ \eta_n \dots \eta_0}} 
\!\!\!\delta_{\xi_n \eta_n}
\bigg(\prod_{l=1}^{n} 
\mathcal{A}^{(\nu_l,\mu_l),(\tilde{\nu}_l,\tilde{\mu}_l)}_
{(\xi_l,\eta_l),(\xi_{l-1},\eta_{l-1})} \bigg)
\rho^{E}_{\xi_0\eta_0}.
\label{eq:PT}
\end{align}
Here, $\rhoS_{\nu_0\mu_0}$ and $\rho^E_{\xi_0\eta_0}$ are the initial
system and environment states, respectively. 
The implicit assumption of a factorisation
of the initial state into system and environment parts, i.e., uncorrelated 
initial states, does not restrict the generality, because
initial states with finite system-environment correlations 
can always be rewritten as sums of product states using Schmidt decomposition.

By combining pairs of Hilbert space indices into Liouville space indices 
\mbox{$\alpha_{l}=(\nu_l,\mu_l)$}, 
$\tilde{\alpha}_l=(\tilde{\nu}_l, \tilde{\mu}_l)$ 
and $d_l=(\xi_l,\eta_l)$, 
Eq.~\eqref{eq:rho_numu} becomes Eq.~\eqref{eq:rho_alpha}
and Eq.~\eqref{eq:PT} takes the form of Eq.~\eqref{eq:PT_MPR}.
The matrices $\mathcal{Q}$ can be obtained by comparison with 
Eq.~\eqref{eq:PT} as
\begin{align}
\mathcal{Q}^{(\alpha_l,\tilde{\alpha}_l)}_{d_l d_{l-1}}=
\begin{cases}
\delta_{d_0,1}\sum\limits_{d'_0}
\mathcal{A}^{\alpha_1,\tilde{\alpha}_l}_
{d_1,d'_0}\rho^E_{d'_0}&l=1, 
\\
\mathcal{A}^{\alpha_l,\tilde{\alpha}_l}_
{d_l,d_{l-1}}& 1<l<n,
\\
\delta_{d_n,1}\sum\limits_{d'_n} \mathfrak{I}_{d'_n}
\mathcal{A}^{\alpha_n,\tilde{\alpha}_n}_
{d'_n,d_{n-1}}& l=n.
\end{cases}
\end{align}
where $\mathfrak{I}_{d'_n=(\xi,\eta)}=\delta_{\xi,\eta}$.

\subsection{Network summation}
\label{network_summation}
The network structure determining the reduced system density matrix, 
visualised in Fig.~\ref{fig:sketchMPS}d, 
can be most easily evaluated by propagating the quantity
$\mathcal{R}_{\alpha_l d_l}$ defined recursively via
\begin{subequations}
\label{eq:general}
\begin{align}
\mathcal{R}_{\alpha_0 1}=&\rhoS_{\alpha_0}=\rhoS_{\nu_0\mu_0},
\\
\mathcal{R}_{\alpha_l d_l}=&\sum_{\tilde{\alpha}_l\alpha_{l-1}}\sum_{d_{l-1}}
\mathcal{Q}^{(\alpha_l,\tilde{\alpha}_l)}_{d_l d_{l-1}}
\mathcal{M}^{\tilde{\alpha}_l \alpha_{l-1}} \mathcal{R}_{\alpha_{l-1} d_{l-1}}.
\end{align}
\end{subequations}
Comparing with Eqs.~\eqref{eq:rho_alpha} and \eqref{eq:PT_MPR}, it can be
seen that the density matrix at the last time step is given by
$\rhoS_{\alpha_n}=\mathcal{R}_{\alpha_n 1}$.

When the environment time evolution operator is unitary, %
the reduced density matrix $\rhoS_{\alpha_l}$ at intermediate time 
steps $t_l$ can be easily obtained from $\mathcal{R}_{\alpha_l d_l}$ as
\begin{align}
\rhoS_{\alpha_l}=\sum_{d_l} q_{d_l} \mathcal{R}_{\alpha_l d_l}
\label{eq:close}
\end{align}
using the closures $q_{d_l}$ defined by the recursion 
(cf. Supplementary Material~S.1 %
for a detailed derivation)
\begin{align}
q_{d_n=1}=&1 \\
q_{d_{l-1}}=&\sum_{d_{l}} q_{d_{l}}
\sum_{\alpha_l}\mathfrak{I}_{\alpha_l}  %
\mathcal{Q}^{(\alpha_{l}0)}_{d_{l}d_{l-1}}.
\end{align}
Thus, in practice one needs to calculate only a single PT MPO with $n$ time
steps, where $n\Delta t=t_\textrm{final}$ is the final time one is interested 
in, and obtains the density matrix at all intermediate time steps $l\Delta t$
at marginal numerical extra cost.

\subsection{PT combination rule}
In order to combine the influences of multiple environments or 
of independent environments into a single PT, consider a system
coupled to multiple environmental degrees of freedom (which we henceforth call modes) via
\begin{align}
H_E=\sum_{k=1}^{N_E} H_E^k.
\end{align}
We define the partial sum of the Hamiltonians from modes $1,2,\dots K$ as
\begin{align}
H_E[K]=\sum_{k=1}^K H_E^k
\end{align}
and denote by $\mathcal{Q}^{(\alpha_l,\tilde{\alpha}_l)}_{d_l d_{l-1}}[K]$ 
the $l$-th MPO matrix of the PT including the influences of the modes
$1,2,\dots K$. Then, by means of the symmetric Trotter decomposition
\begin{align}
&e^{-\frac i\hbar H_E[K]\Delta t}
=e^{-\frac i\hbar\big(H_E[K-1]+H_E^K\big)\Delta t}
\nn&= e^{-\frac i\hbar H_E^{K}\frac{\Delta t}2}
e^{-\frac i\hbar H_E[K-1] \Delta t}
e^{-\frac i\hbar H_E^{K}\frac{\Delta t}2}
+\mathcal{O}(\Delta t^3)
\end{align}
the influence of mode $K$ can be combined with the PT containing already the
influences of the first $K-1$ modes by
\begin{align}
&\mathcal{Q}^{(\alpha_l,\tilde{\alpha}_l)}_{(d'_{l}, d_{l}) 
(d'_{l-1}, d_{l-1})} \big[K\big]
\nn&\approx
\sum_{\gamma_l,\tilde{\gamma}_l,\tilde{d}_l}
\mathcal{B}^{(\alpha_l, \gamma_l)}_{d_{l} \tilde{d}_l}(K)\;\,
\mathcal{Q}^{(\gamma_l,\tilde{\gamma}_l)}_{d'_{l} d'_{l-1}}
\big[K-1] \;\,
\mathcal{B}^{(\tilde{\gamma}_l, \tilde{\alpha}_l)}_{\tilde{d}_{l} d_{l-1}}(K),
\label{eq:combine}
\end{align}
where
\begin{align}
&\mathcal{B}^{((\nu_{l},\mu_l), (\tilde{\nu}_l,\tilde{\mu}_l))}
_{(\xi_{l},\eta_{l}),(\xi_{l-1},\eta_{l-1})}(K)
\nn&=
\langle \nu_{l}, \xi_{l}| e^{-\frac i\hbar H_E^{K}\frac{\Delta t}2}| 
\tilde{\nu}_{l}, \xi_{l-1}\rangle
\langle \tilde{\mu}_{l}, \eta_{l-1}| 
e^{\frac i\hbar H_E^{K}\frac{\Delta t}2}| {\mu}_{l}, \eta_{l}\rangle.
\end{align}
This step is visualised in Fig.~\ref{fig:sketchMPS}e.

In practice, we start with the trivial PT MPO with matrices
$\mathcal{Q}^{(\alpha_l,\tilde{\alpha}_l)}_{d_l d_{l-1}}[0]=
\delta_{d_l,1}\delta_{d_{l-1},1}\delta_{\alpha_l, \tilde{\alpha}_l}$ and
add the influence of all environment modes by recursively applying 
Eq.~\eqref{eq:combine} until $K=N_E$. 
After each combination step, the PT MPO is compressed using 
the SVD-based compression as described in the next section.

\subsection{MPO Compression}
\label{app:compression}
In order to reduce the inner dimension of the MPO representing the PT, we 
perform sweeps of singular value decompositions (SVDs) across the MPO chain.
Any matrix $A\in \mathbb{C}^{n\times m}$ can be factorised into a product
\begin{align}
A=&U\Sigma V^\dagger,
\end{align}
where $U\in \mathbb{C}^{n\times k}$ and $V\in \mathbb{C}^{m\times k}$ 
are matrices with orthogonal column vectors
and $\Sigma$ is a diagonal matrix containing the 
$k=\textrm{min}(n,m)$ real and non-negative singular values $\sigma_i$ 
in descending order. 
Here, we start with the first MPO matrix, we define
\begin{align}
A_{d_1,(\alpha_1,\tilde{\alpha}_1)}
=\mathcal{Q}^{(\alpha_1,\tilde{\alpha}_1)}_{d_11},
\end{align}
and we calculate a SVD of the matrix $A$. 
In order to reduce the inner dimension,
we truncate the matrices $U, \Sigma$, and $V$, keeping only the 
$k_\textrm{eff}\le k$ singular values with $\sigma_i > \epsilon\sigma_1$, where 
$\sigma_1$ is the largest singular value of $A$ and $\epsilon$ is a 
predefined threshold. Then, we replace 
$\mathcal{Q}^{(\alpha_1,\tilde{\alpha}_1)}_{d_11}$ 
by $\big(V^\dagger\big)_{k_\textrm{eff} (\alpha_1,\tilde{\alpha}_1)}$
and multiply the next matrix 
$\mathcal{Q}^{(\alpha_2,\tilde{\alpha}_2)}_{d_2d_1}$
from the right by $U_{d_1 k_\textrm{eff}}\sigma_{k_\textrm{eff}}$ and
perform a SVD of
\begin{align}
A_{d_2,(\alpha_2,\tilde{\alpha}_2, k_\textrm{eff})}
=\sum_{d_1} \mathcal{Q}^{(\alpha_2,\tilde{\alpha}_2)}_{d_2d_1}
U_{d_1 k_\textrm{eff}}\sigma_{k_\textrm{eff}}.
\end{align}
The reduction is continued until the end of the MPO is reached. 
Then, another line sweep is performed in the opposite direction. 
Note that sweeps along the whole chain are required between each 
PT combination step, because information necessary to effectively compress the 
MPO, such as the initial environment state, needs to be propagated from the
ends throughout the whole MPO.

In the overall process, the inner dimensions $d_i$ are reduced to the
respective effective ranks $k_\textrm{eff}$, where the latter are 
controlled by the threshold $\epsilon$. 

\subsection{Parameters for QD, QD-phonon, and QD-photon Hamiltonians}
\label{app:QDPhonon_details}
The effects of the dot-phonon coupling are 
completely defined by the phonon spectral density 
\begin{align}
J(\omega)=\sum_\mathbf{q} \gamma^2_\mathbf{q}\delta(\omega-\omega_\mathbf{q}).
\end{align}
Using established parameters~\cite{Krummheuer} 
for a GaAs quantum dot with electron radius 
$a_e=3.0$ nm and hole radius $a_h=a_e/1.15$
\begin{align}
J(\omega)=\frac{\omega^3}{4\pi^2\rho\hbar c_s^5}
\bigg(D_e e^{-\omega^2a_e^2/(4c_s^2)} - D_he^{-\omega^2a_h^2/(4c_s^2)}\bigg)^2
\end{align}
with mass density $\rho=5370$ kg/m$^3$, speed of sound $c_s=5110$ m/s and
electron and hole deformation potential constants 
$D_e=7.0$ eV and $D_h=-3.5$ eV.
We discretise the phonon continuum using steps of equal width, so that
$\omega_q=q d\omega $ with 
$d\omega=\omega_\textrm{max}/N_E$, $N_E=100$ and
$\omega_\textrm{max}=5~\textrm{meV}/\hbar$
and we obtain the coupings $\gamma_q$ from the phonon density of states using
$\gamma_q=\sqrt{J(\omega_q) d\omega}$.
The phonon modes are initially assumed to be in thermal equilibrium with
temperature $T=4$ K. We have checked that for these parameters 
it is enough to consider up to two excitations per mode.

We use a radiative decay rate of $\kappa=0.1$ ps$^{-1}$.
When the electromagnetic environment is treated micro\-scopically 
we assume a constant density of states with bandwidth 
$\omega_{BW}=10$ ps$^{-1}$, discretised using
$N_E=100$ equally spaced modes. The coupling constants $g_k$ are taken to be constant and the value is chosen such that Fermi's golden rule reproduces the radiative decay rate $\kappa$.
The PTs for the phonon and photon environments are calculated separately 
and combined using Eq.~\eqref{eq:combine} without 
performing a final SVD sweep.
For both baths, we use time steps $\Delta t=0.1$ ps and an MPO compression 
threshold $\epsilon=5\times 10^{-8}$.

The Gaussian excitation pulse is detuned $\hbar\delta=1.5$~meV 
above the quantum dot resonance and the envelope is described by
\begin{align}
\Omega(t)=&\frac{A}{\sqrt{2\pi}\sigma}
\exp\bigg(-\frac{(t-t_0)^2}{2\sigma^2}\bigg),
\end{align}
where we use the pulse area $A=3\pi$, pulse centre $t_0=7$~ps, and 
$\sigma=\tau_\textrm{FWHM}/\big(2\sqrt{2\ln 2}\big)$ with 
$\tau_\textrm{FWHM}=5$ ps.

\subsection{Numerical implementation}
We have implemented ACE in a C++ code using the Eigen library to calculate matrix exponentials and singular value decompositions.
All calculations have been performed on a conventional laptop computer with Intel Core i5-8265U processor and 16~GB of RAM.
The computation times for the presented examples are listed in the Supplementary Material S.3.

\secnl{Data availability}
The data presented in the figures as well as the computer code including documentation is available online at \url{https://doi.org/10.5281/zenodo.5214128}

\secnl{Acknowledgement}
M.\,Co.\ and V.\,M.\,A. are grateful for funding by the Deutsche Forschungsgemeinschaft (DFG, German Research Foundation) under project No. 419036043. A.\,V.\ acknowledges the support from the Russian Science Foundation under the Project 18-12-00429. M.\,Cy. and E.\,M.\,G.\ acknowledge funding from EPSRC grant no.\,EP/T01377X/1. B.\,W.\,L. and J.\,K. were supported by EPSRC grant no.\,EP/T014032/1.

\secnl{Author contributions}
M.\,Cy., M.\,Co., A.\,V., and V.\,M.\,A.\ developed the concept of explicitly constructing the PT 
to simulate open quantum systems with arbitrary
system-environment couplings. M.\,Cy., B.\,W.\,L., J.\,K.\ and E.\,M.\,G.\ contributed the
idea of using MPO representations for efficient storage 
and evaluation of the PT.
M.\,Cy.\ is responsible for the details of the algorithm, 
the implementation in the form of the C++-code, and the generation of data.
All authors analysed and discussed the results and contributed to writing the article.

\secnl{Competing interests}
The authors declare no competing interests.

\sect{Additional information}
\subsection{Supplementary information} is available for this article at $\dots$

\end{document}


\title{Supplementary material: Numerically exact open quantum systems simulations for arbitrary environments using automated compression of environments}
\maketitle

\section{Calculation of intermediate-time closures}
\label{app:closure}

In Eq.~(18) %
we gave an expression 
for the reduced system density matrix at time $t_n$ using
the process tensor (PT) for $n$ time steps. In practical applications, 
it is desirable to also be able to calculate the reduced system density matrix 
 at intermediate times $t_l$ to extract the full dynamics of the system.  Here we show how this can be extracted.  As a reminder, in the following we use the symbols $\nu$ or $\mu$ to enumerate system states,  and $\xi$ or $\eta$ for environment states.

The calculation of $\rhoS_{\alpha_l}$ for $l<n$ requires the knowledge 
of a PT for $l$ time steps.
From Eq.~(19) %
it is clear that, before any matrix product operator (MPO) compression,
the PT for $l$ time steps can be obtained from the PT for $n>l$ time 
steps by tracing over the environment at that step, $\sum_{\xi_l,\eta_l} \delta_{\xi_l,\eta_l}$.
After compression, it is less clear how this trace
is to be executed on the inner indices $d_l$.
In principle, it is possible to track how the trace operation 
transforms under the individual MPO compression steps. 
A more practical alternative
is to make use of the unitarity of the 
environment evolution and recursively obtain the PT for $n-1$ time steps
from the PT for $n$ time steps. 

Consider the terms corresponding to the last time step 
in Eq.~(19): %
\begin{align}
&\sum_{\xi_n\eta_n}\delta_{\xi_n \eta_n} 
A^{\nu_n\tilde{\nu}_n}_{\xi_n\xi_{n-1}}
A^{\mu_n\tilde{\mu}_n*}_{\eta_n\eta_{n-1}}
=\sum_{\xi}
\langle \tilde{\mu}_{n}, \eta_{n-1}| e^{\frac i\hbar H_E\Delta t}| 
\mu_{n}, \xi\rangle
\langle \nu_{n}, \xi| e^{-\frac i\hbar H_E\Delta t}| 
\tilde{\nu}_{n}, \xi_{n-1}\rangle.
\end{align}
Performing the trace over the system states 
$\sum_{\nu_n,\mu_n}\delta_{\nu_n\mu_n}$ in addition to the trace over
the environment states yields
\begin{align}
\sum_{\nu_n\mu_n}\delta_{\nu_n\mu_n}
\sum_{\xi_n\eta_n}\delta_{\xi_n \eta_n}
A^{\nu_n\tilde{\nu}_n}_{\xi_n\xi_{n-1}}
A^{\mu_n\tilde{\mu}_n*}_{\eta_n\eta_{n-1}}
=&
\langle \tilde{\mu}_{n}, \eta_{n-1}| e^{\frac i\hbar H_E\Delta t} 
\bigg[\sum_{\nu,\xi} |\nu, \xi\rangle \langle \nu, \xi| \bigg]
e^{-\frac i\hbar H_E\Delta t}| 
\tilde{\nu}_{n}, \xi_{n-1}\rangle
\nonumber\\=&
\langle \tilde{\mu}_{n}, \eta_{n-1}| 
e^{\frac i\hbar H_E(\Delta t-\Delta t)} 
|\tilde{\nu}_{n}, \xi_{n-1}\rangle
=
\langle \tilde{\mu}_{n}, \eta_{n-1}|\tilde{\nu}_{n}, \xi_{n-1}\rangle
=\delta_{\tilde{\mu}_{n}\tilde{\nu}_{n}}
\delta_{\eta_{n-1} \xi_{n-1}}.
\end{align}
Together with the sum over the $\eta_{n-1}$ and $\xi_{n-1}$ in the $(n-1)$-th 
time step in the PT,
the term $\delta_{\eta_{n-1} \xi_{n-1}}$ again becomes equivalent to 
calculating the trace over the environment modes, but at time step $n-1$.
Therefore, the PT for $n-1$ time steps can be related to the
PT for $n$ time steps by
\begin{align}
I^{(\nu_{n-1}\tilde{\nu}_{n-1})\dots(\nu_1\tilde{\nu}_1)}
_{(\mu_{n-1}\tilde{\mu}_{n-1})\dots(\mu_1\tilde{\mu}_1)}
=\sum_{\nu_n\mu_n}
\delta_{\nu_n\mu_n}
I^{(\nu_{n}\tilde{\nu})(\nu_{n-1}\tilde{\nu}_{n-1})\dots(\nu_1\tilde{\nu}_1)}
_{(\mu_{n}\tilde{\nu})(\mu_{n-1}\tilde{\mu}_{n-1})\dots(\mu_1\tilde{\mu}_1)},
\end{align}
where $\tilde{\nu}$ is an arbitrary system state, which we choose as $\tilde{\nu}=0$.
As this expression only involves outer indices and is independent of the 
inner indices, it applies equally to the PT after MPO compression.
Thus, given the PT for $n$ time steps in MPO form in Liouville space
\begin{align}
&\mathcal{I}^{(\alpha_{n},\tilde{\alpha}_{n})(\alpha_{n-1},\tilde{\alpha}_{n-1})
\dots (\alpha_{1},\tilde{\alpha}_{1})}
= 
\sum_{d_{n-1}\dots d_1}
\mathcal{Q}_{1 d_{n-1}}^{(\alpha_{n},\tilde{\alpha}_{n})}
\mathcal{Q}_{d_{n-1} d_{n-2}}^{(\alpha_{n-1},\tilde{\alpha}_{n-1})}\dots
\mathcal{Q}_{d_1 1}^{(\alpha_{1},\tilde{\alpha}_{1})},
\end{align}
we can obtain the PT for $l$ time steps as
\begin{align}
&\mathcal{I}^{(\alpha_{l},\tilde{\alpha}_{l})(\alpha_{l-1},\tilde{\alpha}_{l-1})
\dots (\alpha_{1},\tilde{\alpha}_{1})}
= 
\sum_{d_{l}\dots d_1} q_{d_l}
\mathcal{Q}_{d_l d_{l-1}}^{(\alpha_{l},\tilde{\alpha}_{l})}
\mathcal{Q}_{d_{l-1} d_{l-2}}^{(\alpha_{l-1},\tilde{\alpha}_{l-1})}\dots
\mathcal{Q}_{d_1 1}^{(\alpha_{1},\tilde{\alpha}_{1})},
\end{align}
where the closures $q_{d_l}$ are calculated recursively via
\begin{align}
q_{d_n=1}=&1 \\
q_{d_{l-1}}=&\sum_{d_{l}} q_{d_{l}} 
\sum_{\alpha_l\nu_{l}} \delta_{\alpha_{l}, (\nu_{l},\nu_{l})}
\mathcal{Q}^{(\alpha_{l}0)}_{d_{l}d_{l-1}}.
\end{align}
With the closures $q_{d_l}$ the reduced system density matrix 
$\rhoS_{\alpha_l}$ at time step $t_l$ can be extracted
from the propagated quantities $\mathcal{R}_{\alpha_l d_l}$ defined 
in Eq.~(19) as %
\begin{align}
\rhoS_{\alpha_l}=\sum_{d_l} q_{d_l} \mathcal{R}_{\alpha_l d_l}.
\end{align}

\newpage

\section{Numerical convergence of the ACE algorithm}
\setcounter{figure}{0}

The ACE algorithm, as described in the main text, is numerically exact in the following sense:
 Every step in the derivation that involves an approximation is controlled by convergence parameters, such that in principle the error can be made arbitrarily small as the corresponding convergence parameters are taken to zero or infinity as appropriate. 
Thus, in principle, exact results can be approximated to arbitrary precision given enough computational resources.
In this section we first review the sources of numerical error that exist---time discretisation, MPO compression, and discretisation of a continuum of environment modes. We then present a study of the tradeoff between accuracy and the  computational cost of a calculation.

\subsection{Sources of numerical error}

\subsubsection{Time discretization}
The starting point of the derivation of ACE is the introduction of an 
equidistant time grid $t_n=n \Delta t$, defined by a time step width $\Delta t$. 
The maximal number of time steps $n_\textnormal{max}$ then determines the simulation end time $t_e= n_\textnormal{max} \Delta t$. Decomposing the total time evolution operator into system and environment parts 
for a time step $\Delta t$ introduces numerical Trotter errors. For the system--environment decomposition we use a first-order expansion
\begin{align}
e^{-\frac i\hbar (H_S+H_E) \Delta t} = 
e^{-\frac i\hbar H_E \Delta t } e^{-\frac i\hbar H_S \Delta t} + 
\mathcal{E}_\textnormal{Trotter}^{SE}
\end{align}
while between different environment modes we use a second-order  expansion
\begin{align}
e^{-\frac i\hbar\big( H_E[K-1]+H_E^K\big)\Delta t}
= 
e^{-\frac i\hbar H_E[K-1]\frac{\Delta t}2 }
e^{-\frac i\hbar H_E^K\Delta t}
e^{-\frac i\hbar H_E[K-1]\frac{\Delta t}2 }
+\mathcal{E}_\textnormal{Trotter}^{K}.
\end{align}

While there has been considerable work on finding rigorous bounds for Trotter errors (see e.g.~\citet{Childs2021:Theory} and references therein), here we limit our discussion to a simple analysis in terms of a Taylor expansion orders. 
This yields single-step error terms of the order
$\mathcal{E}_\textnormal{Trotter}^{SE}=\mathcal{O}(\Delta t^2)$ and
$\mathcal{E}_\textnormal{Trotter}^{K}=\mathcal{O}(\Delta t^3)$, respectively.
Regarding the system--environment decoupling we may
note however, that when the full time evolution up to the final time $t_e$ is 
considered, the product
\begin{align}
P_1:=\big(e^{-\frac i\hbar H_E \Delta t } e^{-\frac i\hbar H_S \Delta t} \big)
\big(e^{-\frac i\hbar H_E \Delta t } e^{-\frac i\hbar H_S \Delta t} \big)
\dots 
\big(e^{-\frac i\hbar H_E \Delta t } e^{-\frac i\hbar H_S \Delta t} \big) 
\end{align}
is related to the product obtained by symmetric Trotter decomposition
\begin{align}
P_2:=&
\big(e^{-\frac i\hbar H_S \frac{\Delta t}2 } 
e^{-\frac i\hbar H_E \Delta t} e^{-\frac i\hbar H_S \frac{\Delta t}2 }\big)
\big(e^{-\frac i\hbar H_S \frac{\Delta t}2 } 
e^{-\frac i\hbar H_E \Delta t} e^{-\frac i\hbar H_S \frac{\Delta t}2 }\big)
\dots
\big(e^{-\frac i\hbar H_S \frac{\Delta t}2 } 
e^{-\frac i\hbar H_E \Delta t} e^{-\frac i\hbar H_S \frac{\Delta t}2 }\big),
\end{align}
by the relation $P_2=e^{-\frac i\hbar  H_S \frac{\Delta t}2 } P_1 
e^{+\frac i\hbar  H_S \frac{\Delta t}2 }$. Thus, the results obtained by first-order Trotter decomposition converge identically to those obtained by a second-order Trotter decomposition,
up to evolving the initial and final states by a half time step.

By keeping the final time $t_e$ fixed and expressing the time step width 
$\Delta t=t_e/n_\textnormal{max}$, the total error accumulated can be written in terms of the total number of time steps $n_\textnormal{max}$.
For the second-order Trotter decomposition the total error is 
$\big| e^{-\frac i\hbar(H_S+H_E) t} - P_2\big|= 
n_\textnormal{max} \, \mathcal{O}(1/n_\textnormal{max}^3)=
 \mathcal{O}(1/n_\textnormal{max}^2)$.
As the environment propagator $e^{-\frac i\hbar H_E \Delta t}$ itself
is also approximated up to an error $\mathcal{O}(1/n_\textnormal{max}^3)$--- 
arising from decomposing it into different modes---
the overall Trotter error accumulated during the simulation scales as
$\mathcal{O}(1/n_\textnormal{max}^2)$. This error can thus be made arbitrarily small by choosing a fine enough time discretisation.

\subsubsection{MPO compression}
A second source of numerical error occurs when the MPO
representing the process tensor is compressed.
This compression is done using a singular value decomposition (SVD), and truncation by neglecting singular values below a given threshold.  A sequential sweep of SVDs is performed across the MPO.

For a single SVD step the Eckart--Young--Mirsky theorem~\cite{eckart1936approximation} provides concrete error bounds:
Given the SVD of a matrix $A$, we define
\begin{align}
A=&U\Sigma V^\dagger = \sum_{i=1}^n \sigma_i u_i v_i^\dagger
=\underbrace{\sum_{i=1}^k \sigma_i u_i v_i^\dagger}_{=:\tilde{A}}
+\underbrace{\sum_{i=k+1}^n \sigma_i u_i v_i^\dagger}_{=:\delta A}, 
\end{align}
where $\sigma_i$ are the singular values in descending order, $u_i, v_i$ the corresponding singular vectors, and $k$ is
the smallest number such that $\sigma_i<\epsilon\sigma_1$ for all $i>k$.
$\tilde{A}$ represents the relevant part of the matrix $A$, 
whereas $\delta{A}$ is considered irrelevant and is therefore neglected.
The Eckart--Young--Mirsky theorem states that the matrix $\tilde{A}$
provides the best approximation to $A$ of all matrices with rank $k$.
In particular, the error in the spectral norm is 
$\|A-\tilde{A}\|_2=\sigma_{k+1}$
while for the Frobenius norm 
$\|A-\tilde{A}\|_F=\sqrt{\sigma_{k+1}^2 + \sigma_{k+2}^2 + \dots +\sigma_n^2}$.
In any case, for $\epsilon\to 0$ one finds  $\|A-\tilde{A}\| \to 0$ and
the low-rank approximation $\tilde{A}\to A$ becomes exact.

Exact bounds for the accumulated error of a full line sweep 
are more difficult to assess. 
This is because, e.g., in a sweep from right to left, the 
next matrix is multiplied with vectors $\sigma_i u_i$ ($i=1,2,\dots,k$) 
from the SVD of the previous matrix, so the result depends on the overlap
between $u_i$ and the row vectors of the next matrix. 
Furthermore, it is a priori not clear how strongly a given 
matrix element of the PT MPO contributes to 
the reduced system density matrix at a given point in time, in particular 
because this also depends on the concrete system Hamiltonian $H_S$. 
It also remains an open question how numerical errors propagate, e.g., 
if small deviations from Hermitianity and positivity of the density 
matrices grow exponentially or behave more advantageously.

A rigorous mathematical analysis of the error bounds for tensor network 
methods is beyond the scope of the present article. Yet, it is clear that the
MPO compressed object
 turns into an exact reformulation of the original uncompressed PT
in the limit $\epsilon\to 0$.
As such, in this work, we restrict further analysis to numerical convergence tests.

\subsubsection{Continuum discretisation}
While some open quantum systems, such as the example of spin baths
in the main text, contain a finite number of environment modes, others involve
a continuum of modes that require discretisation in order to apply the ACE algorithm. 
For this discretisation to converge numerically, one has to additionally
demand piecewise continuity of the environment initial state as well as
the environment Hamiltonian with respect to the index $\mathbf{k}$ describing
the continuum.

\subsection{Numerical convergence and computational cost}

To numerically test the convergence of ACE with respect to different 
convergence parameters, we consider again the example of the resonant 
level model (first example in the main text). 
In particular, we focus on the case of  
band width $\omega_{BW}=10 \gamma$, where visible deviations from
the Markovian result can be seen. 

First, using $N_E=10$ modes to discretise the continuum and
fixed final time $\gamma t_e=2.5$, we investigate
the numerical error for different time step widths $\Delta t$ 
and MPO compression thresholds $\epsilon$.
Here, we define the error as
\begin{align}
\label{eq:errordef}
\textnormal{Error}= 
\max_{i} \big| n_S\big(t_i, \{ \Delta t, \epsilon\} \big) 
 - n_S\big(t_i, \{ \Delta t, \epsilon_\textnormal{min}\} \big)\big|,
\end{align}
where $n_S\big(t_i, \{ \Delta t, \epsilon\} \big)$ is the system site
occupation at time $t_i = i\Delta t$ calculated using the set of 
convergence parameters $\{ \Delta t, \epsilon\}$.
Because  the computation time
and the convergence with respect to the threshold $\epsilon$ differ vastly for different time discretisations $\Delta t$,
we use reference calculations with different $\epsilon_\textnormal{min}$
for each $\Delta t$. Specifically we use
$\epsilon_\textnormal{min}(\gamma\Delta t=0.1)=3\times 10^{-9}$, 
$\epsilon_\textnormal{min}(\gamma\Delta t=0.05)=3\times 10^{-10}$,
and
$\epsilon_\textnormal{min}(\gamma\Delta t=0.01)=
\epsilon_\textnormal{min}(\gamma\Delta t=0.005)=
\epsilon_\textnormal{min}(\gamma\Delta t=0.001)=10^{-11}$. 

\begin{figure}[t]
\begin{centering}
\includegraphics[width=0.99\textwidth]{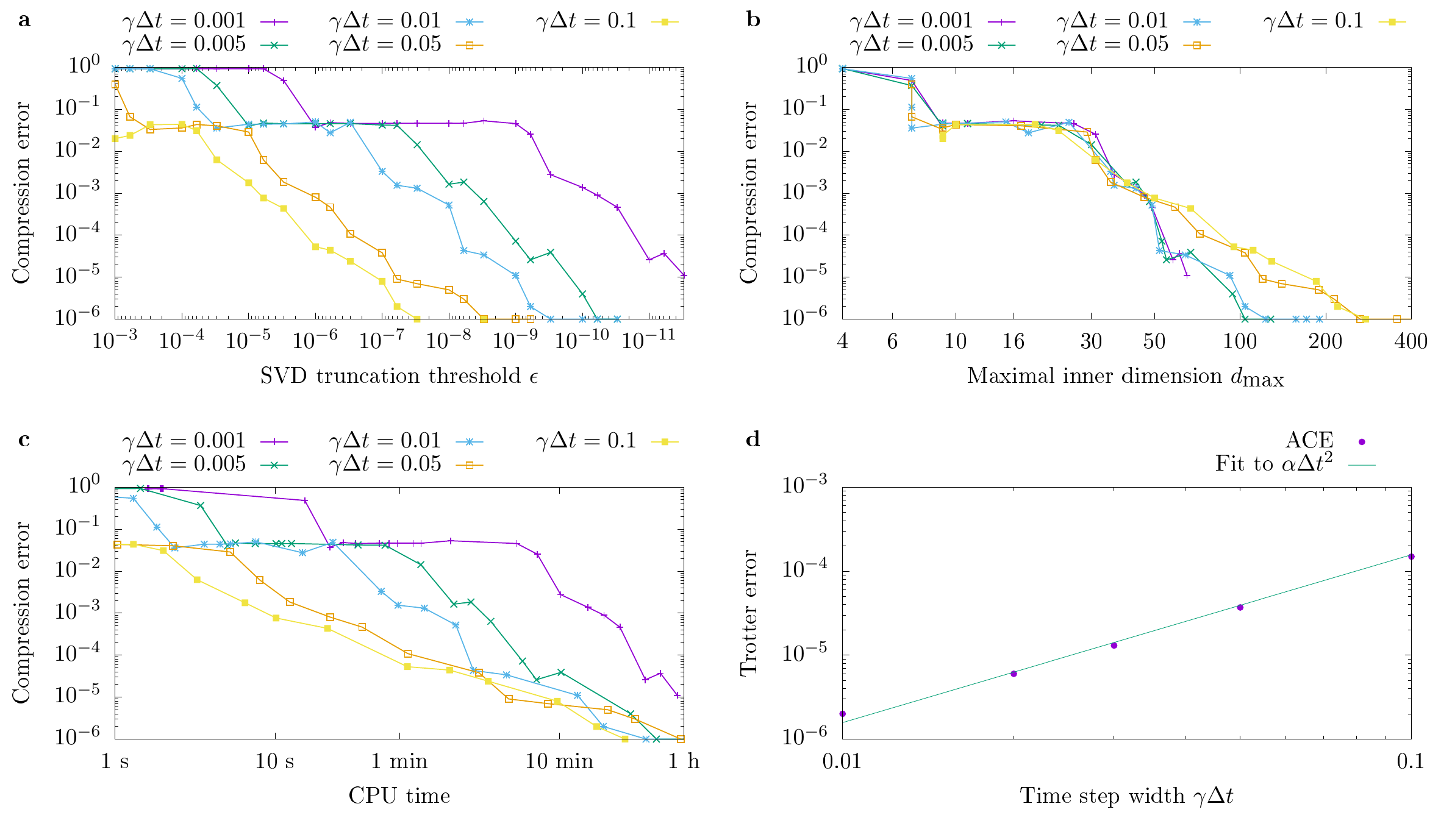}
\caption{\label{fig:conv}Convergence of ACE simulations with respect 
to SVD truncation threshold $\epsilon$ (panel \textbf{a}). Note the horizontal axis goes from largest $\epsilon$ to smallest. 
The same results are also plotted as a function of the maximal inner dimension 
$d_\textrm{max}$ (panel \textbf{b}) and the total CPU time (panel \textbf{c}) 
needed for the calculations on a conventional laptop
computer with Intel Core i5-8265U. The error associated with the
time discretisation (Trotter error) is depicted in panel \textbf{d}.}
\end{centering}
\end{figure}

These numerical errors vs compression threshold 
$\epsilon$ are depicted in Fig.~\ref{fig:conv}\textbf{a} on 
a double logarithmic scale.
As expected, the numerical error generally decreases as the 
threshold is reduced. 
The threshold needed to obtain a given numerical accuracy 
is found to depend strongly on the time step width $\Delta t$. Furthermore,
a common feature in all curves is that there exists a plateau where for a
broad range of thresholds no significant gain in accuracy is observed.

These facts can be explained by the distribution of singular values:
The uncompressed PT MPO matrices are directly related to the environment
propagator, which for very small time steps can be approximated as
$e^{-\frac i\hbar H_E \Delta t}\approx 1 - \frac i\hbar H_E \Delta t$ and
therefore possesses matrix elements (diagonals) of order 1 as well as 
contributions (off-diagonals) of order $\|H_E\| \Delta t/\hbar$, but hardly any
elements in the intermediate range. Matrix elements of vastly different orders
of magnitude translate into singular values of different orders of magnitude,
leading to a corresponding gap in the SVD spectrum.
This analysis also demonstrates that smaller time steps require 
smaller convergence thresholds to produce the same level of accuracy, to avoid
terms of the order $\|H_E\| \Delta t/\hbar \lesssim \epsilon$ being disregarded. 

Figure~\ref{fig:conv}\textbf{b} depicts the results of the same calculations
plotted against the maximal inner dimension $d_\textnormal{max}$ of the PT.
Displayed in this way, the curves calculated for different time step widths
$\Delta t$ all nearly overlap, suggesting that the maximal inner dimension
is a more reliable measure of the numerical accuracy than the compression threshold $\epsilon$ itself. 
Note, however, that in our fixed-precision algorithm, $d_\textnormal{max}$ is not known a priori.

For practical applications it is important to relate these parameters controlling precision to the computation time needed to perform ACE simulations. 
We therefore also present the numerical 
error of the above calculations as a function of the CPU time in 
Fig.~\ref{fig:conv}\textbf{c}. These were obtained on a conventional laptop
computer with Intel Core i5-8265U processor.
For a given accuracy, the fastest 
computation is achieved for the largest time steps because the PT MPO has
fewer sites and, thus, fewer SVDs to perform.
In particular, for this problem, we find that very accurate results are achievable 
within minutes of computation time. 

Figure~\ref{fig:conv}\textbf{d}, shows the Trotter error
$\max_{i} \big| n_S\big(t_i, \{ \Delta t, \epsilon_\textnormal{min}\} \big) 
 - n_S\big(t_i, \{ \Delta t_\textnormal{min}, \epsilon_\textnormal{min}\} \big)\big|$ 
defined as the difference with respect to reference calculations with
time step width $\gamma\Delta t_\textnormal{min}=0.005$, 
where for given $\Delta t$ 
the corresponding best converged results with smallest threshold 
$\epsilon_\textnormal{min}$ is used.
The theoretical expectation that the accumulated Trotter error is 
proportional to $\Delta t^2$ (or $1/n_\textnormal{max}^{2}$) 
is corroborated by a fit of the data points to this trend.

\begin{figure}
\begin{centering}
\includegraphics[width=\textwidth]{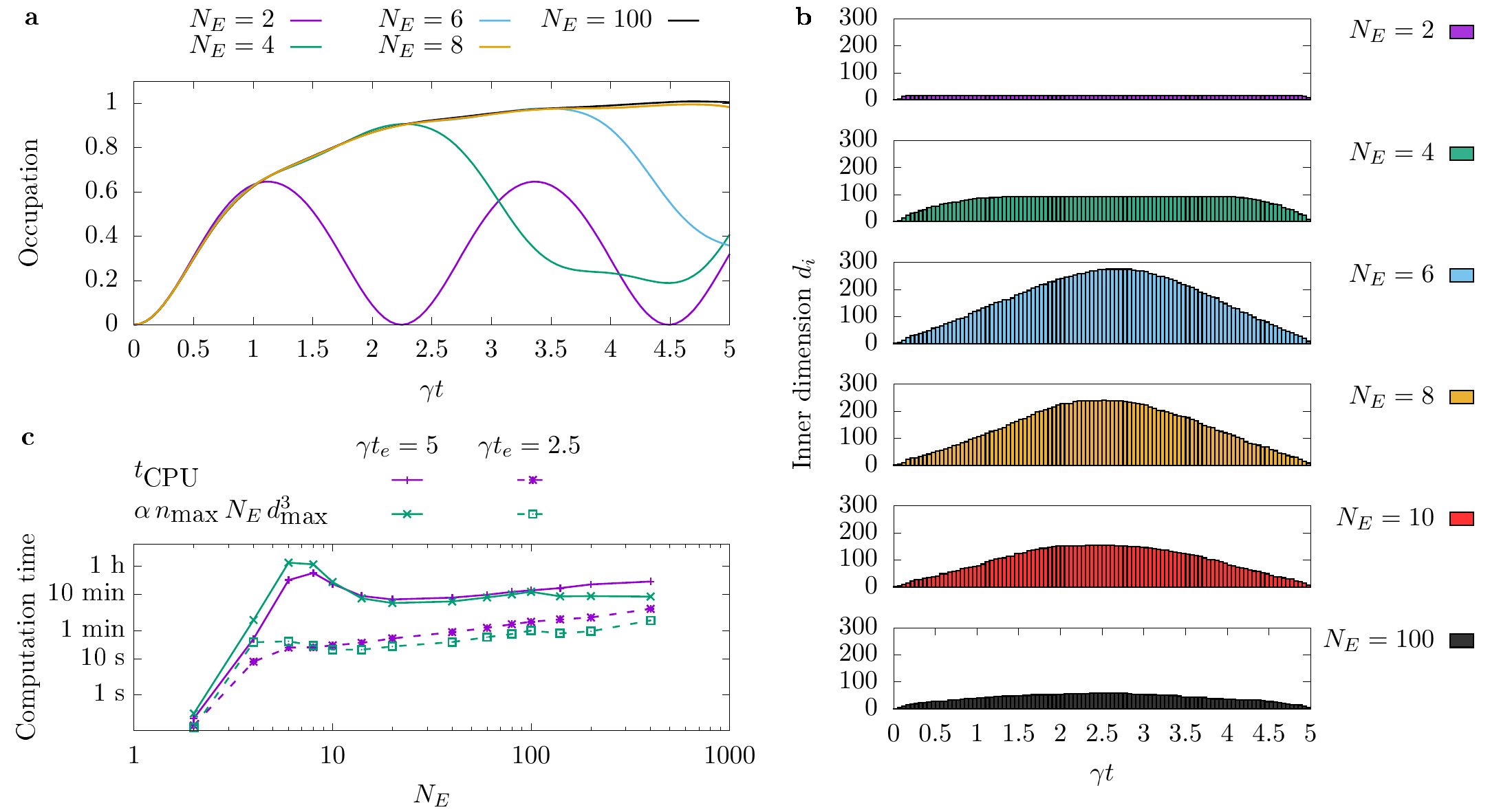}
\caption{\label{fig:vsN}Convergence of ACE simulations with respect to the
number of modes $N_E$ used to discretise a continuum. Panel \textbf{(a)} compares the time dependence of occupation for different values of $N_E$.  Panel \textbf{(b)} shows how $N_E$ affects the inner dimensions of the MPO, while panel \textbf{(c)} shows how this affects the computation time required.
}
\end{centering}
\end{figure}

Finally, we numerically investigate the convergence of the PT 
with respect to the mode discretisation of a continuum of environment 
modes. Figure~\ref{fig:vsN}\textbf{a} shows the time evolution of the 
system occupation for the same Fermionic open quantum system as discussed
above.  Results are shown up to final time $\gamma t_e=5$  using a fixed compression threshold 
$\epsilon=10^{-6}$ and a fixed time step width $\gamma\Delta t=0.05$, 
comparing the results for different numbers of environment modes $N_E$ spanning the total band with $\omega_{BW}=10 \gamma$. 
For a large number of modes $N_E=100$ the results shown 
in the main text are reproduced, i.e. system occupation grows roughly as predicted by the Markovian limit $1-e^{-\gamma t}$, albeit with visible deviations. However, if the continuum discretisation is
too coarse as in the case $N_E=2$, the description is only accurate for
a short time, after which strong deviations occur. For small $N_E$ this is inevitable, because there is a limited set of degrees of freedom, corresponding to a limited set of frequencies controlling the dynamics of the system. 
With increasing number of modes $N_E$, the 
time at which these strong deviations appear becomes later.
Here, for $N_E=8$ this point is almost beyond the final time $t_e$, 
so a good description of environment effects is retained for the full 
simulation. 

The mode discretisation has a significant impact on the 
structure of the process tensor. In  Fig.~\ref{fig:vsN}\textbf{b},
the inner dimension $d_i$ at every time step $t_i$ is depicted. 
The typical shape of the $d_i$ distribution is roughly trapezoidal;  this reflects the constraint in the bond dimension near the ends of the MPO, so that the maximum inner dimension $d_\textnormal{max}$ occurs near the centre of the MPO.
Notably, we find that increasing the number of modes leads first to an increase in $d_\textnormal{max}$ up to a certain $N_E$ after which $d_\textnormal{max}$ 
begins to decrease. The initial increase can be explained by the PT 
including more and more degrees of freedom. The eventual decrease is due to dephasing between modes with similar frequencies.
This result is consistent with recent results by \citet{ye2021constructing}.
The largest values of $d_\textnormal{max}$ are found at similar values of $N_E$ to the conditions where convergence of the result with $N_E$ is first reached, as seen in Fig.~\ref{fig:vsN}\textbf{a}.

The dependence of the inner dimensions of the PT on the mode discretisation
has a significant impact on the computational resources required for the
ACE method. Figure~\ref{fig:vsN}\textbf{c} shows the CPU time 
needed for carrying out the ACE simulation for two different final times
$\gamma t_e=5$ and $\gamma t_e=2.5$. The computation time is found to
increase rapidly with $N_E$ when $N_E$ is small, but then reach an approximate plateau at large $N_E$. (The required computation time can even have a minimum, as seen around $N_E=20$ in the case $\gamma t_e=5$.) 
These trends can be explained 
by a simple scaling argument: Exact SVD routines scale as $d^3$ where $d$ is the
matrix dimension. The number of SVDs performed in total is
proportional to the number of time steps $n_\textnormal{max}$ times the 
number of environment modes $N_E$. Using the maximal inner dimension 
$d_\textnormal{max}$ as a proxy for the typical dimension, one expects 
the computation time to scale as
$t_\textnormal{estimate}=\alpha\, n_\textnormal{max} N_E d_\textnormal{max}^3$.
With $d_\textnormal{max}$ extracted from the simulations, 
$t_\textnormal{estimate}$
is fit against the CPU times of the curve for $\gamma t_e=5$.
This is depicted in Fig.~\ref{fig:vsN}\textbf{c}. We find that this formula, with a constant $\alpha$, indeed captures the trends in the computation time well.

\section{Run times and comparison to Gaussian methods}
\setcounter{figure}{0}

\subsection{Run time for examples provided}

\begin{table}[H]
\newcommand\Tstrut{\rule{0mm}{3.5mm}}  
\newcommand\Bstrut{\rule[-2mm]{0mm}{0mm}}  
\newcommand{\tabtitle}[1]{\underline{\emph{#1}} & \phantom{.} \Tstrut\Bstrut\\}
\small
\begin{tabularx}{0.49\textwidth}[t]{|X|r|}
\hline
Example  & Run time \\
\hline \hline 
\tabtitle{Resonant level model:}
$N_E=2$   & $<$ 1 s \\
$N_E=4$   &  4 s \\
$N_E=10$  & 56 s \\
$N_E=100$ & 13 h 3 min \\
\hline
\hline
\tabtitle{Phonons \& photons:}
Construct PT phonons  & 1 h 14 min \\
Construct PT photons, $\omega_{BW}=10$ ps$^{-1}$  & 10 h 47 min \\
Construct PT photons, $\omega_{BW}=0.4$ ps$^{-1}$  & 17 s \\
Contraction of PT (combined) & 40 s  \\
{}[iQUAPI: phonons] & [1 min 0 s]  \\
\hline
\hline
\tabtitle{Spins, fully polarised:}
$N=10$, $\epsilon=10^{-10}$  & 1 min 10 s \\
$N=100$, $\epsilon=10^{-10}$ & 4 min 09 s\\
$N=1000$, $\epsilon=10^{-10}$ & 13 min 50 s\\
\tabtitle{Spins, partially polarised:}
$N=10$, $\epsilon=10^{-10}$ & 3 min 32 s  \\
$N=10$, $\epsilon=10^{-13}$ & 19 min 59 s\\
$N=10$, $\epsilon=10^{-16}$ & 1 h 45 min\\
$N=100$, $\epsilon=10^{-10}$ & 5 min 30 s\\
$N=100$, $\epsilon=10^{-13}$ & 33 min 12 s\\
$N=100$, $\epsilon=10^{-16}$ & 2 h 28 min \\
\tabtitle{Spins, unpolarised:}
$N=10$, $\epsilon=10^{-10}$  & 5 min 21 s\\
$N=10$, $\epsilon=10^{-13}$  & 34 min 2 s\\
$N=10$, $\epsilon=10^{-16}$  & 4 h 40 min \\
$N=100$, $\epsilon=10^{-10}$ & 8 min 27 s \\
$N=100$, $\epsilon=10^{-13}$ & 40 min 52 s\\
$N=100$, $\epsilon=10^{-16}$ & 3 h 8 min \\
\hline
\end{tabularx}\hfill\begin{tabularx}{0.49\textwidth}[t]{|X|r|}
\hline
Example  & Run time \\
\hline \hline
\tabtitle{Morse potential:}
SBM, $M=5$  & 20 min 17 s\\
HO, $M=5$   & 23 min 30 s\\
$\Lambda=2$, $M=2$  & 3 min 39 s\\ 
$\Lambda=2$, $M=2$, renorm. & 3 min 48 s\\ 
$\Lambda=3$, $M=3$  & 22 min 39 s \\
$\Lambda=3$, $M=3$, renorm.  & 21 min 58 s \\
$\Lambda=4$, $M=4$  & 1 h 37 min \\
$\Lambda=4$, $M=4$, renorm.  & 1 h 36 min \\
$\Lambda=5$, $M=5$  & 4 h 4 min \\
$\Lambda=5$, $M=5$, renorm.  & 3 h 48 min \\
$\Lambda=10$, $M=5$  & 1 h 6 min \\
$\Lambda=10$, $M=5$, renorm.  & 1 h 7 min\\
$\Lambda=100$, $M=5$  & 33 min 15 s\\
$\Lambda=100$, $M=5$, renorm.  & 32 min 15 s \\
\hline
\hline
\tabtitle{Superradiance:}
Construct PT  & 31 min 39 s\\ 
Contraction of PT & 2 s \\
\hline
\hline
\tabtitle{Dispersive coupling:}
single mode  & 7 s\\
instant. Fock & 2 s \\
pulsed, no losses & 23 min 36 s \\
pulsed, with losses  & 13 min 2 s \\
\hline 
\end{tabularx}
\caption{\label{tab:runtimes}Run times for the examples
discussed in the main text and the Supplementary Material.  In some examples (``phonons \& photons'' and ``superradiance''), we use the fact that the construction of the process tensor (``Construct PT'') using ACE and the subsequent contraction to determine time evolution can be separated.  This separation is useful when one environment is used multiple times with different system Hamiltonians.
}
\end{table}

The simulations for this article are performed on a conventional laptop computer with Intel i5-8265U processor and 16 GB of RAM. 
The ACE code is available at Ref.~\cite{aceCode}.
The numerically most demanding part, the MPO compression using SVDs, is done using the \texttt{JacobiSVD} routine provided by the Eigen library (version 3.4-rc1), which calls the corresponding LAPACK routines when compiled and linked appropriately. Here, we use the LAPACK implementation provided by the Intel MKL (version 2021.3.0). The C++ code is compiled and linked using the GCC compiler (version 9.3.0).

The run times of the simulations for the examples discussed in the main text as well as in other sections of the Supplemental Material are listed in Tab.~\ref{tab:runtimes}. 
As can be seen, typical calculation times for these examples range between minutes to several hours, demonstrating the efficiency and practicability of ACE over a broad range of different physical systems. The challenging simulations of spin baths with tiny thresholds $\epsilon=10^{-16}$ require more than the physical 16 GB of RAM, and so the times observed here are affected by swapping to disk. Note that swapping is efficient for ACE simulations because a single MPO compression step only modifies a single MPO matrix at a time.

\newpage

\subsection{Comparison with numerically exact methods for Gaussian baths}

\begin{figure}
\includegraphics[width=0.99\textwidth]{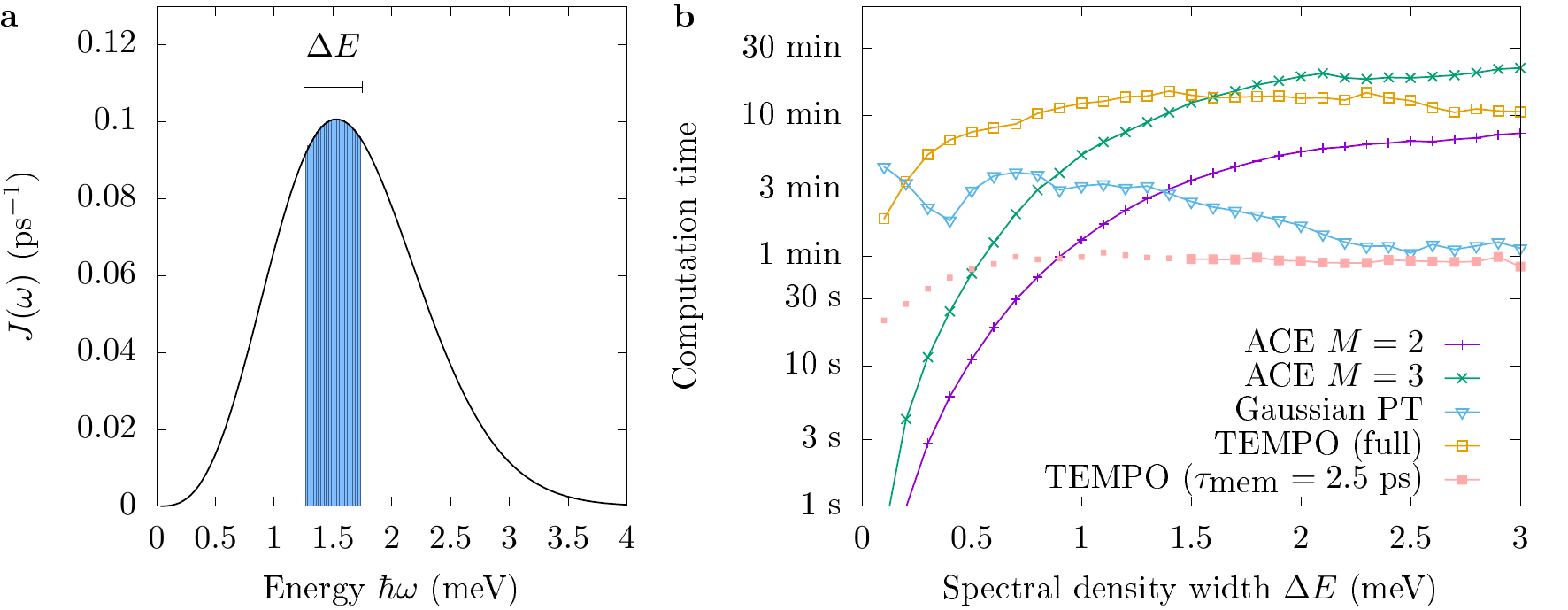}
\caption{\label{fig:ACE_vs_Gauss} 
Comparison of different numerically exact 
methods for Gaussian environments: 
ACE, Gaussian PT calculation, and 
TEMPO for the example of an off-resonantly driven quantum dot coupled to 
a bath of phonons. 
To control the non-Markovianity of the bath, the spectral density is
manually restricted to a finite spectral range $\Delta E$ centred around
the energy $\hbar\delta=1.5$ meV corresponding to the detuning of the 
excitation, as depicted in panel \textbf{a}.
The corresponding computation times are shown in panel \textbf{b}.
}
\end{figure}

In the special case of Gaussian environments, other numerically exact methods have been established such as
the calculation of PTs for Gaussian baths devised by J{\o}rgensen and Pollock~\cite{ProcessTensor} building on the TEMPO algorithm~\cite{TEMPO}, which itself is a reformulation of the iterative path integral approach iQUAPI~\cite{Makri}.  In such methods, the augmented density matrix is presented, compressed, and propagated in MPO representation.
An implementation of these two methods is incorporated into our ACE computer code~\cite{aceCode}.

These methods rely on the fact that for Gaussian baths, the path integral over the environment can be solved analytically, giving explicit expressions for the Feynman-Vernon influence functional. 
As all these methods are numerically exact, it is an interesting question which method performs best (requires least computation time) in which situation---restricting to Gaussian cases where all methods are available. 
To this end, we consider again the example of the quantum dot coupled to phonons and subject to radiative losses discussed in the main text,  focusing on the numerically exact modelling of phonon effects while losses are accounted for by Lindblad terms. 
Anticipating that the performance of the different methods strongly depends on the memory time of the environment, we perform calculations for various widths of the spectral density. 
As depicted in Fig.~\ref{fig:ACE_vs_Gauss}\textbf{a}, we use the envelope of the spectral density as in the main text, but restrict it to a finite support of $\Delta E$ centred around $\hbar\delta=1.5$ meV, 
corresponding to the detuning of the excitation from the quantum dot transition.
For large values of $\Delta E\approx$ 3 meV, the memory time of the environment is shortest, while in the limiting case for small widths $\Delta E\to 0$ only a single environment mode energy exists and the memory time becomes infinite. 

In all methods, we use a time step width of $\Delta t=0.1$ ps, compression threshold $\epsilon=10^{-7}$ and initial bath temperature $T=0$ K. For ACE, we discretise the continuum $\Delta E$ on $N_E$ intervals with a density of states $N_E/\Delta E=20$ meV$^{-1}$ and we truncate the environment Hilbert space per Boson mode to dimension $M=2$ or $M=3$.
For this set of parameters, the typical relative difference between exciton populations in the different methods is $\approx 2 \times 10^{-3}$. 

In Fig.~\ref{fig:ACE_vs_Gauss}\textbf{b}, the computation times needed for the different methods are depicted as a function of the spectral density width $\Delta E$. 
For large widths $\Delta E\gtrsim 1.4$ meV, we find the Gaussian PT calculation to be faster than both ACE and TEMPO without memory truncation. Note, however, that TEMPO can benefit significantly from memory truncation, as this reduces the length of the MPO chain to be compressed and propagated. Fixing the memory time to $\tau_\textrm{mem}=2.5$ ps in TEMPO leads to the fastest results of all considered methods in the regime of large $\Delta E$. Yet, for smaller widths $\Delta E \lesssim 1.5$ meV, the bath memory time starts to exceed 2.5 ps, which leads to visible deviations in the occupations (not shown). 
The smaller (disconnected) pink dots in Fig.~\ref{fig:ACE_vs_Gauss}\textbf{b} indicate data points where the relative error with respect to Gaussian PT calculations exceeds $1\%$. 
For narrow spectral densities, we observe a cross-over rendering ACE faster than all other methods. As shown, the required dimension $M$ per boson mode has a large influence on the run time for ACE.

To summarize, in the special case of Gaussian baths, alternative methods can benefit from the existence of analytical expressions for the influence functional describing the environment, and therefore perform faster than ACE in cases where broad spectral densities lead to short memory times. On the other hand, for narrowly peaked spectral densities that can be well described in terms of a few environment modes, the general method ACE can even outperform specialised methods for Gaussian baths. 
Finally, it is noteworthy that our computer code produces Gaussian PTs that are completely compatible with the PTs utilised by ACE, paving the way for prospective hybrid approaches for spectral densities with sharp peaks on top of broad continua.

\newpage
\section{Environment modes with anharmonic potentials}
\setcounter{figure}{0}
In the main text, we present ACE simulations for an open quantum system coupled to a bath of anharmonic environment modes whose free evolution is governed by the Morse potential. 
Here, we lay out the full numerical treatment starting from the Schr{\"o}dinger equation of a single environment mode for an arbitrary potential $V(r)$ directly from a numerical representation of the potential on a real space grid. 

\subsection{Finite differences to find environment states}
We start from the one-dimensional  Schr{\"o}dinger equation for a given potential $V(r)$:
\begin{align}
H=& -\frac{\hbar^2}{2m}  \frac{\partial^2}{\partial r^2} + V(r).
\end{align}
We first map this onto a dimensionless ordinary differential equation by introducing a characteristic length scale $a_0$
and energy scale $\epsilon=\hbar^2/(2m a^2_0)$, and defining the dimensionless coordinate $x=r/a_0$.
We then define the dimensionless Hamiltonian
\begin{align}
\label{eq:h}
h:=&\frac 1\epsilon H = -\frac{\partial^2}{\partial x^2} + v(x),\\
v(x):=&\frac 1\epsilon V(a_0 x),
\end{align}
where $v(x)$ is the dimensionless potential.
The dimensionless problem is solved by a finite differences method, where a real space grid $x_j= x_0 + j \Delta x$ with width $\Delta x$ and $N_x$ sample points is introduced and the second derivative is approximated by
\begin{align}
 \frac{\partial^2}{\partial x^2} f(x_i)= 
\frac{ f(x_{i-1}) - 2 f(x_i) + f(x_{i+1})}{\Delta x^2}.
\end{align}
The ODE  in Eq.~\eqref{eq:h} then takes the form of a  symmetric tridiagonal matrix, which is diagonalized numerically.

For simulations in ACE, we work in the truncated energy eigenbasis 
accounting for only the $M$ lowest energy eigenstates of a given mode.
The energy eigenvalues $E_i$ of the original problem are obtained by multiplying
the eigenvalues of $h$ with $\epsilon$.
The operators describing the system-environment coupling are evaluated in 
the truncated basis depending on the concrete details of the model.
For example, if the system couples to the environment modes via the 
position operator $\hat{r}$, one has to numerically evaluate matrix elements
$\langle i| \hat{r} |j\rangle=a_0 \langle i| \hat{x}|j\rangle$
with $i,j\in 0,1,\dots,M-1$.

\subsection{Example: Harmonic oscillator}
As a reference, we first consider the example of the harmonic oscillator
potential  $V(r)=\frac{m\omega^2}2 r^2$. 
Defining length and energy scales $a_0=\sqrt{\frac{\hbar}{m\omega}}$ and 
$\epsilon=\frac{\hbar\omega}2$, the corresponding dimensionless 
Schr{\"o}dinger equation is
\begin{align}
h:=&-\frac{\partial^2}{\partial x^2} + x^2.
\end{align}

Back-transforming the numerically obtained eigenvalues of $h:$ $1,3,5,\dots$, by multiplying with $\epsilon$, one recovers the series $E_n=\hbar\omega\big(n+\frac 12\big)$ with $n=0,1,2,\dots$.
From the conventional definition of the harmonic oscillator climbing operators it follows that
$a_0\hat{x}=\hat{r}=\sqrt{\frac{\hbar}{2m\omega}}\big(a^\dagger + a\big)=\frac{a_0}{\sqrt{2}}\big(a^\dagger + a\big)$. 
Consequently, to enable a comparison with the spin Boson model, we consider an environment Hamiltonian for ACE simulations of the form 
\begin{align}
H_E=&\sum_k \hbar \omega_k \left(a^\dagger_k a_k+\frac 12\right)+ 
\sum_k \hbar g_k (a^\dagger_k +a_k)|e\rangle\langle e|
\nonumber\\=&
\sum_k\sum_{j=0}^{M-1}  \hbar \omega_k  \frac{E_j}{2\epsilon}
\sigma^k_{jj} + 
\sum_k \hbar g_k   \sum_{i,j=0}^{M-1}
\Big(\sqrt{2}\langle i|\hat{x}|j\rangle\Big)
\sigma^k_{ij}\, |e\rangle\langle e|,
\label{eq:HEHO}
\end{align}
where $\sigma^k_{ij}$ describes the effect of the single-particle operator $|i\rangle\langle j|$ for the $k$-th environment mode.
As shown in Fig.~5 of the main text, this procedure perfectly reproduces the results of the spin Boson model.

\subsection{Example: Morse potential}
The Morse potential~\cite{Morse1936} is an asymmetric anharmonic potential with a finite
number of bound states below a continuum of unconfined states. It
is often used to describe molecular vibrations with a finite dissociation 
energy~\cite{Morse_deVega}. It takes the form
\begin{align}
V(r)=&D_e \Big( e^{-2(r-r_e)/a_0 } - 2e^{-(r-r_e)/a_0}\Big),
\end{align}
where $D_e$ is the well depth, $r_e$ is the position of the minimum of the
potential, and $a_0$ defines its spatial extent. 
Here, we use $a_0$ as the length scale and shift the coordinate system
such that $r_e=0$. The Morse potential is made dimensionless
\begin{align}
v(x)=&\Lambda^2\Big( e^{-2x} - 2e^{-x}\Big)
\end{align}
by introducing the parameter
$\Lambda=\sqrt{{D_e}/{\epsilon}}=\sqrt{{2 m a_0^2 D_e}/{\hbar^2}}$.
The Morse potential is known to have $M$ bound states~\cite{Morse1936}, where $M$ is the largest integer smaller than $\Lambda+\frac12$, with energies
\begin{align}
E_n=-\epsilon \Big(\Lambda - n -\frac 12\Big)^2=
\epsilon\bigg[-\Lambda^2 +2\Lambda \Big(n +\frac12\Big) 
- \Big(n +\frac12\Big)^2 \bigg].
\end{align}
For deep potentials $\Lambda\to\infty$, the spectrum of the lowest states
becomes equivalent to that of a harmonic oscillator with 
$\hbar\omega=2\epsilon \Lambda$, which is consistent with the second-order 
Taylor expansion around $r=r_e$  being
$V(r) \approx - D_e + \frac{m\omega^2}{2}(r-r_e)^2 $.
For general $\Lambda$, the level spacings between confined states are
\begin{align}
\Delta E_n=E_{n+1}-E_n
=\hbar\omega \Big( 1-\frac{n+1}{\Lambda}\Big).
\end{align}
The energy difference between first excited state and ground state is
$\Delta E_g = \hbar\omega \big(1-\Lambda^{-1}\big)=
\sqrt{\frac{2\hbar^2 D_e}{ma_0^2}} - \frac{\hbar^2}{ma_0^2}$.

In Fig.~5\textbf{a} in the main article, the five bound eigenstates of 
the Morse potential with $\Lambda=5$ obtained from numerical finite-differences 
calculations are depicted.
In contrast to harmonic oscillator wave functions, the anharmonicity of 
the Morse potential manifests itself in the decreasing level spacings for higher states. 
Furthermore, the wavefunctions are strongly asymmetric leading to nonzero
values of the average position operator $\langle i| \hat{x}|i\rangle$ for
the $i$-th state.  This non-zero expectation has a significant impact on the system-environment 
coupling.

Note that the matrix element of the dimensionless position operator 
$\hat{x}$ between subsequent eigenstates of the Morse potential behaves as
$\langle i+1|\hat{x}|i\rangle \to \sqrt{i+1}/\sqrt{2\Lambda}$ for 
$\Lambda \to \infty$, so that a situation comparable with the independent-boson
model in this limit requires an environment Hamiltonian of the form
\begin{align}
\label{eq:HEmorse}
H_E=& \sum_k\sum_j\hbar\omega_k \tilde{E}_j \sigma^k_{jj}+ 
\sum_k \hbar g_k   \sum_{i,j=0}^{M-1}
\Big(\sqrt{2}\langle i|\tilde{x}|j\rangle\Big)
\sigma^k_{ij} \,  |e\rangle\langle e|,
\end{align}
with $\tilde{E}_j=E_j/\Delta E_g$ and $\langle i|\tilde{x}|j\rangle=\sqrt{\Lambda}\langle i|\hat{x}|j\rangle$.

With nonzero diagonals $\langle i|\hat{x}|i\rangle$ for finite $\Lambda$, the asymmetry of the potential leads to the additional effect of a renormalisation of the system excited state energy by a value of
$\Delta E=\sum_k \hbar g_k \sqrt{2\Lambda}\, \langle \hat{x}\rangle_E^k=\sum_k \hbar g_k \sqrt{2\Lambda}\,\textrm{Tr}\big(\hat{x}\rho^k_E\big)$, which depends on the state of the environment.

\subsection{Details of the calculation and parameters}
For the ACE simulation depicted in Fig.~5 in the main article, we consider a continuously driven two-level system with system Hamiltonian 
$H_S=\tfrac{\hbar}{2}\Omega \big(|e\rangle\langle g|+|g\rangle\langle e|\big)$.
The environment is described by the Hamiltonian $H_E$ in Eq.~\eqref{eq:HEmorse} with parameters $\omega_k$ and $g_k$ sampling a Lorentzian spectral density
\begin{align}
J(\omega)=  C \frac{1}{\pi}\frac{\gamma}{(\omega-\omega_c)^2+\gamma^2}.
\end{align}
We do this by discretising $\omega_k/\Omega$ equidistantly in the range $[0,7.5]$ with $N_E=100$ modes, and setting $g_k= \sqrt{J(\omega_k) \Delta \omega}$, where $\Delta\omega$ is the distance between subsequent $\omega_k$ sample points.
Here, we set $\hbar=1$, and fix the parameters of the spectral density to $C=0.1 \Omega^2$, $\gamma=0.1 \Omega$, and $\omega_c=\Omega$. The corresponding values $g_k/\Omega$ are plotted in Fig.~5\textbf{b} in the main article. 
The environment modes are initialised with thermal states at temperature $k_B T=0.5 \Omega$. %

\newpage
\section{Superradiance of two quantum emitters}
\setcounter{figure}{0}
\label{app:superrad}
In this section we present an additional illustration of the potential of ACE.
We consider the radiative decay of 
two quantum emitters coherently coupled to the same radiation
field as depicted in Fig.~\ref{fig:superrad}a.
If the distance $d$ between the emitters is much smaller
than the wavelength $\lambda$ associated with the
fundamental transitions of the emitters, both emitters couple with 
the same phase to the radiation field. This gives the Hamiltonian of photon mode $k$ as
\begin{align}
H_E^k=&\hbar \omega_k a^\dagger_k a_k
+ \hbar g_k \Big[
a^\dagger_k \big(|g_1\rangle\langle e_1|+|g_2\rangle\langle e_2|\big)
+ h.c.\Big],
\label{eq:Hsup}
\end{align}
where $|g_i\rangle$ and $|e_i\rangle$ denote the ground and excited state
of emitter $i=1,2$ and $a_k^\dagger$ creates a photon in mode $k$.
In analogy to the first example in the main article, our environment is a discretised quasi-continuum of 
electromagnetic modes with a density of states which would correspond to keeping the Fermi's golden rule rate 
for the decay of a single emitter, $\kappa$, fixed.
We admit in general a detuning $\delta$ between
the transitions of the two emitters, which enters the system Hamiltonian
$H_S=\frac{\hbar\delta}2\big(|e_1\rangle\langle e_1|-|e_2\rangle\langle e_2|\big)$.
The initial conditions are chosen such that both emitters are excited 
at time $t=0$.
\begin{figure}[h]
\includegraphics[width=0.7\linewidth]{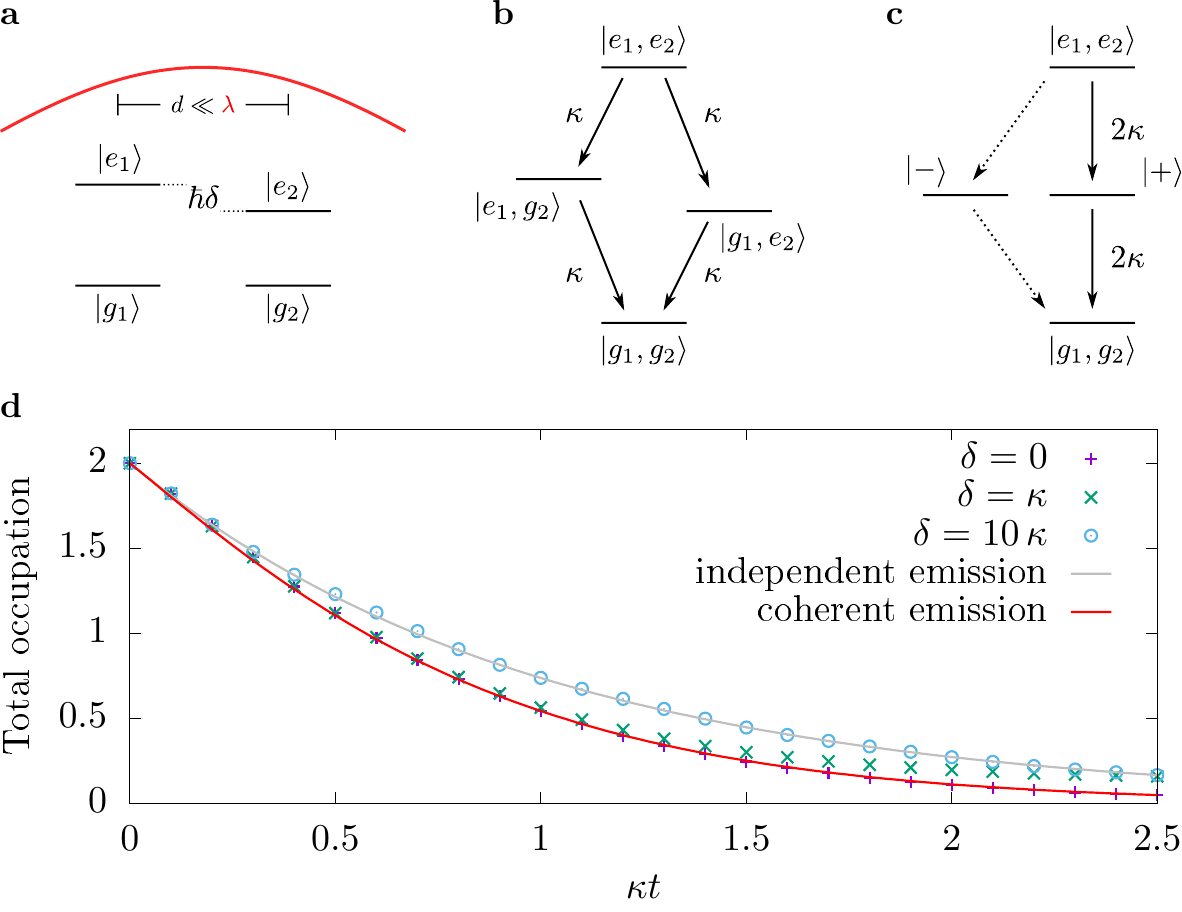}
\caption{\label{fig:superrad}
\textbf{Transition between independent and superradiant emission of two proximal optical dipoles.}
\textbf{a,} Two quantum emitters detuned by an energy $\hbar\delta$ and 
separated by a distance $d$ much smaller than the wavelength $\lambda$ 
corresponding to the emitter transitions. 
\textbf{b,} Radiative decay of two independent emitters.
\textbf{c,} Superradiance of two coherently coupled emitters. The transition
through the symmetric state $|+\rangle$ has twice the rate compared to
that of independent emitters whereas the transition through the antisymmetric
state $|-\rangle$ is forbidden.
\textbf{d,} ACE simulations for different detunings $\delta$ and analytic results
for independent emission and for coherent emission in the superradiant 
regime of two emitters.
}
\end{figure}

This situation is interesting as it constitutes a minimal setup for 
superradiance: If the emitters are distinguishable, e.g., if the detuning
$\delta$ is large, both emitters radiatively decay with a rate $\kappa$,
as depicted in Fig.~\ref{fig:superrad}b, so that the sum of the occupations
decays as $2\exp(-\kappa t)$. If, however, the emitters are indistinguishable
$\delta=0$, the coherent coupling makes it necessary to derive the 
respective decay rates in the symmetrised basis including the states
$|\pm\rangle=\big(|e_1,g_2\rangle\pm|g_1,e_2\rangle\big)/\sqrt{2}$.
The dipole for transitions involving the symmetrised state $|+\rangle$ is 
larger than that of a single emitter, whereas it is zero of the 
antisymmetrised state $|-\rangle$. Consequently, $|-\rangle$ is dark and
the decay takes place from $|e_1,e_2\rangle$ to $|+\rangle$ and 
from $|+\rangle$ to $|g_1,g_2\rangle$ with the rate $2\kappa$, 
as depicted in Fig.~\ref{fig:superrad}c.
Taking into account the dynamics of the intermediate state occupations,
the total occupation of the two indistinguishable emitters 
is $n_\textrm{tot}=2(1+\kappa t)\exp(-2\kappa t)$~\cite{suppl_Superradiance_Haroche}.

Figure \ref{fig:superrad}d shows the total occupation of the two emitters
for $\delta=0$, $\delta=\kappa$, and $\delta=10\kappa$ 
obtained using the ACE method as well as the 
analytic results corresponding to the case of distinguishable and 
indistinguishable emitters. For $\delta=10\kappa$, 
the ACE simulation agrees with the exact result for independent emitters, 
while for $\delta=0$ the result for coherently coupled 
indistinguishable emitters is reproduced.
In the intermediate regime $\delta=\kappa$, the dynamics can be 
understood qualitatively by interpreting $\delta$ as a perturbation
facilitating a rotation from the symmetric $|+\rangle$ to the antisymmetric
$|-\rangle$ state. As the latter is dark, the total occupation at long times
is found to be even slower than the decay of independent emitters.

This example demonstrates that in a situation
where rate equations crucially depend on the basis in which
they are derived, the ACE reproduces correct results 
independent of the basis. Thus, even in Markovian scenarios, 
ACE simulations can have an advantage over conventional techniques in that 
it can be applied straightforwardly in an arbitrary basis.

On the technical side, we have solved the dynamics of a four-level system, 
showing that the method is not restricted two-level systems. 
To achieve this, we have made use
of the fact that the matrices 
$\mathcal{Q}^{(\alpha_l,\tilde{\alpha}_l)}_{d_ld_{l-1}}$
are identical for some combinations of $(\alpha_l,\tilde{\alpha}_l)$. 
This can be done by analogy with the method devised for iQUAPI~\cite{suppl_PI_cQED}, using a decomposition into groups with identical couplings.
With this, we only compute non-redundant values, which reduces the 256 possible
combinations of $(\alpha_l,\tilde{\alpha}_l)$ to 18.
These groups can be identified numerically from the specified 
environment Hamiltonians $H_E^k$, so that this step is also automated.

\newpage
\section{Dispersive coupling}
\setcounter{figure}{0}
Here, we consider a toy model of a TLS dispersively coupled to a multi-mode 
microcavity, in order to demonstrate several remaining aspects of the 
generality of ACE
that were not covered in previous examples.  Specifically:
non-Gaussian interactions 
due to non-linear system-environment coupling,
time-dependent driving of the environment, and non-unitary evolution of the environment modes.
 
A common situation in which non-Gaussian environments emerge is 
when the coupling to the environment is non-linear in environment mode
creation and annihilation operators. The simplest case is that of quadratic
system-environment coupling, as in the case of dispersive coupling 
described by an interaction Hamiltonian
$H_I = \sum_k \hbar g_k a_k^\dagger a_k \sigma_z$. 
Such a coupling arises, e.g., in an effective description of a 
two-level system (TLS) coupled to a microcavity in the limit where
the detuning between TLS and cavity is much larger than the coupling strength~\cite{Blais2004}.
The main effect of dispersive coupling is that the TLS transition energy
experiences a shift depending on the cavity photon number and, vice versa,
the cavity mode energy is modified by the excited state population $n_e$ of
the TLS. The former paves the way for quantum non-demolition measurement
of the cavity photons by probing the TLS~\cite{Blais2004}.

\begin{figure}
\begin{centering}
\includegraphics[width=0.95\textwidth]{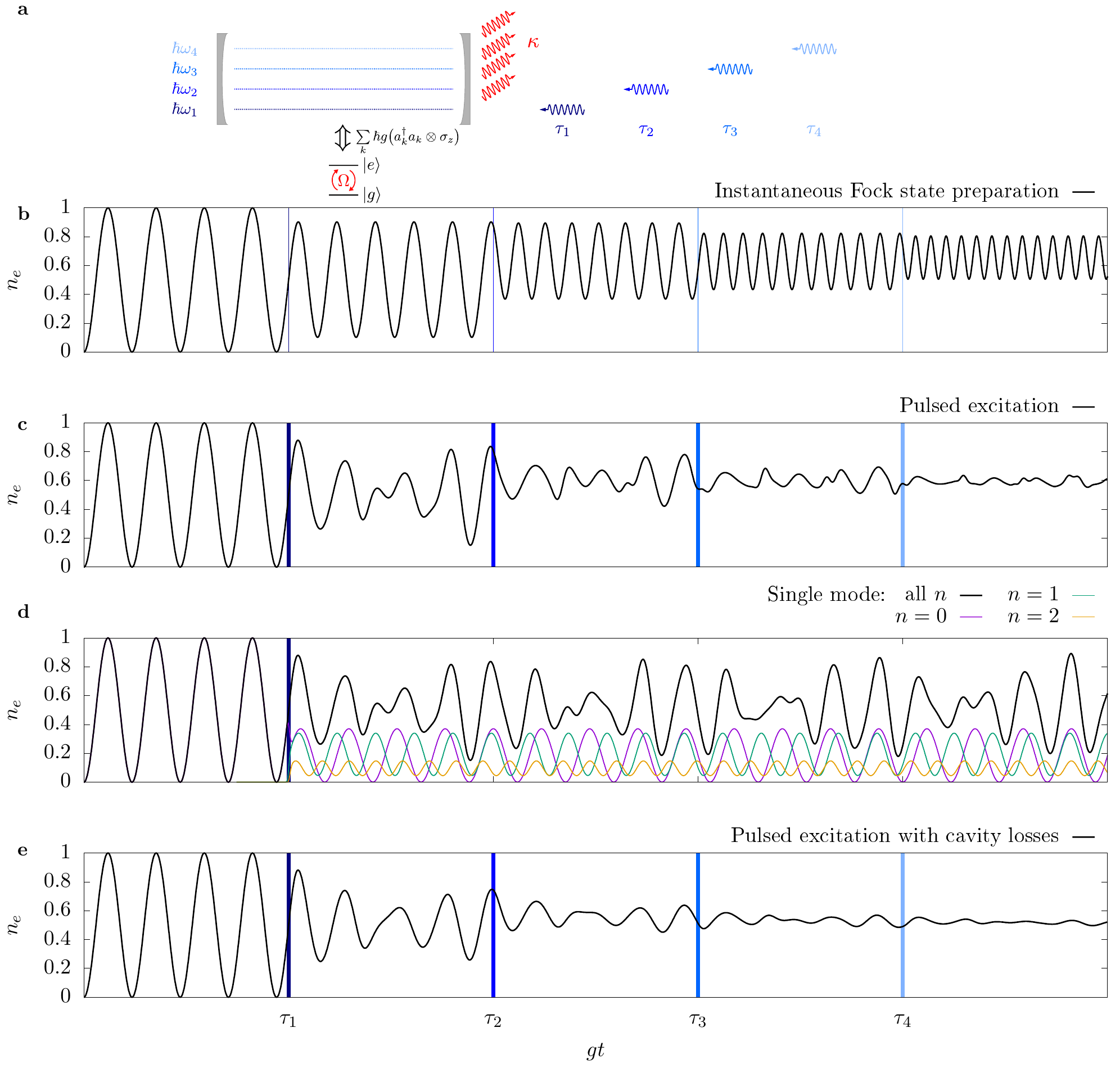}
\caption{\label{fig:dispersive} \textbf{a}: Sketch of a TLS dispersively coupled to a multi-mode microcavity, resulting in changes of the effective transition frequency of the TLS when the cavity modes are driven externally by pulses arriving at times $\tau_k$. 
The TLS itself is continuously driven with bare Rabi frequency $\Omega$, allowing one to detect signatures brought about by changes of the transition frequency. 
Photons are lost from the cavity modes with rate $\kappa$ (but $\kappa=0$ for panels \textbf{b}-\textbf{d})
\textbf{b}:
Evolution of the excited state population $n_e$ when the pulses exciting the
cavity modes are replaced by an instantaneous change of the state of the $i$-th
cavity mode from the vacuum to the one-photon Fock state at time $\tau_i$ indicated by blue vertical lines.
\textbf{c}: 
Excited state population when the cavity modes are excited by Gaussian pulses.
\textbf{d}:
Numerical simulation accounting for a single cavity mode as part of the system.
The total TLS excitation is presented as well as the photon-number-resolved
TLS excitations. %
\textbf{e}:
Like \textbf{c} but additionally accounting for non-zero photon loss rate $\kappa=0.1g$.}
\end{centering}
\end{figure}

We consider the setting depicted in Fig.~\ref{fig:dispersive}\textbf{a}: 
A TLS is coupled to a microcavity that supports multiple discrete photon
modes which can be individually addressed by external driving. 
The cavity modes are assumed to be detuned far enough from the fundamental
TLS transition that the TLS-cavity coupling is well described by 
a dispersive interaction. 
The initially empty cavity modes are then driven one-by-one by 
short external pulses centred around times $\tau_k$, $k=1,2,3,4$.
As a result of the dispersive coupling, this leads to a shift of the
TLS transition energy, which is probed by driving the TLS directly 
and continuously with a driving field (driving strength $\Omega$)
that is resonant with the bare TLS transition frequency. 

To this end, we apply ACE to the  Hamiltonian
$H=H_S+\sum_k H_E^k$ with
\begin{align}
H_S=&\frac{\hbar}2\Omega \big(|e\rangle\langle g| + |g\rangle\langle e|\big),\\
H_E^k=& \hbar g a^\dagger_k a_k \sigma_z
+ \hbar \omega_k a^\dagger_k a_k 
+  \frac \hbar 2 G_k(t) 
\big( a^\dagger_k e^{-i\omega_k t} + a_k e^{i\omega_k t}\big),
\end{align}
using model parameters  $\Omega/g={8.5 \pi}/{10}$, and $\omega_k/g=10+k$, 
as well as convergence parameters
$g \Delta t=0.01$ and $\epsilon=10^{-9}$ accounting for up to four 
bosons per cavity mode.

In a first step, to aid understanding of the general case, instead of modelling the cavity mode excitation explicitly 
by pulses with envelope $G_k(t)$, we consider the instantaneous preparation of one-photon 
Fock states. This corresponds to applying creation operators $a^\dagger_k$ to the 
forward propagating part of the environment mode Liouville propagator, and the annihilation operators $a_k$ to the backward propagating part at time
$g\tau_k=10 k$ for the $k$-th mode.
The results are depicted in Fig.~\ref{fig:dispersive}\textbf{b} and indeed
show that whenever a photon is added to a cavity mode, the observed
Rabi oscillations of the TLS become more and more off-resonant as indicated 
by their decreasing amplitude and increasing frequency.

Next, we consider explicit time-dependent driving of the cavity modes
by Gaussian pulses with 
\begin{align}
G_k(t)= \frac{A_k}{\sqrt{2\pi}\sigma}
\exp\left[-\frac{(t-\tau_k)^2}{2\sigma^2}\right],
\end{align}
with $\sigma=\tau_\textnormal{FWHM}/(2\sqrt{2\ln 2})$,
$g\tau_\textnormal{FWHM}=0.2$, and $A_k=2$.  These parameters are chosen such that
after the $k$-th pulse the cavity photon number 
$\langle a^\dagger_k a_k\rangle\approx 1$.
As can be seen in Fig.~\ref{fig:dispersive}\textbf{c}, the TLS dynamics is now more complicated. 
As seen for instantaneous Fock state preparation, the oscillation amplitudes are reduced when another cavity mode is excited, however the signal now involves more than a single frequency. 
This is due to the fact that the external driving as described
by $H_E^k$ induces coherent states as opposed to one-photon Fock states,
so that now contributions corresponding to Fock states 
with $n=0$, $n=2$, $n=3$, and $n=4$ are also excited with a finite probability. 
The joint state of TLS and cavity is therefore best discussed in terms of
sectors with fixed photon and excitation numbers $n$ and $n_e$, respectively.
This  can be illustrated by considering a single cavity mode, where the total system plus
environment is tractable without compression, so that the total system can be propagated as a single, closed quantum system.  This
provides access to the full state including photon-number-resolved TLS populations $|e, n\rangle$, where $|e\rangle$ refers to the excited state of the TLS and $n$ is the cavity photon number. These are shown in Fig.~\ref{fig:dispersive}\textbf{d}.  This calculation agrees
with the ACE simulations in Fig.~\ref{fig:dispersive}\textbf{c} up to time $\tau_2$, when a second mode becomes involved. This demonstrates that the complicated evolution of the TLS occupation is just a sum of contributions from individual $n$-photon sectors, each evolving with a single, well-defined frequency.

Finally, because the starting point of ACE are the propagators of the environment modes in Liouville space, it is straightforward to include non-unitary evolution of the environment, such as loss terms that directly affect the dynamics of the environment modes. 
Including Lindblad terms 
\begin{align}
\kappa\bigg[ a_k \rho a^\dagger_k -\frac 12 \big(a^\dagger_k a_k \rho 
+\rho a^\dagger_k a_k\big) \bigg]
\end{align}
describing the loss of photons with rate $\kappa=0.1g$ to the environment propagator, one obtains the results depicted in Fig.~\ref{fig:dispersive}\textbf{e}.
While generally very similar to the behaviour in the case without losses shown in Fig.~\ref{fig:dispersive}\textbf{c}, losses with a finite rate $\kappa$ are found to lead to more efficient dephasing as they intermix sectors with different photon numbers.

\bibliography{supplementary_bib}